\documentclass[longauth]{aaEC}
\usepackage{amsmath,amssymb,amsfonts}
\usepackage{graphicx}
\usepackage{txfonts}
\usepackage{enumerate} % advanced enumerate environment
\usepackage{colordvi} % for color texthttps://fr.overleaf.com/project/5f92c102fc740c0001d90f85
\usepackage{bm}% bold math
\usepackage{epsfig}
\usepackage[dvipsnames]{xcolor}
\usepackage{color, colortbl}
\usepackage[T1]{fontenc} % if needed
\usepackage{braket}
\usepackage{placeins}
\usepackage{multirow}
\usepackage{tabularx}
\usepackage{comment}
\usepackage[normalem]{ulem}
\graphicspath{{./Figures/}}
\usepackage[pdfencoding=auto,psdextra]{hyperref}
\hypersetup{
    colorlinks=true,
    linkcolor=blue,
    filecolor=magenta,      
    urlcolor=blue,
    citecolor=blue
}
\usepackage[dvipsnames]{xcolor}
\usepackage{color, colortbl}
\usepackage[T1]{fontenc} % if neededhttps://www.overleaf.com/project/5f92c102fc740c0001d90f85
\usepackage{braket}
\usepackage{placeins}
\usepackage{tabularx}

\usepackage[switch, mathlines]{lineno}
\makeatletter
\renewcommand*\aa@pageof{, page \thepage{} of \pageref*{LastPage}}
\makeatother

\renewcommand{\linenumbers}[0]{}

\usepackage[utf8]{inputenc}

\newcommand{\orcid}[1]{} %% if already defined in aa.cls: comment, or use renewcommand	

\definecolor{amaranth}{rgb}{0.9, 0.17, 0.31}
\definecolor{forestgreen(web)}{rgb}{0.13, 0.55, 0.13}
\definecolor{lavender(web)}{rgb}{0.9, 0.9, 0.98}
\definecolor{cosmiclatte}{rgb}{1.0, 0.97, 0.91}
\definecolor{jonquil}{rgb}{0.98, 0.85, 0.37}
\definecolor{khaki(x11)(lightkhaki)}{rgb}{0.94, 0.9, 0.55}
\definecolor{thistle}{rgb}{0.85, 0.75, 0.85}
\definecolor{cuteblue}{RGB}{66, 135, 245}

\newcommand{\GCsp}{\text{GC}\ensuremath{_\mathrm{sp}}}
\newcommand{\GCph}{\text{GC}\ensuremath{_\mathrm{ph}}}
\newcommand{\Omegam}{\ensuremath{\Omega_{\mathrm{m},0}}}

\newcommand{\OmegaDE}{\ensuremath{\Omega_{\mathrm{DE},0}}}

\newcommand{\lcdm}{\ensuremath{\Lambda\mathrm{CDM}}\xspace}
\newcommand{\logfr}{\ensuremath{\log_{10}(|f_{R0}|)}}
\newcommand{\fr}{\ensuremath{|f_{R0}|}}

\newcommand{\de}{\mathrm{d}}
\newcommand{\XCph}{\text{XC}\ensuremath{_\mathrm{ph}}}

\newcommand{\aB}{\alpha_{\rm B}}
\newcommand{\aM}{\alpha_{\rm M}}
\newcommand{\aT}{\alpha_{\rm T}}
\newcommand{\aK}{\alpha_{\rm K}}

\newcommand{\MPl}{M_{\rm Pl}}

\newcommand{\commentout}[1]{}

\defcitealias{Euclid:2019clj}{EC20}
\defcitealias{Euclid:2025tpw}{MG25}
\defcitealias{EUCLID:KP_nu}{NU25}

\usepackage[nameinlink,capitalise]{cleveref}
\Crefname{section}{Section}{Sections}
\crefname{section}{Sect.}{Sects.}
\Crefname{figure}{Figure}{Figures}
\crefname{figure}{Fig.}{Figs.}
\Crefname{equation}{Equation}{Equations}
\crefname{equation}{Eq.}{Eqs.}

\usepackage{euclid}

\begin{document}

\title{\Euclid preparation}
\subtitle{Review of forecast constraints on dark energy and modified gravity}   

%%%% Version Monday 24th of November 2025 01:13:54 PM UT												
%%%% Please do not edit the author list -- contact ECEB Bureau for changes
\author{Euclid Collaboration: N.~Frusciante\orcid{0000-0002-7375-1230}\thanks{\email{noemi.frusciante@unina.it}}\inst{\ref{aff1}}
\and M.~Martinelli\orcid{0000-0002-6943-7732}\inst{\ref{aff2},\ref{aff3}}
\and L.~Lombriser\inst{\ref{aff4}}
\and A.~Silvestri\orcid{0000-0001-6904-5061}\inst{\ref{aff5}}
\and M.~Archidiacono\orcid{0000-0003-4952-9012}\inst{\ref{aff6},\ref{aff7}}
\and M.~Baldi\orcid{0000-0003-4145-1943}\inst{\ref{aff8},\ref{aff9},\ref{aff10}}
\and M.~Ballardini\orcid{0000-0003-4481-3559}\inst{\ref{aff11},\ref{aff9},\ref{aff12}}
\and N.~Bartolo\inst{\ref{aff13},\ref{aff14},\ref{aff15}}
\and E.~Bellini \inst{\ref{aff16},\ref{aff17},\ref{aff18}}
\and G.~Benevento\orcid{0000-0002-6999-2429}\inst{\ref{aff19},\ref{aff20}}
\and D.~Bertacca\orcid{0000-0002-2490-7139}\inst{\ref{aff13},\ref{aff15},\ref{aff14}}
\and C.~Bonvin\orcid{0000-0002-5318-4064}\inst{\ref{aff4}}
\and B.~Bose\orcid{0000-0003-1965-8614}\inst{\ref{aff21}}
\and P.~Brax\inst{\ref{aff22},\ref{aff23}}
\and V.~F.~Cardone\inst{\ref{aff2},\ref{aff3}}
\and S.~Casas\orcid{0000-0002-4751-5138}\inst{\ref{aff24},\ref{aff25}}
\and M.~Y.~Elkhashab\orcid{0000-0001-9306-2603}\inst{\ref{aff18},\ref{aff17},\ref{aff26},\ref{aff27}}
\and P.~G.~Ferreira\orcid{0000-0002-3021-2851}\inst{\ref{aff28}}
\and F.~Finelli\orcid{0000-0002-6694-3269}\inst{\ref{aff9},\ref{aff29}}
\and F.~Hassani\orcid{0000-0003-2640-4460}\inst{\ref{aff30}}
\and S.~Ili\'c\orcid{0000-0003-4285-9086}\inst{\ref{aff31},\ref{aff32}}
\and K.~Koyama\orcid{0000-0001-6727-6915}\inst{\ref{aff25}}
\and M.~Kunz\orcid{0000-0002-3052-7394}\inst{\ref{aff4}}
\and F.~Lepori\inst{\ref{aff33}}
\and J.~Lesgourgues\orcid{0000-0001-7627-353X}\inst{\ref{aff24}}
\and C.~J.~A.~P.~Martins\orcid{0000-0002-4886-9261}\inst{\ref{aff34},\ref{aff35}}
\and D.~F.~Mota\orcid{0000-0003-3141-142X}\inst{\ref{aff30}}
\and J.~Noller\orcid{0000-0003-2210-775X}\inst{\ref{aff36},\ref{aff25}}
\and F.~Pace\orcid{0000-0001-8039-0480}\inst{\ref{aff37},\ref{aff38},\ref{aff39}}
\and D.~Paoletti\orcid{0000-0003-4761-6147}\inst{\ref{aff9},\ref{aff29}}
\and G.~Parimbelli\orcid{0000-0002-2539-2472}\inst{\ref{aff40},\ref{aff41},\ref{aff42}}
\and V.~Pettorino\inst{\ref{aff43}}
\and Z.~Sakr\orcid{0000-0002-4823-3757}\inst{\ref{aff44},\ref{aff32},\ref{aff45}}
\and S.~Srinivasan\orcid{0000-0003-1539-3276}\inst{\ref{aff46}}
\and E.~M.~Teixeira\orcid{0000-0001-7417-0780}\inst{\ref{aff47}}
\and I.~Tutusaus\orcid{0000-0002-3199-0399}\inst{\ref{aff40},\ref{aff48},\ref{aff32}}
\and P.~Valageas\inst{\ref{aff22}}
\and H.-A.~Winther\orcid{0000-0002-6325-2710}\inst{\ref{aff30}}
\and J.~Adamek\orcid{0000-0002-0723-6740}\inst{\ref{aff33}}
\and I.~S.~Albuquerque\orcid{0000-0002-5369-1226}\inst{\ref{aff49}}
\and L.~Atayde\orcid{0000-0001-6373-9193}\inst{\ref{aff49}}
\and M.-A.~Breton\inst{\ref{aff40},\ref{aff50},\ref{aff51}}
\and S.~Camera\orcid{0000-0003-3399-3574}\inst{\ref{aff37},\ref{aff38},\ref{aff39}}
\and C.~Carbone\orcid{0000-0003-0125-3563}\inst{\ref{aff52}}
\and E.~Carella\inst{\ref{aff52}}
\and P.~Carrilho\orcid{0000-0003-1339-0194}\inst{\ref{aff21}}
\and F.~J.~Castander\orcid{0000-0001-7316-4573}\inst{\ref{aff40},\ref{aff48}}
\and R.~Durrer\orcid{0000-0001-9833-2086}\inst{\ref{aff4}}
\and B.~Fiorini\orcid{0000-0002-0092-4321}\inst{\ref{aff25}}
\and P.~Fosalba\orcid{0000-0002-1510-5214}\inst{\ref{aff48},\ref{aff40}}
\and M.~Marinucci\orcid{0000-0003-1159-3756}\inst{\ref{aff13},\ref{aff14}}
\and C.~Moretti\orcid{0000-0003-3314-8936}\inst{\ref{aff42},\ref{aff53},\ref{aff18},\ref{aff27},\ref{aff17}}
\and M.~Pietroni\orcid{0000-0001-5480-5996}\inst{\ref{aff54},\ref{aff55}}
\and L.~Piga\orcid{0000-0003-2221-7406}\inst{\ref{aff54},\ref{aff55},\ref{aff52}}
\and G.~R\'acz\orcid{0000-0003-3906-5699}\inst{\ref{aff56},\ref{aff57}}
\and F.~Sorrenti\orcid{0000-0001-7141-9659}\inst{\ref{aff4}}
\and F.~Vernizzi\orcid{0000-0003-3426-2802}\inst{\ref{aff22}}
\and C.~Viglione\inst{\ref{aff48},\ref{aff40}}
\and L.~Amendola\orcid{0000-0002-0835-233X}\inst{\ref{aff44}}
\and S.~Andreon\orcid{0000-0002-2041-8784}\inst{\ref{aff58}}
\and C.~Baccigalupi\orcid{0000-0002-8211-1630}\inst{\ref{aff27},\ref{aff18},\ref{aff17},\ref{aff42}}
\and S.~Bardelli\orcid{0000-0002-8900-0298}\inst{\ref{aff9}}
\and R.~Bender\orcid{0000-0001-7179-0626}\inst{\ref{aff59},\ref{aff46}}
\and A.~Biviano\orcid{0000-0002-0857-0732}\inst{\ref{aff18},\ref{aff27}}
\and D.~Bonino\orcid{0000-0002-3336-9977}\inst{\ref{aff39}}
\and E.~Branchini\orcid{0000-0002-0808-6908}\inst{\ref{aff60},\ref{aff61},\ref{aff58}}
\and M.~Brescia\orcid{0000-0001-9506-5680}\inst{\ref{aff1},\ref{aff62},\ref{aff63}}
\and J.~Brinchmann\orcid{0000-0003-4359-8797}\inst{\ref{aff35},\ref{aff64}}
\and A.~Caillat\inst{\ref{aff65}}
\and G.~Ca\~nas-Herrera\orcid{0000-0003-2796-2149}\inst{\ref{aff43},\ref{aff5}}
\and V.~Capobianco\orcid{0000-0002-3309-7692}\inst{\ref{aff39}}
\and J.~Carretero\orcid{0000-0002-3130-0204}\inst{\ref{aff66},\ref{aff67}}
\and M.~Castellano\orcid{0000-0001-9875-8263}\inst{\ref{aff2}}
\and G.~Castignani\orcid{0000-0001-6831-0687}\inst{\ref{aff9}}
\and S.~Cavuoti\orcid{0000-0002-3787-4196}\inst{\ref{aff62},\ref{aff63}}
\and K.~C.~Chambers\orcid{0000-0001-6965-7789}\inst{\ref{aff68}}
\and A.~Cimatti\inst{\ref{aff69}}
\and C.~Colodro-Conde\inst{\ref{aff70}}
\and G.~Congedo\orcid{0000-0003-2508-0046}\inst{\ref{aff21}}
\and C.~J.~Conselice\orcid{0000-0003-1949-7638}\inst{\ref{aff71}}
\and L.~Conversi\orcid{0000-0002-6710-8476}\inst{\ref{aff72},\ref{aff73}}
\and Y.~Copin\orcid{0000-0002-5317-7518}\inst{\ref{aff74}}
\and F.~Courbin\orcid{0000-0003-0758-6510}\inst{\ref{aff75},\ref{aff76},\ref{aff77}}
\and H.~M.~Courtois\orcid{0000-0003-0509-1776}\inst{\ref{aff78}}
\and M.~Cropper\orcid{0000-0003-4571-9468}\inst{\ref{aff79}}
\and A.~Da~Silva\orcid{0000-0002-6385-1609}\inst{\ref{aff80},\ref{aff49}}
\and H.~Degaudenzi\orcid{0000-0002-5887-6799}\inst{\ref{aff81}}
\and G.~De~Lucia\orcid{0000-0002-6220-9104}\inst{\ref{aff18}}
\and A.~M.~Di~Giorgio\orcid{0000-0002-4767-2360}\inst{\ref{aff82}}
\and J.~Dinis\orcid{0000-0001-5075-1601}\inst{\ref{aff80},\ref{aff49}}
\and H.~Dole\orcid{0000-0002-9767-3839}\inst{\ref{aff83}}
\and F.~Dubath\orcid{0000-0002-6533-2810}\inst{\ref{aff81}}
\and X.~Dupac\inst{\ref{aff73}}
\and S.~Dusini\orcid{0000-0002-1128-0664}\inst{\ref{aff14}}
\and A.~Ealet\orcid{0000-0003-3070-014X}\inst{\ref{aff74}}
\and S.~Escoffier\orcid{0000-0002-2847-7498}\inst{\ref{aff84}}
\and M.~Farina\orcid{0000-0002-3089-7846}\inst{\ref{aff82}}
\and S.~Farrens\orcid{0000-0002-9594-9387}\inst{\ref{aff85}}
\and F.~Faustini\orcid{0000-0001-6274-5145}\inst{\ref{aff20},\ref{aff2}}
\and S.~Ferriol\inst{\ref{aff74}}
\and S.~Fotopoulou\orcid{0000-0002-9686-254X}\inst{\ref{aff86}}
\and M.~Frailis\orcid{0000-0002-7400-2135}\inst{\ref{aff18}}
\and E.~Franceschi\orcid{0000-0002-0585-6591}\inst{\ref{aff9}}
\and M.~Fumana\orcid{0000-0001-6787-5950}\inst{\ref{aff52}}
\and S.~Galeotta\orcid{0000-0002-3748-5115}\inst{\ref{aff18}}
\and B.~Gillis\orcid{0000-0002-4478-1270}\inst{\ref{aff21}}
\and C.~Giocoli\orcid{0000-0002-9590-7961}\inst{\ref{aff9},\ref{aff87}}
\and J.~Gracia-Carpio\inst{\ref{aff59}}
\and A.~Grazian\orcid{0000-0002-5688-0663}\inst{\ref{aff15}}
\and F.~Grupp\inst{\ref{aff59},\ref{aff46}}
\and L.~Guzzo\orcid{0000-0001-8264-5192}\inst{\ref{aff6},\ref{aff58}}
\and S.~V.~H.~Haugan\orcid{0000-0001-9648-7260}\inst{\ref{aff30}}
\and H.~Hoekstra\orcid{0000-0002-0641-3231}\inst{\ref{aff88}}
\and W.~Holmes\inst{\ref{aff56}}
\and I.~Hook\orcid{0000-0002-2960-978X}\inst{\ref{aff89}}
\and F.~Hormuth\inst{\ref{aff90}}
\and A.~Hornstrup\orcid{0000-0002-3363-0936}\inst{\ref{aff91},\ref{aff92}}
\and P.~Hudelot\inst{\ref{aff93}}
\and K.~Jahnke\orcid{0000-0003-3804-2137}\inst{\ref{aff94}}
\and M.~Jhabvala\inst{\ref{aff95}}
\and B.~Joachimi\orcid{0000-0001-7494-1303}\inst{\ref{aff36}}
\and E.~Keih\"anen\orcid{0000-0003-1804-7715}\inst{\ref{aff96}}
\and S.~Kermiche\orcid{0000-0002-0302-5735}\inst{\ref{aff84}}
\and A.~Kiessling\orcid{0000-0002-2590-1273}\inst{\ref{aff56}}
\and B.~Kubik\orcid{0009-0006-5823-4880}\inst{\ref{aff74}}
\and H.~Kurki-Suonio\orcid{0000-0002-4618-3063}\inst{\ref{aff57},\ref{aff97}}
\and O.~Lahav\orcid{0000-0002-1134-9035}\inst{\ref{aff36}}
\and A.~M.~C.~Le~Brun\orcid{0000-0002-0936-4594}\inst{\ref{aff51}}
\and S.~Ligori\orcid{0000-0003-4172-4606}\inst{\ref{aff39}}
\and P.~B.~Lilje\orcid{0000-0003-4324-7794}\inst{\ref{aff30}}
\and V.~Lindholm\orcid{0000-0003-2317-5471}\inst{\ref{aff57},\ref{aff97}}
\and I.~Lloro\orcid{0000-0001-5966-1434}\inst{\ref{aff98}}
\and G.~Mainetti\orcid{0000-0003-2384-2377}\inst{\ref{aff99}}
\and D.~Maino\inst{\ref{aff6},\ref{aff52},\ref{aff7}}
\and E.~Maiorano\orcid{0000-0003-2593-4355}\inst{\ref{aff9}}
\and O.~Mansutti\orcid{0000-0001-5758-4658}\inst{\ref{aff18}}
\and O.~Marggraf\orcid{0000-0001-7242-3852}\inst{\ref{aff100}}
\and K.~Markovic\orcid{0000-0001-6764-073X}\inst{\ref{aff56}}
\and N.~Martinet\orcid{0000-0003-2786-7790}\inst{\ref{aff65}}
\and F.~Marulli\orcid{0000-0002-8850-0303}\inst{\ref{aff101},\ref{aff9},\ref{aff10}}
\and R.~Massey\orcid{0000-0002-6085-3780}\inst{\ref{aff102}}
\and S.~Maurogordato\inst{\ref{aff103}}
\and E.~Medinaceli\orcid{0000-0002-4040-7783}\inst{\ref{aff9}}
\and S.~Mei\orcid{0000-0002-2849-559X}\inst{\ref{aff104}}
\and M.~Melchior\inst{\ref{aff105}}
\and Y.~Mellier\inst{\ref{aff106},\ref{aff93}}
\and M.~Meneghetti\orcid{0000-0003-1225-7084}\inst{\ref{aff9},\ref{aff10}}
\and E.~Merlin\orcid{0000-0001-6870-8900}\inst{\ref{aff2}}
\and G.~Meylan\inst{\ref{aff75}}
\and A.~Mora\orcid{0000-0002-1922-8529}\inst{\ref{aff107}}
\and M.~Moresco\orcid{0000-0002-7616-7136}\inst{\ref{aff101},\ref{aff9}}
\and L.~Moscardini\orcid{0000-0002-3473-6716}\inst{\ref{aff101},\ref{aff9},\ref{aff10}}
\and C.~Neissner\orcid{0000-0001-8524-4968}\inst{\ref{aff108},\ref{aff67}}
\and R.~C.~Nichol\orcid{0000-0003-0939-6518}\inst{\ref{aff109}}
\and S.-M.~Niemi\inst{\ref{aff43}}
\and J.~W.~Nightingale\orcid{0000-0002-8987-7401}\inst{\ref{aff110}}
\and C.~Padilla\orcid{0000-0001-7951-0166}\inst{\ref{aff108}}
\and S.~Paltani\orcid{0000-0002-8108-9179}\inst{\ref{aff81}}
\and F.~Pasian\orcid{0000-0002-4869-3227}\inst{\ref{aff18}}
\and K.~Pedersen\inst{\ref{aff111}}
\and W.~J.~Percival\orcid{0000-0002-0644-5727}\inst{\ref{aff112},\ref{aff113},\ref{aff114}}
\and S.~Pires\orcid{0000-0002-0249-2104}\inst{\ref{aff85}}
\and G.~Polenta\orcid{0000-0003-4067-9196}\inst{\ref{aff20}}
\and M.~Poncet\inst{\ref{aff115}}
\and L.~A.~Popa\inst{\ref{aff116}}
\and L.~Pozzetti\orcid{0000-0001-7085-0412}\inst{\ref{aff9}}
\and F.~Raison\orcid{0000-0002-7819-6918}\inst{\ref{aff59}}
\and R.~Rebolo\inst{\ref{aff70},\ref{aff117},\ref{aff118}}
\and A.~Renzi\orcid{0000-0001-9856-1970}\inst{\ref{aff13},\ref{aff14}}
\and J.~Rhodes\orcid{0000-0002-4485-8549}\inst{\ref{aff56}}
\and G.~Riccio\inst{\ref{aff62}}
\and E.~Romelli\orcid{0000-0003-3069-9222}\inst{\ref{aff18}}
\and M.~Roncarelli\orcid{0000-0001-9587-7822}\inst{\ref{aff9}}
\and R.~Saglia\orcid{0000-0003-0378-7032}\inst{\ref{aff46},\ref{aff59}}
\and A.~G.~S\'anchez\orcid{0000-0003-1198-831X}\inst{\ref{aff59}}
\and D.~Sapone\orcid{0000-0001-7089-4503}\inst{\ref{aff119}}
\and B.~Sartoris\orcid{0000-0003-1337-5269}\inst{\ref{aff46},\ref{aff18}}
\and R.~Scaramella\orcid{0000-0003-2229-193X}\inst{\ref{aff2}}
\and J.~A.~Schewtschenko\orcid{0000-0002-4913-6393}\inst{\ref{aff21}}
\and M.~Schirmer\orcid{0000-0003-2568-9994}\inst{\ref{aff94}}
\and T.~Schrabback\orcid{0000-0002-6987-7834}\inst{\ref{aff120}}
\and A.~Secroun\orcid{0000-0003-0505-3710}\inst{\ref{aff84}}
\and E.~Sefusatti\orcid{0000-0003-0473-1567}\inst{\ref{aff18},\ref{aff27},\ref{aff17}}
\and G.~Seidel\orcid{0000-0003-2907-353X}\inst{\ref{aff94}}
\and M.~Seiffert\orcid{0000-0002-7536-9393}\inst{\ref{aff56}}
\and S.~Serrano\orcid{0000-0002-0211-2861}\inst{\ref{aff48},\ref{aff121},\ref{aff40}}
\and C.~Sirignano\orcid{0000-0002-0995-7146}\inst{\ref{aff13},\ref{aff14}}
\and G.~Sirri\orcid{0000-0003-2626-2853}\inst{\ref{aff10}}
\and A.~Spurio~Mancini\orcid{0000-0001-5698-0990}\inst{\ref{aff122},\ref{aff79}}
\and L.~Stanco\orcid{0000-0002-9706-5104}\inst{\ref{aff14}}
\and J.~Steinwagner\orcid{0000-0001-7443-1047}\inst{\ref{aff59}}
\and P.~Tallada-Cresp\'{i}\orcid{0000-0002-1336-8328}\inst{\ref{aff66},\ref{aff67}}
\and D.~Tavagnacco\orcid{0000-0001-7475-9894}\inst{\ref{aff18}}
\and A.~N.~Taylor\inst{\ref{aff21}}
\and I.~Tereno\inst{\ref{aff80},\ref{aff123}}
\and N.~Tessore\orcid{0000-0002-9696-7931}\inst{\ref{aff36}}
\and S.~Toft\orcid{0000-0003-3631-7176}\inst{\ref{aff124},\ref{aff125}}
\and R.~Toledo-Moreo\orcid{0000-0002-2997-4859}\inst{\ref{aff126}}
\and F.~Torradeflot\orcid{0000-0003-1160-1517}\inst{\ref{aff67},\ref{aff66}}
\and E.~A.~Valentijn\inst{\ref{aff127}}
\and L.~Valenziano\orcid{0000-0002-1170-0104}\inst{\ref{aff9},\ref{aff29}}
\and J.~Valiviita\orcid{0000-0001-6225-3693}\inst{\ref{aff57},\ref{aff97}}
\and T.~Vassallo\orcid{0000-0001-6512-6358}\inst{\ref{aff46},\ref{aff18}}
\and G.~Verdoes~Kleijn\orcid{0000-0001-5803-2580}\inst{\ref{aff127}}
\and A.~Veropalumbo\orcid{0000-0003-2387-1194}\inst{\ref{aff58},\ref{aff61},\ref{aff41}}
\and Y.~Wang\orcid{0000-0002-4749-2984}\inst{\ref{aff128}}
\and J.~Weller\orcid{0000-0002-8282-2010}\inst{\ref{aff46},\ref{aff59}}
\and A.~Zacchei\orcid{0000-0003-0396-1192}\inst{\ref{aff18},\ref{aff27}}
\and G.~Zamorani\orcid{0000-0002-2318-301X}\inst{\ref{aff9}}
\and E.~Zucca\orcid{0000-0002-5845-8132}\inst{\ref{aff9}}
\and E.~Bozzo\orcid{0000-0002-8201-1525}\inst{\ref{aff81}}
\and C.~Burigana\orcid{0000-0002-3005-5796}\inst{\ref{aff129},\ref{aff29}}
\and M.~Calabrese\orcid{0000-0002-2637-2422}\inst{\ref{aff130},\ref{aff52}}
\and D.~Di~Ferdinando\inst{\ref{aff10}}
\and J.~A.~Escartin~Vigo\inst{\ref{aff59}}
\and G.~Fabbian\orcid{0000-0002-3255-4695}\inst{\ref{aff131},\ref{aff132}}
\and M.~Maturi\orcid{0000-0002-3517-2422}\inst{\ref{aff44},\ref{aff133}}
\and N.~Mauri\orcid{0000-0001-8196-1548}\inst{\ref{aff69},\ref{aff10}}
\and A.~Pezzotta\orcid{0000-0003-0726-2268}\inst{\ref{aff59}}
\and M.~P\"ontinen\orcid{0000-0001-5442-2530}\inst{\ref{aff57}}
\and C.~Porciani\orcid{0000-0002-7797-2508}\inst{\ref{aff100}}
\and V.~Scottez\inst{\ref{aff106},\ref{aff134}}
\and M.~Tenti\orcid{0000-0002-4254-5901}\inst{\ref{aff10}}
\and M.~Viel\orcid{0000-0002-2642-5707}\inst{\ref{aff27},\ref{aff18},\ref{aff42},\ref{aff17},\ref{aff53}}
\and M.~Wiesmann\orcid{0009-0000-8199-5860}\inst{\ref{aff30}}
\and Y.~Akrami\orcid{0000-0002-2407-7956}\inst{\ref{aff135},\ref{aff136}}
\and V.~Allevato\orcid{0000-0001-7232-5152}\inst{\ref{aff62}}
\and S.~Anselmi\orcid{0000-0002-3579-9583}\inst{\ref{aff14},\ref{aff13},\ref{aff51}}
\and F.~Atrio-Barandela\orcid{0000-0002-2130-2513}\inst{\ref{aff137}}
\and A.~Balaguera-Antolinez\orcid{0000-0001-5028-3035}\inst{\ref{aff70},\ref{aff117}}
\and A.~Blanchard\orcid{0000-0001-8555-9003}\inst{\ref{aff32}}
\and L.~Blot\orcid{0000-0002-9622-7167}\inst{\ref{aff138},\ref{aff51}}
\and H.~B\"ohringer\orcid{0000-0001-8241-4204}\inst{\ref{aff59},\ref{aff139},\ref{aff140}}
\and S.~Borgani\orcid{0000-0001-6151-6439}\inst{\ref{aff26},\ref{aff27},\ref{aff18},\ref{aff17},\ref{aff53}}
\and M.~L.~Brown\orcid{0000-0002-0370-8077}\inst{\ref{aff71}}
\and S.~Bruton\orcid{0000-0002-6503-5218}\inst{\ref{aff141}}
\and R.~Cabanac\orcid{0000-0001-6679-2600}\inst{\ref{aff32}}
\and A.~Calabro\orcid{0000-0003-2536-1614}\inst{\ref{aff2}}
\and B.~Camacho~Quevedo\orcid{0000-0002-8789-4232}\inst{\ref{aff48},\ref{aff40}}
\and A.~Cappi\inst{\ref{aff9},\ref{aff103}}
\and F.~Caro\inst{\ref{aff2}}
\and C.~S.~Carvalho\inst{\ref{aff123}}
\and T.~Castro\orcid{0000-0002-6292-3228}\inst{\ref{aff18},\ref{aff17},\ref{aff27},\ref{aff53}}
\and F.~Cogato\orcid{0000-0003-4632-6113}\inst{\ref{aff101},\ref{aff9}}
\and S.~Contarini\orcid{0000-0002-9843-723X}\inst{\ref{aff59}}
\and A.~R.~Cooray\orcid{0000-0002-3892-0190}\inst{\ref{aff142}}
\and S.~Davini\orcid{0000-0003-3269-1718}\inst{\ref{aff61}}
\and G.~Desprez\orcid{0000-0001-8325-1742}\inst{\ref{aff143}}
\and A.~D\'iaz-S\'anchez\orcid{0000-0003-0748-4768}\inst{\ref{aff144}}
\and S.~Di~Domizio\orcid{0000-0003-2863-5895}\inst{\ref{aff60},\ref{aff61}}
\and A.~G.~Ferrari\orcid{0009-0005-5266-4110}\inst{\ref{aff10}}
\and I.~Ferrero\orcid{0000-0002-1295-1132}\inst{\ref{aff30}}
\and A.~Finoguenov\orcid{0000-0002-4606-5403}\inst{\ref{aff57}}
\and F.~Fornari\orcid{0000-0003-2979-6738}\inst{\ref{aff29}}
\and L.~Gabarra\orcid{0000-0002-8486-8856}\inst{\ref{aff28}}
\and K.~Ganga\orcid{0000-0001-8159-8208}\inst{\ref{aff104}}
\and J.~Garc\'ia-Bellido\orcid{0000-0002-9370-8360}\inst{\ref{aff135}}
\and T.~Gasparetto\orcid{0000-0002-7913-4866}\inst{\ref{aff18}}
\and V.~Gautard\inst{\ref{aff145}}
\and E.~Gaztanaga\orcid{0000-0001-9632-0815}\inst{\ref{aff40},\ref{aff48},\ref{aff25}}
\and F.~Giacomini\orcid{0000-0002-3129-2814}\inst{\ref{aff10}}
\and F.~Gianotti\orcid{0000-0003-4666-119X}\inst{\ref{aff9}}
\and G.~Gozaliasl\orcid{0000-0002-0236-919X}\inst{\ref{aff146}}
\and A.~Gregorio\orcid{0000-0003-4028-8785}\inst{\ref{aff26},\ref{aff18},\ref{aff17}}
\and M.~Guidi\orcid{0000-0001-9408-1101}\inst{\ref{aff8},\ref{aff9}}
\and C.~M.~Gutierrez\orcid{0000-0001-7854-783X}\inst{\ref{aff147}}
\and A.~Hall\orcid{0000-0002-3139-8651}\inst{\ref{aff21}}
\and H.~Hildebrandt\orcid{0000-0002-9814-3338}\inst{\ref{aff148}}
\and J.~Hjorth\orcid{0000-0002-4571-2306}\inst{\ref{aff111}}
\and A.~Jimenez~Mu\~noz\orcid{0009-0004-5252-185X}\inst{\ref{aff149}}
\and J.~J.~E.~Kajava\orcid{0000-0002-3010-8333}\inst{\ref{aff150},\ref{aff151}}
\and Y.~Kang\orcid{0009-0000-8588-7250}\inst{\ref{aff81}}
\and V.~Kansal\orcid{0000-0002-4008-6078}\inst{\ref{aff152},\ref{aff153}}
\and D.~Karagiannis\orcid{0000-0002-4927-0816}\inst{\ref{aff154},\ref{aff155}}
\and C.~C.~Kirkpatrick\inst{\ref{aff96}}
\and S.~Kruk\orcid{0000-0001-8010-8879}\inst{\ref{aff73}}
\and F.~Lacasa\orcid{0000-0002-7268-3440}\inst{\ref{aff156},\ref{aff83}}
\and M.~Lattanzi\inst{\ref{aff12}}
\and S.~Lee\orcid{0000-0002-8289-740X}\inst{\ref{aff56}}
\and J.~Le~Graet\orcid{0000-0001-6523-7971}\inst{\ref{aff84}}
\and L.~Legrand\orcid{0000-0003-0610-5252}\inst{\ref{aff157},\ref{aff131}}
\and M.~Lembo\orcid{0000-0002-5271-5070}\inst{\ref{aff11},\ref{aff12}}
\and T.~I.~Liaudat\orcid{0000-0002-9104-314X}\inst{\ref{aff158}}
\and S.~J.~Liu\orcid{0000-0001-7680-2139}\inst{\ref{aff82}}
\and A.~Loureiro\orcid{0000-0002-4371-0876}\inst{\ref{aff159},\ref{aff160}}
\and J.~Macias-Perez\orcid{0000-0002-5385-2763}\inst{\ref{aff149}}
\and M.~Magliocchetti\orcid{0000-0001-9158-4838}\inst{\ref{aff82}}
\and F.~Mannucci\orcid{0000-0002-4803-2381}\inst{\ref{aff161}}
\and R.~Maoli\orcid{0000-0002-6065-3025}\inst{\ref{aff162},\ref{aff2}}
\and J.~Mart\'{i}n-Fleitas\orcid{0000-0002-8594-569X}\inst{\ref{aff107}}
\and L.~Maurin\orcid{0000-0002-8406-0857}\inst{\ref{aff83}}
\and R.~B.~Metcalf\orcid{0000-0003-3167-2574}\inst{\ref{aff101},\ref{aff9}}
\and M.~Migliaccio\inst{\ref{aff163},\ref{aff19}}
\and M.~Miluzio\inst{\ref{aff73},\ref{aff164}}
\and P.~Monaco\orcid{0000-0003-2083-7564}\inst{\ref{aff26},\ref{aff18},\ref{aff17},\ref{aff27}}
\and A.~Montoro\orcid{0000-0003-4730-8590}\inst{\ref{aff40},\ref{aff48}}
\and G.~Morgante\inst{\ref{aff9}}
\and S.~Nadathur\orcid{0000-0001-9070-3102}\inst{\ref{aff25}}
\and K.~Naidoo\orcid{0000-0002-9182-1802}\inst{\ref{aff36}}
\and P.~Natoli\orcid{0000-0003-0126-9100}\inst{\ref{aff11},\ref{aff12}}
\and Nicholas~A.~Walton\orcid{0000-0003-3983-8778}\inst{\ref{aff132}}
\and L.~Pagano\orcid{0000-0003-1820-5998}\inst{\ref{aff11},\ref{aff12}}
\and K.~Paterson\orcid{0000-0001-8340-3486}\inst{\ref{aff94}}
\and L.~Patrizii\inst{\ref{aff10}}
\and A.~Pisani\orcid{0000-0002-6146-4437}\inst{\ref{aff84},\ref{aff165}}
\and V.~Popa\orcid{0000-0002-9118-8330}\inst{\ref{aff116}}
\and D.~Potter\orcid{0000-0002-0757-5195}\inst{\ref{aff33}}
\and P.~Reimberg\orcid{0000-0003-3410-0280}\inst{\ref{aff106}}
\and I.~Risso\orcid{0000-0003-2525-7761}\inst{\ref{aff166}}
\and P.-F.~Rocci\inst{\ref{aff83}}
\and M.~Sahl\'en\orcid{0000-0003-0973-4804}\inst{\ref{aff167}}
\and E.~Sarpa\orcid{0000-0002-1256-655X}\inst{\ref{aff42},\ref{aff53},\ref{aff17}}
\and A.~Schneider\orcid{0000-0001-7055-8104}\inst{\ref{aff33}}
\and M.~Schultheis\inst{\ref{aff103}}
\and D.~Sciotti\orcid{0009-0008-4519-2620}\inst{\ref{aff2},\ref{aff3}}
\and M.~Sereno\orcid{0000-0003-0302-0325}\inst{\ref{aff9},\ref{aff10}}
\and L.~C.~Smith\orcid{0000-0002-3259-2771}\inst{\ref{aff132}}
\and K.~Tanidis\inst{\ref{aff28}}
\and C.~Tao\orcid{0000-0001-7961-8177}\inst{\ref{aff84}}
\and G.~Testera\inst{\ref{aff61}}
\and R.~Teyssier\orcid{0000-0001-7689-0933}\inst{\ref{aff165}}
\and S.~Tosi\orcid{0000-0002-7275-9193}\inst{\ref{aff60},\ref{aff61}}
\and A.~Troja\orcid{0000-0003-0239-4595}\inst{\ref{aff13},\ref{aff14}}
\and M.~Tucci\inst{\ref{aff81}}
\and C.~Valieri\inst{\ref{aff10}}
\and D.~Vergani\orcid{0000-0003-0898-2216}\inst{\ref{aff9}}
\and G.~Verza\orcid{0000-0002-1886-8348}\inst{\ref{aff168},\ref{aff169}}
\and P.~Vielzeuf\orcid{0000-0003-2035-9339}\inst{\ref{aff84}}}

%%%% please do not edit the affiliation list -- contact ECEB Bureau for changes
\institute{Department of Physics "E. Pancini", University Federico II, Via Cinthia 6, 80126, Napoli, Italy\label{aff1}
\and
INAF-Osservatorio Astronomico di Roma, Via Frascati 33, 00078 Monteporzio Catone, Italy\label{aff2}
\and
INFN-Sezione di Roma, Piazzale Aldo Moro, 2 - c/o Dipartimento di Fisica, Edificio G. Marconi, 00185 Roma, Italy\label{aff3}
\and
Universit\'e de Gen\`eve, D\'epartement de Physique Th\'eorique and Centre for Astroparticle Physics, 24 quai Ernest-Ansermet, CH-1211 Gen\`eve 4, Switzerland\label{aff4}
\and
Institute Lorentz, Leiden University, Niels Bohrweg 2, 2333 CA Leiden, The Netherlands\label{aff5}
\and
Dipartimento di Fisica "Aldo Pontremoli", Universit\`a degli Studi di Milano, Via Celoria 16, 20133 Milano, Italy\label{aff6}
\and
INFN-Sezione di Milano, Via Celoria 16, 20133 Milano, Italy\label{aff7}
\and
Dipartimento di Fisica e Astronomia, Universit\`a di Bologna, Via Gobetti 93/2, 40129 Bologna, Italy\label{aff8}
\and
INAF-Osservatorio di Astrofisica e Scienza dello Spazio di Bologna, Via Piero Gobetti 93/3, 40129 Bologna, Italy\label{aff9}
\and
INFN-Sezione di Bologna, Viale Berti Pichat 6/2, 40127 Bologna, Italy\label{aff10}
\and
Dipartimento di Fisica e Scienze della Terra, Universit\`a degli Studi di Ferrara, Via Giuseppe Saragat 1, 44122 Ferrara, Italy\label{aff11}
\and
Istituto Nazionale di Fisica Nucleare, Sezione di Ferrara, Via Giuseppe Saragat 1, 44122 Ferrara, Italy\label{aff12}
\and
Dipartimento di Fisica e Astronomia "G. Galilei", Universit\`a di Padova, Via Marzolo 8, 35131 Padova, Italy\label{aff13}
\and
INFN-Padova, Via Marzolo 8, 35131 Padova, Italy\label{aff14}
\and
INAF-Osservatorio Astronomico di Padova, Via dell'Osservatorio 5, 35122 Padova, Italy\label{aff15}
\and
Center for Astrophysics and Cosmology, University of Nova Gorica, Nova Gorica, Slovenia\label{aff16}
\and
INFN, Sezione di Trieste, Via Valerio 2, 34127 Trieste TS, Italy\label{aff17}
\and
INAF-Osservatorio Astronomico di Trieste, Via G. B. Tiepolo 11, 34143 Trieste, Italy\label{aff18}
\and
INFN, Sezione di Roma 2, Via della Ricerca Scientifica 1, Roma, Italy\label{aff19}
\and
Space Science Data Center, Italian Space Agency, via del Politecnico snc, 00133 Roma, Italy\label{aff20}
\and
Institute for Astronomy, University of Edinburgh, Royal Observatory, Blackford Hill, Edinburgh EH9 3HJ, UK\label{aff21}
\and
Institut de Physique Th\'eorique, CEA, CNRS, Universit\'e Paris-Saclay 91191 Gif-sur-Yvette Cedex, France\label{aff22}
\and
CERN, Theoretical Physics Department, Geneva, Switzerland\label{aff23}
\and
Institute for Theoretical Particle Physics and Cosmology (TTK), RWTH Aachen University, 52056 Aachen, Germany\label{aff24}
\and
Institute of Cosmology and Gravitation, University of Portsmouth, Portsmouth PO1 3FX, UK\label{aff25}
\and
Dipartimento di Fisica - Sezione di Astronomia, Universit\`a di Trieste, Via Tiepolo 11, 34131 Trieste, Italy\label{aff26}
\and
IFPU, Institute for Fundamental Physics of the Universe, via Beirut 2, 34151 Trieste, Italy\label{aff27}
\and
Department of Physics, Oxford University, Keble Road, Oxford OX1 3RH, UK\label{aff28}
\and
INFN-Bologna, Via Irnerio 46, 40126 Bologna, Italy\label{aff29}
\and
Institute of Theoretical Astrophysics, University of Oslo, P.O. Box 1029 Blindern, 0315 Oslo, Norway\label{aff30}
\and
Universit\'e Paris-Saclay, CNRS/IN2P3, IJCLab, 91405 Orsay, France\label{aff31}
\and
Institut de Recherche en Astrophysique et Plan\'etologie (IRAP), Universit\'e de Toulouse, CNRS, UPS, CNES, 14 Av. Edouard Belin, 31400 Toulouse, France\label{aff32}
\and
Department of Astrophysics, University of Zurich, Winterthurerstrasse 190, 8057 Zurich, Switzerland\label{aff33}
\and
Centro de Astrof\'{\i}sica da Universidade do Porto, Rua das Estrelas, 4150-762 Porto, Portugal\label{aff34}
\and
Instituto de Astrof\'isica e Ci\^encias do Espa\c{c}o, Universidade do Porto, CAUP, Rua das Estrelas, PT4150-762 Porto, Portugal\label{aff35}
\and
Department of Physics and Astronomy, University College London, Gower Street, London WC1E 6BT, UK\label{aff36}
\and
Dipartimento di Fisica, Universit\`a degli Studi di Torino, Via P. Giuria 1, 10125 Torino, Italy\label{aff37}
\and
INFN-Sezione di Torino, Via P. Giuria 1, 10125 Torino, Italy\label{aff38}
\and
INAF-Osservatorio Astrofisico di Torino, Via Osservatorio 20, 10025 Pino Torinese (TO), Italy\label{aff39}
\and
Institute of Space Sciences (ICE, CSIC), Campus UAB, Carrer de Can Magrans, s/n, 08193 Barcelona, Spain\label{aff40}
\and
Dipartimento di Fisica, Universit\`a degli studi di Genova, and INFN-Sezione di Genova, via Dodecaneso 33, 16146, Genova, Italy\label{aff41}
\and
SISSA, International School for Advanced Studies, Via Bonomea 265, 34136 Trieste TS, Italy\label{aff42}
\and
European Space Agency/ESTEC, Keplerlaan 1, 2201 AZ Noordwijk, The Netherlands\label{aff43}
\and
Institut f\"ur Theoretische Physik, University of Heidelberg, Philosophenweg 16, 69120 Heidelberg, Germany\label{aff44}
\and
Universit\'e St Joseph; Faculty of Sciences, Beirut, Lebanon\label{aff45}
\and
Universit\"ats-Sternwarte M\"unchen, Fakult\"at f\"ur Physik, Ludwig-Maximilians-Universit\"at M\"unchen, Scheinerstr.~1, 81679 M\"unchen, Germany\label{aff46}
\and
Laboratoire univers et particules de Montpellier, Universit\'e de Montpellier, CNRS, 34090 Montpellier, France\label{aff47}
\and
Institut d'Estudis Espacials de Catalunya (IEEC),  Edifici RDIT, Campus UPC, 08860 Castelldefels, Barcelona, Spain\label{aff48}
\and
Instituto de Astrof\'isica e Ci\^encias do Espa\c{c}o, Faculdade de Ci\^encias, Universidade de Lisboa, Campo Grande, 1749-016 Lisboa, Portugal\label{aff49}
\and
Institut de Ciencies de l'Espai (IEEC-CSIC), Campus UAB, Carrer de Can Magrans, s/n Cerdanyola del Vall\'es, 08193 Barcelona, Spain\label{aff50}
\and
Laboratoire Univers et Th\'eorie, Observatoire de Paris, Universit\'e PSL, Universit\'e Paris Cit\'e, CNRS, 92190 Meudon, France\label{aff51}
\and
INAF-IASF Milano, Via Alfonso Corti 12, 20133 Milano, Italy\label{aff52}
\and
ICSC - Centro Nazionale di Ricerca in High Performance Computing, Big Data e Quantum Computing, Via Magnanelli 2, Bologna, Italy\label{aff53}
\and
Dipartimento di Scienze Matematiche, Fisiche e Informatiche, Universit\`a di Parma, Viale delle Scienze 7/A 43124 Parma, Italy\label{aff54}
\and
INFN Gruppo Collegato di Parma, Viale delle Scienze 7/A 43124 Parma, Italy\label{aff55}
\and
Jet Propulsion Laboratory, California Institute of Technology, 4800 Oak Grove Drive, Pasadena, CA, 91109, USA\label{aff56}
\and
Department of Physics, P.O. Box 64, University of Helsinki, 00014 Helsinki, Finland\label{aff57}
\and
INAF-Osservatorio Astronomico di Brera, Via Brera 28, 20122 Milano, Italy\label{aff58}
\and
Max Planck Institute for Extraterrestrial Physics, Giessenbachstr. 1, 85748 Garching, Germany\label{aff59}
\and
Dipartimento di Fisica, Universit\`a di Genova, Via Dodecaneso 33, 16146, Genova, Italy\label{aff60}
\and
INFN-Sezione di Genova, Via Dodecaneso 33, 16146, Genova, Italy\label{aff61}
\and
INAF-Osservatorio Astronomico di Capodimonte, Via Moiariello 16, 80131 Napoli, Italy\label{aff62}
\and
INFN section of Naples, Via Cinthia 6, 80126, Napoli, Italy\label{aff63}
\and
Faculdade de Ci\^encias da Universidade do Porto, Rua do Campo de Alegre, 4150-007 Porto, Portugal\label{aff64}
\and
Aix-Marseille Universit\'e, CNRS, CNES, LAM, Marseille, France\label{aff65}
\and
Centro de Investigaciones Energ\'eticas, Medioambientales y Tecnol\'ogicas (CIEMAT), Avenida Complutense 40, 28040 Madrid, Spain\label{aff66}
\and
Port d'Informaci\'{o} Cient\'{i}fica, Campus UAB, C. Albareda s/n, 08193 Bellaterra (Barcelona), Spain\label{aff67}
\and
Institute for Astronomy, University of Hawaii, 2680 Woodlawn Drive, Honolulu, HI 96822, USA\label{aff68}
\and
Dipartimento di Fisica e Astronomia "Augusto Righi" - Alma Mater Studiorum Universit\`a di Bologna, Viale Berti Pichat 6/2, 40127 Bologna, Italy\label{aff69}
\and
Instituto de Astrof\'{\i}sica de Canarias, E-38205 La Laguna, Tenerife, Spain\label{aff70}
\and
Jodrell Bank Centre for Astrophysics, Department of Physics and Astronomy, University of Manchester, Oxford Road, Manchester M13 9PL, UK\label{aff71}
\and
European Space Agency/ESRIN, Largo Galileo Galilei 1, 00044 Frascati, Roma, Italy\label{aff72}
\and
ESAC/ESA, Camino Bajo del Castillo, s/n., Urb. Villafranca del Castillo, 28692 Villanueva de la Ca\~nada, Madrid, Spain\label{aff73}
\and
Universit\'e Claude Bernard Lyon 1, CNRS/IN2P3, IP2I Lyon, UMR 5822, Villeurbanne, F-69100, France\label{aff74}
\and
Institute of Physics, Laboratory of Astrophysics, Ecole Polytechnique F\'ed\'erale de Lausanne (EPFL), Observatoire de Sauverny, 1290 Versoix, Switzerland\label{aff75}
\and
Institut de Ci\`{e}ncies del Cosmos (ICCUB), Universitat de Barcelona (IEEC-UB), Mart\'{i} i Franqu\`{e}s 1, 08028 Barcelona, Spain\label{aff76}
\and
Instituci\'o Catalana de Recerca i Estudis Avan\c{c}ats (ICREA), Passeig de Llu\'{\i}s Companys 23, 08010 Barcelona, Spain\label{aff77}
\and
UCB Lyon 1, CNRS/IN2P3, IUF, IP2I Lyon, 4 rue Enrico Fermi, 69622 Villeurbanne, France\label{aff78}
\and
Mullard Space Science Laboratory, University College London, Holmbury St Mary, Dorking, Surrey RH5 6NT, UK\label{aff79}
\and
Departamento de F\'isica, Faculdade de Ci\^encias, Universidade de Lisboa, Edif\'icio C8, Campo Grande, PT1749-016 Lisboa, Portugal\label{aff80}
\and
Department of Astronomy, University of Geneva, ch. d'Ecogia 16, 1290 Versoix, Switzerland\label{aff81}
\and
INAF-Istituto di Astrofisica e Planetologia Spaziali, via del Fosso del Cavaliere, 100, 00100 Roma, Italy\label{aff82}
\and
Universit\'e Paris-Saclay, CNRS, Institut d'astrophysique spatiale, 91405, Orsay, France\label{aff83}
\and
Aix-Marseille Universit\'e, CNRS/IN2P3, CPPM, Marseille, France\label{aff84}
\and
Universit\'e Paris-Saclay, Universit\'e Paris Cit\'e, CEA, CNRS, AIM, 91191, Gif-sur-Yvette, France\label{aff85}
\and
School of Physics, HH Wills Physics Laboratory, University of Bristol, Tyndall Avenue, Bristol, BS8 1TL, UK\label{aff86}
\and
Istituto Nazionale di Fisica Nucleare, Sezione di Bologna, Via Irnerio 46, 40126 Bologna, Italy\label{aff87}
\and
Leiden Observatory, Leiden University, Einsteinweg 55, 2333 CC Leiden, The Netherlands\label{aff88}
\and
Department of Physics, Lancaster University, Lancaster, LA1 4YB, UK\label{aff89}
\and
Felix Hormuth Engineering, Goethestr. 17, 69181 Leimen, Germany\label{aff90}
\and
Technical University of Denmark, Elektrovej 327, 2800 Kgs. Lyngby, Denmark\label{aff91}
\and
Cosmic Dawn Center (DAWN), Denmark\label{aff92}
\and
Institut d'Astrophysique de Paris, UMR 7095, CNRS, and Sorbonne Universit\'e, 98 bis boulevard Arago, 75014 Paris, France\label{aff93}
\and
Max-Planck-Institut f\"ur Astronomie, K\"onigstuhl 17, 69117 Heidelberg, Germany\label{aff94}
\and
NASA Goddard Space Flight Center, Greenbelt, MD 20771, USA\label{aff95}
\and
Department of Physics and Helsinki Institute of Physics, Gustaf H\"allstr\"omin katu 2, University of Helsinki, 00014 Helsinki, Finland\label{aff96}
\and
Helsinki Institute of Physics, Gustaf H{\"a}llstr{\"o}min katu 2, University of Helsinki, 00014 Helsinki, Finland\label{aff97}
\and
SKAO, Jodrell Bank, Lower Withington, Macclesfield SK11 9FT, UK\label{aff98}
\and
Centre de Calcul de l'IN2P3/CNRS, 21 avenue Pierre de Coubertin 69627 Villeurbanne Cedex, France\label{aff99}
\and
Universit\"at Bonn, Argelander-Institut f\"ur Astronomie, Auf dem H\"ugel 71, 53121 Bonn, Germany\label{aff100}
\and
Dipartimento di Fisica e Astronomia "Augusto Righi" - Alma Mater Studiorum Universit\`a di Bologna, via Piero Gobetti 93/2, 40129 Bologna, Italy\label{aff101}
\and
Department of Physics, Institute for Computational Cosmology, Durham University, South Road, Durham, DH1 3LE, UK\label{aff102}
\and
Universit\'e C\^{o}te d'Azur, Observatoire de la C\^{o}te d'Azur, CNRS, Laboratoire Lagrange, Bd de l'Observatoire, CS 34229, 06304 Nice cedex 4, France\label{aff103}
\and
Universit\'e Paris Cit\'e, CNRS, Astroparticule et Cosmologie, 75013 Paris, France\label{aff104}
\and
University of Applied Sciences and Arts of Northwestern Switzerland, School of Engineering, 5210 Windisch, Switzerland\label{aff105}
\and
Institut d'Astrophysique de Paris, 98bis Boulevard Arago, 75014, Paris, France\label{aff106}
\and
Aurora Technology for European Space Agency (ESA), Camino bajo del Castillo, s/n, Urbanizacion Villafranca del Castillo, Villanueva de la Ca\~nada, 28692 Madrid, Spain\label{aff107}
\and
Institut de F\'{i}sica d'Altes Energies (IFAE), The Barcelona Institute of Science and Technology, Campus UAB, 08193 Bellaterra (Barcelona), Spain\label{aff108}
\and
School of Mathematics and Physics, University of Surrey, Guildford, Surrey, GU2 7XH, UK\label{aff109}
\and
School of Mathematics, Statistics and Physics, Newcastle University, Herschel Building, Newcastle-upon-Tyne, NE1 7RU, UK\label{aff110}
\and
DARK, Niels Bohr Institute, University of Copenhagen, Jagtvej 155, 2200 Copenhagen, Denmark\label{aff111}
\and
Waterloo Centre for Astrophysics, University of Waterloo, Waterloo, Ontario N2L 3G1, Canada\label{aff112}
\and
Department of Physics and Astronomy, University of Waterloo, Waterloo, Ontario N2L 3G1, Canada\label{aff113}
\and
Perimeter Institute for Theoretical Physics, Waterloo, Ontario N2L 2Y5, Canada\label{aff114}
\and
Centre National d'Etudes Spatiales -- Centre spatial de Toulouse, 18 avenue Edouard Belin, 31401 Toulouse Cedex 9, France\label{aff115}
\and
Institute of Space Science, Str. Atomistilor, nr. 409 M\u{a}gurele, Ilfov, 077125, Romania\label{aff116}
\and
Universidad de La Laguna, Dpto. Astrof\'\i sica, E-38206 La Laguna, Tenerife, Spain\label{aff117}
\and
Consejo Superior de Investigaciones Cientificas, Calle Serrano 117, 28006 Madrid, Spain\label{aff118}
\and
Departamento de F\'isica, FCFM, Universidad de Chile, Blanco Encalada 2008, Santiago, Chile\label{aff119}
\and
Universit\"at Innsbruck, Institut f\"ur Astro- und Teilchenphysik, Technikerstr. 25/8, 6020 Innsbruck, Austria\label{aff120}
\and
Satlantis, University Science Park, Sede Bld 48940, Leioa-Bilbao, Spain\label{aff121}
\and
Department of Physics, Royal Holloway, University of London, Surrey TW20 0EX, UK\label{aff122}
\and
Instituto de Astrof\'isica e Ci\^encias do Espa\c{c}o, Faculdade de Ci\^encias, Universidade de Lisboa, Tapada da Ajuda, 1349-018 Lisboa, Portugal\label{aff123}
\and
Cosmic Dawn Center (DAWN)\label{aff124}
\and
Niels Bohr Institute, University of Copenhagen, Jagtvej 128, 2200 Copenhagen, Denmark\label{aff125}
\and
Universidad Polit\'ecnica de Cartagena, Departamento de Electr\'onica y Tecnolog\'ia de Computadoras,  Plaza del Hospital 1, 30202 Cartagena, Spain\label{aff126}
\and
Kapteyn Astronomical Institute, University of Groningen, PO Box 800, 9700 AV Groningen, The Netherlands\label{aff127}
\and
Infrared Processing and Analysis Center, California Institute of Technology, Pasadena, CA 91125, USA\label{aff128}
\and
INAF, Istituto di Radioastronomia, Via Piero Gobetti 101, 40129 Bologna, Italy\label{aff129}
\and
Astronomical Observatory of the Autonomous Region of the Aosta Valley (OAVdA), Loc. Lignan 39, I-11020, Nus (Aosta Valley), Italy\label{aff130}
\and
Kavli Institute for Cosmology Cambridge, Madingley Road, Cambridge, CB3 0HA, UK\label{aff131}
\and
Institute of Astronomy, University of Cambridge, Madingley Road, Cambridge CB3 0HA, UK\label{aff132}
\and
Zentrum f\"ur Astronomie, Universit\"at Heidelberg, Philosophenweg 12, 69120 Heidelberg, Germany\label{aff133}
\and
ICL, Junia, Universit\'e Catholique de Lille, LITL, 59000 Lille, France\label{aff134}
\and
Instituto de F\'isica Te\'orica UAM-CSIC, Campus de Cantoblanco, 28049 Madrid, Spain\label{aff135}
\and
CERCA/ISO, Department of Physics, Case Western Reserve University, 10900 Euclid Avenue, Cleveland, OH 44106, USA\label{aff136}
\and
Departamento de F{\'\i}sica Fundamental. Universidad de Salamanca. Plaza de la Merced s/n. 37008 Salamanca, Spain\label{aff137}
\and
Center for Data-Driven Discovery, Kavli IPMU (WPI), UTIAS, The University of Tokyo, Kashiwa, Chiba 277-8583, Japan\label{aff138}
\and
University Observatory, LMU Faculty of Physics, Scheinerstr.~1, 81679 Munich, Germany\label{aff139}
\and
Max-Planck-Institut f\"ur Physik, Boltzmannstr. 8, 85748 Garching, Germany\label{aff140}
\and
California Institute of Technology, 1200 E California Blvd, Pasadena, CA 91125, USA\label{aff141}
\and
Department of Physics \& Astronomy, University of California Irvine, Irvine CA 92697, USA\label{aff142}
\and
Department of Astronomy \& Physics and Institute for Computational Astrophysics, Saint Mary's University, 923 Robie Street, Halifax, Nova Scotia, B3H 3C3, Canada\label{aff143}
\and
Departamento F\'isica Aplicada, Universidad Polit\'ecnica de Cartagena, Campus Muralla del Mar, 30202 Cartagena, Murcia, Spain\label{aff144}
\and
CEA Saclay, DFR/IRFU, Service d'Astrophysique, Bat. 709, 91191 Gif-sur-Yvette, France\label{aff145}
\and
Department of Computer Science, Aalto University, PO Box 15400, Espoo, FI-00 076, Finland\label{aff146}
\and
 Instituto de Astrof\'{\i}sica de Canarias, E-38205 La Laguna; Universidad de La Laguna, Dpto. Astrof\'\i sica, E-38206 La Laguna, Tenerife, Spain\label{aff147}
\and
Ruhr University Bochum, Faculty of Physics and Astronomy, Astronomical Institute (AIRUB), German Centre for Cosmological Lensing (GCCL), 44780 Bochum, Germany\label{aff148}
\and
Univ. Grenoble Alpes, CNRS, Grenoble INP, LPSC-IN2P3, 53, Avenue des Martyrs, 38000, Grenoble, France\label{aff149}
\and
Department of Physics and Astronomy, Vesilinnantie 5, University of Turku, 20014 Turku, Finland\label{aff150}
\and
Serco for European Space Agency (ESA), Camino bajo del Castillo, s/n, Urbanizacion Villafranca del Castillo, Villanueva de la Ca\~nada, 28692 Madrid, Spain\label{aff151}
\and
ARC Centre of Excellence for Dark Matter Particle Physics, Melbourne, Australia\label{aff152}
\and
Centre for Astrophysics \& Supercomputing, Swinburne University of Technology,  Hawthorn, Victoria 3122, Australia\label{aff153}
\and
School of Physics and Astronomy, Queen Mary University of London, Mile End Road, London E1 4NS, UK\label{aff154}
\and
Department of Physics and Astronomy, University of the Western Cape, Bellville, Cape Town, 7535, South Africa\label{aff155}
\and
Universit\'e Libre de Bruxelles (ULB), Service de Physique Th\'eorique CP225, Boulevard du Triophe, 1050 Bruxelles, Belgium\label{aff156}
\and
DAMTP, Centre for Mathematical Sciences, Wilberforce Road, Cambridge CB3 0WA, UK\label{aff157}
\and
IRFU, CEA, Universit\'e Paris-Saclay 91191 Gif-sur-Yvette Cedex, France\label{aff158}
\and
Oskar Klein Centre for Cosmoparticle Physics, Department of Physics, Stockholm University, Stockholm, SE-106 91, Sweden\label{aff159}
\and
Astrophysics Group, Blackett Laboratory, Imperial College London, London SW7 2AZ, UK\label{aff160}
\and
INAF-Osservatorio Astrofisico di Arcetri, Largo E. Fermi 5, 50125, Firenze, Italy\label{aff161}
\and
Dipartimento di Fisica, Sapienza Universit\`a di Roma, Piazzale Aldo Moro 2, 00185 Roma, Italy\label{aff162}
\and
Dipartimento di Fisica, Universit\`a di Roma Tor Vergata, Via della Ricerca Scientifica 1, Roma, Italy\label{aff163}
\and
HE Space for European Space Agency (ESA), Camino bajo del Castillo, s/n, Urbanizacion Villafranca del Castillo, Villanueva de la Ca\~nada, 28692 Madrid, Spain\label{aff164}
\and
Department of Astrophysical Sciences, Peyton Hall, Princeton University, Princeton, NJ 08544, USA\label{aff165}
\and
INAF-Osservatorio Astronomico di Brera, Via Brera 28, 20122 Milano, Italy, and INFN-Sezione di Genova, Via Dodecaneso 33, 16146, Genova, Italy\label{aff166}
\and
Theoretical astrophysics, Department of Physics and Astronomy, Uppsala University, Box 516, 751 37 Uppsala, Sweden\label{aff167}
\and
Center for Cosmology and Particle Physics, Department of Physics, New York University, New York, NY 10003, USA\label{aff168}
\and
Center for Computational Astrophysics, Flatiron Institute, 162 5th Avenue, 10010, New York, NY, USA\label{aff169}}

\date{\today }
  
\authorrunning{Euclid Collaboration: N.~Frusciante et al.}

\titlerunning{\Euclid: Review of forecast constraints on dark energy and modified gravity}

\abstract{
The \Euclid mission has been designed to provide, as one of its main deliverables, information on the nature of the gravitational interaction, which determines the expansion of the Universe and the formation of structures. Thus, \Euclid has the potential to test deviations from general relativity that will allow us to shed light on long-lasting problems in the standard cosmological model, \lcdm. \Euclid will mainly do this by using two complementary probes: weak gravitational lensing and galaxy clustering. 
In this paper we review pre-launch \Euclid analyses for dark energy and modified gravity.  These include forecast constraints with future \Euclid data on cosmological parameters for different cosmological models, such as a time-varying dark energy component, phenomenological modifications of the perturbation sector and specific modified gravity models, with further extensions that include neutrino physics and the coupling to the electromagnetic sector through the fine-structure constant.  We review the study of the impact of nonlinear clustering  methods on beyond-\lcdm constraints with \Euclid.  This is of fundamental importance to efficiently predict the large-scale clustering of matter and dark matter halos, given that we will have access to a wealth of information on scales beyond the linear regime. We inspect the extension of theoretical predictions for observable quantities in alternative cosmologies to \lcdm at fully nonlinear scales by means of $N$-body simulations. We discuss the impact of relativistic corrections in extended cosmological models. Overall,  this review highlights the significant potential of the \Euclid mission to tightly constrain parameters of dark energy and modified gravity models, or perhaps to detect possible signatures of a \lcdm failure.}

\keywords{Gravitational lensing: weak --- large-scale structure of Universe --- cosmological parameters}

\maketitle
  
\section{Introduction}\label{Sec:Intro}
The recently launched \Euclid mission is set to deliver an extensive catalogue of data on the positions and redshifts of billions of galaxies, providing an unparalleled view of the large-scale structure of the Universe \citep{Redbook,2022arXiv221109668T,EuclidSkyOverview}. This endeavour represents a remarkable experimental test for the \lcdm model, the prevailing standard model of cosmology, offering a comprehensive means to identify and constrain any deviations or unexpected signatures beyond it.

By analysing extensions to general relativity (GR) as the fundamental description of gravity, \Euclid is expected to provide further insight into the 25-year old problem regarding the origin of the present accelerated expansion of the Universe \citep{SupernovaSearchTeam:1998fmf,SupernovaCosmologyProject:1998vns}. Additionally, it may offer clues about its nature, testing the predictions of theories beyond the standard model and potentially addressing the cosmological observational tensions that arise under the \lcdm framework \citep{Abdalla:2022yfr}. 
From a phenomenological point of view, such extensions can arise from modifications to the matter or gravitational sectors, embodied by effective theories of dark energy (DE) and modified gravity (MG), respectively.

The \Euclid satellite will foster the investigation of such models through the complementary use of both its spectroscopic and photometric (imaging) surveys. Specifically, \Euclid is most effective at probing the clustering of galaxies, effects of weak gravitational lensing, and galaxy-galaxy lensing, which it will use to unravel the intricacies of the Universe on large scales.

The Euclid Consortium has provided several forecasts of the capabilities of the satellite to constrain DE and MG models.  In this paper, our aim is to conduct a review of these analyses, providing a global picture of \Euclid's capabilities in this class of theories, as well as an overview of the methods used to overcome the complications brought to the analysis pipeline by extensions of \lcdm.

With the validated forecast presented in \citet{Euclid:2019clj}, hereafter \citetalias{Euclid:2019clj}, a first set of results was provided on the DE and MG models, focusing mostly on simple parametric extensions of the standard cosmological model. Subsequently, an effort was dedicated to specific models with \citet{Euclid:2023tqw} and \citet{Euclid:2023rjj} considering popular MG models, such as $f(R)$, Dvali--Gabadadze--Porrati (DGP) and Jordan--Brans--Dicke (JBD) theories. The focus on specific models allowed us to investigate the interplay between these models and other effects that could mimic their signatures. An example is given by massive neutrinos, as including their mass $\sum{m_\nu}$ as a free parameter in the analysis can significantly affect theoretical predictions and introduce degeneracies with DE and MG parameters \citep{He:2013qha,Hu:2014sea,Frusciante:2019puu,Ballardini:2020iws,Frusciante:2020gkx,Atayde:2023aoj}. The analysis of \citet{EUCLID:KP_nu}, hereafter \citetalias{EUCLID:KP_nu}, highlighted how the sensitivity of \Euclid will also be crucial to disentangle and constrain such effects, when more exotic possibilities are considered, such as variations in fundamental constants \citep{Euclid:2021cfn}. 
Similarly, the Euclid Collaboration embarked on modelling the phenomenology of late-time acceleration using model-independent methods, with the aim of understanding its fundamental physical nature \citep[][hereafter \citetalias{Euclid:2025tpw}]{Euclid:2025tpw}. These particular approaches are the effective field theories (EFT) of DE and MG \citep{Frusciante:2019xia} and the gravitational couplings, namely the effective gravitational coupling ($\mu$) and the light-deflection parameter \citep[$\Sigma$; see][]{Amendola:2007rr,Silvestri:2013ne}. These allow us to perform a systematic investigation of theories beyond \lcdm and also to interpret current and future data in terms of the physical effects they parameterize.  Such a systematic exploration of the landscape of an alternative cosmological model aims at putting them to test when data will be available, similarly to what was done in \cite{2016A&A...594A..14P}.

The effort of the Euclid Consortium has been limited not only to obtaining forecasts for this class of models, but also to investigating the robustness of the methods used to obtain theoretical predictions. In particular, investigation has focused on the two extreme regimes of large and small scales, where the capabilities of \Euclid require our modelling to overcome the approximations usually adopted for these cases \citep{Lorenz:2017iez,Fang:2019xat}.

In the large-scale regime, we expect the contribution of relativistic effects to become significant and the impact of DE and MG models to play a fundamental role. In \cite{Lombriser:2013aj}, the authors investigated the particular case of theories with a relativistic description of the galaxy clustering process according to the parameterized post-Friedmann (PPF) framework developed in \cite{Hu:2007pj}. 
The study in \cite{Lombriser:2013aj} encompassed various models, including quintessence, $f(R)$ gravity, a Brans--Dicke universe featuring a dynamical Brans--Dicke parameter and scalar-field potential, both branches of the DGP model, and phenomenological modifications of gravity.

Another approach was implemented in \cite{baker_observational_2015}, where the authors explored the impact of modifications to the gravitational field equations on perturbations around the Hubble scale within a generic modified gravity theory. This investigation revealed that the contribution of relativistic corrections to the angular spectrum of number counts entails distinctive scale- and redshift-dependent deviations in MG scenarios.

Focusing on relativistic effects in galaxy number counts in MG theories and the cross-correlation between the large-scale structure (LSS) distribution and other probes, we highlight some relevant works for this study. In \cite{renk_gravity_2016}, the authors aimed to isolate the integrated Sachs--Wolfe (ISW) effect by computing the cross-correlation between LSS and cosmic microwave background (CMB) temperature anisotropies \citep[e.g.][]{2016A&A...594A..14P}. This was achieved for a broad class of general scalar-tensor theories described by the Horndeski Lagrangian \citep{Horndeski:1974wa}, including covariant Galileons, Brans--Dicke theory, and parameterisations rooted in the EFT of dark energy. Importantly, the authors refrained from assuming a quasi-static evolution.
In \cite{Bosi:2023amu}, the focus shifted towards exploring the optimal combination of gravitational-wave and galaxy-survey data sets for constraining MG theories and unveiling the nature of binary black hole events. Finally, in \cite{Abidi:2022zyd}, an estimator for the $E_{\rm G}$ statistic was constructed. Initially proposed in \cite{Zhang:2007nk} and constrained for the first time by \citet{Reyes:2010tr}, this figure measures the evolution of the sum of the potentials relative to the velocity, relying solely on clustering information obtained from two tracers of large-scale structure. However, it is worth noting that the $E_{\rm G}$ estimator could be susceptible to lensing magnification effects, potentially affecting its reliability \citep{Leonard:2015cba,MoradinezhadDizgah:2016pqy}.

In \cite{Duniya_2015_DE}, the focus shifted towards models in which DE interacts non-gravitationally with dark matter (DM). Such interactions within the dark sector can induce considerable effects on the matter power spectrum at large scales. Consequently, neglecting relativistic effects can lead to a misinterpretation of interacting dark energy imprints on horizon scales, potentially introducing biases in the constraints derived for such models on very large scales.

Building on these previous results, the Euclid Collaboration provided a study of the impact of such effects on \Euclid's expected results, both for the photometric \citep{Euclid:2021rez} and spectroscopic \citep{Euclid:2023qyw} surveys.

Similarly to the effect taking place at large scales, the modelling of small scales needs to be reviewed to obtain reliable theoretical predictions. At these scales, the linear approximation used to obtain analytical predictions breaks down, and one needs to rely on different methods to compute the observables. Indeed, it has been shown how the sensitivity of \Euclid and other contemporary galaxy surveys require accurate modelling of these scales, and how relying on common approaches could lead to significant inaccuracies in the final results for both galaxy clustering \citep{Safi:2020ugb} and weak lensing \citep{Euclid:2020tff}.
Indeed, at these scales one often needs to rely on $N$-body simulations to obtain predictions for the evolution of perturbations. While these are available in \lcdm with varying level of detail and also accounting for baryonic effects in addition to the evolution due to gravity \citep{Schneider:2015yka}, in theories that deviate from the standard model these are available only for some specific models, \citep[such as $f(R)$, ][]{Winther:2015wla}. Within the Euclid Collaboration, new numerical methods have been developed to produce simulations in models alternative to \lcdm \citep{Euclid:2024jej} and a suite of available simulations developed to test future \Euclid data analysis will be described in \cite{Euclid:2024guw}.

With these simulations in hand, it is possible to obtain predictions for nonlinear scales for different values of the parameters of the model under study. In fact, one could obtain fitting formulas for the behaviour of the perturbations using simulations performed for different cases \citep{Winther:2019mus}, but, with the development of machine learning techniques, emulators are currently used extensively, in order to obtain fast and reliable predictions \citep{Valogiannis:2016ane, Arnold:2019vpg, Saez-Casares:2023olw}

This review is structured as follows.
\Cref{sec:Method} introduces the \Euclid mission and outlines its scientific goals. It provides an in-depth description of the techniques used to forecast constraints on various cosmological parameters in different models, leveraging the forthcoming \Euclid data. 
\Cref{Sec:DEGrowth} focuses on diverse theoretical frameworks pertaining to time-varying dark energy. We explore these models and their potential extensions for cosmological parameters. This extends to the inclusion of cosmological parameters for massive neutrinos, the effective neutrino number, and the coupling to the electromagnetic sector, encoded in the fine-structure constant. We also consider the signatures of a growth rate different from \lcdm that encompasses possible deviations from \lcdm in the evolution at both the background and perturbation levels. 
\Cref{sec:MGscaleind} summarizes the results within the context of scale-independent gravitational interaction modifications, while \cref{sec:MGscaledep} broadens the discussion to models characterized by scale-dependent growth.
In \cref{Sec:NLmethods} we investigate the pivotal role of nonlinear clustering techniques in shaping the constraints within extended cosmological models using the \Euclid data. This exploration spans both spectroscopic and photometric probes.
Furthermore, in \cref{Sec:RelEff}, we focus on the influence of relativistic effects within such models, including the assessment of potential biases and constraint power.
\Cref{Sec:NSsim} explores predictions for observable quantities on fully nonlinear scales through $N$-body simulations, tailored to incorporate the additional physics characterising each specific class of extended models.
Lastly, we conclude in \cref{Sec:Out}, with a summary of the main results and their implications for the understanding of dark energy and modified gravity signatures with  \Euclid  data.

\section{Forecasting methodology}\label{sec:Method}

In this section, we describe the methodology used to forecast the constraints on the different cosmological parameters for the different models with the future \Euclid data. This includes the spectroscopic survey for galaxy clustering analysis, as well as the photometric survey for galaxy clustering, weak lensing, and galaxy-galaxy lensing analyses (also called 3\texttimes2\,pt. statistics analysis). We first briefly describe the main specifications of both surveys in \cref{Sec:Spec} and then present the method used to obtain the forecast constraints in \cref{Sec:Analysis}.

\subsection{Survey specifications}\label{Sec:Spec}
The \Euclid satellite contains two instruments on-board, namely the Near Infrared Spectrometer and Photometer \citep[NISP,][]{EuclidSkyNISP} and the VISible imager \citep[VIS,][]{EuclidSkyVIS}. The former  
allows us to obtain slitless spectroscopy for tens of millions of galaxies, enabling three-dimensional galaxy clustering analyses. The latter, combined with ground-based observations, provides shape measurements and photometric redshift estimates for about 1.5 billion galaxies. With these, we can perform 3\texttimes2\,pt. statistics analyses. We follow \citetalias{Euclid:2019clj}, for the specifications of both surveys and  summarize them here for completeness.

Starting with the spectroscopic survey, \Euclid will provide measurements over 14\,816\,deg$^2$ from redshift 0.9 to redshift 1.8, and with an expected redshift ($z$) precision of $\sigma_z=0.001(1+z)$ as presented in \cite{EuclidSkyOverview}. The baseline analysis considers four different redshift bins leading to a total of about $2\times 10^{-3}$ galaxies per $h^{-3}\,\rm{Mpc}^{3}$. The exact number densities and galaxy biases per bin can be found in Table 3 of \citetalias{Euclid:2019clj}. It is important to mention that, since the acquisition of the first \Euclid data, some of these specifications have slightly evolved. This includes an area over about 14\,000\,deg$^2$ and more than 25 million galaxies~\citep{EuclidSkyOverview}. However, in the following, we stick to the pre-launch specifications detailed in~\citetalias{Euclid:2019clj} to allow for a comparison with the results obtained for the concordance model.

With respect to the photometric survey, \Euclid will cover the same area from redshift 0 to about 2.5. The expected number density in this case is 30 galaxies ${\rm arcmin}^{-2}$ with an average precision of $\sigma_z=0.05(1+z)$ -- see Eq.~(112) in \citetalias{Euclid:2019clj} for a detailed expression for the photometric redshift precision. We follow \citetalias{Euclid:2019clj} in considering ten tomographic bins with the same number density in each bin. Finally, we consider a total intrinsic ellipticity dispersion given by $\sigma_{\epsilon}=0.30$. Like in the spectroscopic case, we note that the current baseline is 13 tomographic bins and the total intrinsic ellipticity dispersion is considered to be $\sigma_{\epsilon}=0.37$~\citep{EuclidSkyOverview}. However, we consider here the specifications presented in~\citetalias{Euclid:2019clj} for comparison purposes.

We also note that the survey specifications mentioned above describe the so-called Euclid Wide Survey \citep{Scaramella}. We do not consider here the Euclid Deep Survey, which will probe deeper in smaller regions of the sky \citep[see e.g.,][]{2022arXiv221109668T}. These regions will be used for calibration and are needed to ensure the specifications considered in the wide survey.

\subsection{Analysis method}\label{Sec:Analysis}
We consider a Fisher-matrix formalism to estimate the constraints on the different parameters of the models considered. We follow the approach described in \citetalias{Euclid:2019clj}, \cite{Euclid:2023tqw}, and \cite{Euclid:2023rjj}, but we account for the required modifications for each of the models considered. Although most of the results reported in this work are obtained using such an approach, Monte Carlo Markov Chain (MCMC) methods were employed for some cases, given the strong non-Gaussian nature of the parameter distribution. We explicitly state the use of MCMC methods for the cases where this is done.

The main observable that we consider for the spectroscopic survey is the observed galaxy power spectrum. We model it as follows:
\begin{multline}
P_\text{obs}(k_{\rm ref}, \mu_{\theta, \rm ref} ;z) = 
\frac{1}{q_\perp^2(z)\, q_\parallel(z)} %AP
\left\{\frac{\left[b\sigma_8(z)+f\sigma_8(z)\mu_{\theta}^2\right]^2}{1+ \left[ f(z)\ k\ \mu_{\theta}\ \sigma_{\rm p}(z) \right]^2 } \right\} 
\\
\times \frac{P_\text{dw}(k,\mu_{\theta};z)}{\sigma_8^2(z)}  %nonlinear damping
F_z(k,\mu_{\theta};z) %z-error
+ P_\text{s}(z) \,, % shot noise
\label{eq:GC:pk-ext}
\end{multline}
where $f(k,z)$ is the growth rate and all $k=k(k_{\rm ref},\mu_{\theta, \rm ref})$ and $\mu_{\theta}=\mu_{\theta}(\mu_{\theta, \rm ref})$ 
can be expressed as
\begin{align}
k(k_{\rm ref},\mu_{\theta, \rm ref}) &= \frac{k_{\rm ref}}{q_{\perp}}\left[1+\mu_{\theta, \rm ref}^2\left(\frac{q_{\perp}^2}{q_{\parallel}^2}-1\right)\right]^{1/2}\,,\\
\mu_{\theta}(\mu_{\theta, \rm ref}) &= \mu_{\theta, \rm ref}\frac{q_{\perp}}{q_{\parallel}}\left[1+\mu_{\theta, \rm ref}^2\left(\frac{q_{\perp}^2}{q_{\parallel}^2}-1\right)\right]^{-1/2}\,.
\end{align}
%-- see Eq. (78) in \citetalias{Euclid:2019clj} for the detailed expressions. 
$P_{\rm dw}$ is the `de-wiggled' power spectrum, which accounts for the smearing of the baryon acoustic oscillation (BAO) features:
\begin{equation}
P_\text{dw}(k,\mu_{\theta};z) = P_{\delta\delta}^{\rm lin}(k;z)\,\text{e}^{-g_\mu k^2} + P_\text{nw}(k;z)\left(1-\text{e}^{-g_\mu k^2}\right) \,,
\label{eq:pk_dw}
\end{equation}
where $P_{\rm nw}$ stands for a no-wiggle power spectrum with the same broadband as the linear power spectrum, $P_{\delta\delta}^{\rm lin}$, but without the baryon acoustic oscillation wiggles. We note that the de-wiggled power spectrum uses the linear matter power spectrum; therefore, we consider linear theory for our spectroscopic observable. However, we phenomenologically account for the nonlinear behaviour through de-wiggling and a nonlinear damping factor \citep{Eisenstein:2006nj},
\be
g_\mu^{2}(%k,
\mu_\theta,z) = 
\frac{1}{6\pi^2} \int_{k_{\rm min}}^{k_{\rm max}} \de k\, P_{\delta \delta }^{\rm lin} (k,z)\left\{ 1 - \mu_\theta^2 + \mu_\theta^2 [1 + f(k,z)]^2  \right\}\,,
\ee
where we have numerically integrated the linear matter power spectrum from $k_{\rm min}=0.001\,h\,$Mpc$^{-1}$ to $k_{\rm max}\approx 10\,h\,$Mpc$^{-1}$.
 
We also note from \cref{eq:GC:pk-ext} that we account for redshift-space distortions through the term in the curly brackets, which contains the galaxy bias times the root mean square (RMS) of matter fluctuations on scales of 8\,$h^{-1}$\,Mpc, $b\sigma_8$, and the product of the growth rate $f\sigma_8$, and the square of the cosine of the angle between the wavevector $\vec{k}$ and the line-of-sight direction, $\mu_{\theta}^2$. In \cref{eq:GC:pk-ext} we also account for the `finger-of-God' effect with the denominator of the curly brackets \citep{Kaiser:1987qv}. For simplicity, we assume %$\sigma_{\rm v}^2 = \sigma_{\rm p}^2$
$\sigma_{\rm p}=g_{\mu}$, which we evaluate in every redshift bin and keep fixed in the analyses. To account for the redshift uncertainty, we additionally introduce the following
\begin{equation}
    F_z(k, \mu_{\theta};z) = \text{e}^{-k^2\mu_{\theta}^2\sigma_{r}^2(z)}\,,
\end{equation}
with $\sigma_{r}^2(z) = c\,(1+z)\,\sigma_{z}/H(z)$, while we also consider the residual shot noise, $P_{\rm s}$, as a nuisance parameter. Finally, the quantities
\begin{equation}
q_{\perp}(z) = \frac{D_{\rm A}(z)}{D_{\rm A,\, ref}(z)} \quad \text{and} \quad
q_{\parallel}(z) = \frac{H_\text{ref}(z)}{H(z)}\,,
\end{equation}
account for the Alcock--Paczynski effect. We follow \cite{Euclid:2023tqw} and \cite{Euclid:2023rjj} in directly computing the derivatives of the observed power spectrum with respect to the cosmological parameters in the Fisher matrix analysis, in contrast to the projection considered in \citetalias{Euclid:2019clj}.

For the photometric survey, we consider the harmonic-space angular power spectra $C_{ij}^{XY}(\ell)$, where $i$ and $j$ stand for two tomographic bins, and $X$ and $Y$ represent photometric galaxy clustering or weak lensing. Under the Limber approximation, these are given by
\begin{equation}
 C^{XY}_{ij}(\ell) = c\int_{z_{\rm min}}^{z_{\rm max}}\de z\,{\frac{W_i^X(z)W_j^Y(z)}{H(z)r^2(z)}P_{\delta\delta}(k_\ell,z)}\,, \label{eq:ISTrecipe}
\end{equation}
where $k_\ell=(\ell+1/2)/r(z)$, and $r(z)$ stands for the comoving distance, while $P_{\delta\delta}$ represents the nonlinear matter power spectrum. Because small scales are probed by photometric probes, the nonlinear spectrum is usually obtained from emulators or fitting functions calibrated with simulations. Because of that, the spectra depend on the specific model considered and will be detailed in each of the sections below for the different models.

The remaining ingredients in \cref{eq:ISTrecipe} are the kernels for galaxy clustering or weak lensing,
\begin{align}
 W_i^{\rm G}(k,z) =&\; b_i(k,z)\,\frac{n_i(z)}{\bar{n}_i}\frac{H(z)}{c}\,, \label{eq:wg_mg}\\  
 W_i^{\rm L}(k,z) =&\; \frac{3}{2}\Omegam \,\frac{H_0^2}{c^2}\,(1+z)\,r(z)\,\Sigma(k,z)
\int_z^{z_{\rm max}}{\de z'\frac{n_i(z')}{\bar{n}_i}\frac{r(z'-z)}{r(z')}}\nonumber\\
  &+W^{\rm IA}_i(k,z)\,, \label{eq:wl_mg}
\end{align}
where $n_i/\bar{n}_i$ represents the normalized number density of galaxies, $H(z)$ and $H_0$ the Hubble function and constant, respectively, and $\Sigma$ encodes the changes to the lensing potential due to the specific modified gravity model. We also consider the intrinsic alignment of galaxies with the extended nonlinear alignment (eNLA) model, which leads to the kernel
\begin{equation}\label{eq:IA}
 W^{\rm IA}_i(k,z)=-\frac{\mathcal{A}_{\rm IA}\,\mathcal{C}_{\rm IA}\,\Omega_{\rm m,0}\,\mathcal{F}_{\rm IA}(z)}{\delta(k,z)/\delta(k,z=0)}\frac{n_i(z)}{\bar{n}_i(z)}\,\frac{H(z)}{c}\,,
\end{equation}
with 
\begin{equation}
 \mathcal{F}_{\rm IA}(z)=(1+z)^{\eta_{\rm IA}}\left[\frac{\langle L\rangle(z)}{L_\star(z)}\right]^{\beta_{\rm IA}}\,,
\end{equation}
where $\langle L\rangle(z)$ and $L_{\star}(z)$ are the redshift-dependent mean and the characteristic luminosity of the source galaxies, as computed from the luminosity function. We define $\Omega_{\rm m,0}\equiv 8\pi G_{\rm N}\, \bar{\rho}_{\rm m,0}/3H_0^2$ as the matter density parameter, with $G_{\rm N}$ Newton's gravitational constant.
More details on the eNLA model and the origin of the luminosity dependence can be found in \citetalias{Euclid:2019clj}.

 One could wonder whether the eNLA model which has been developed in the GR framework can still be used when deviations from GR are considered. To our knowledge, there are no dedicated studies to the  IA modelling for the specific MG models we have considered in this paper. As a general remark, we note that IA is expected to originate from tidal forces on typical scales of galaxies and clusters. As far as the gravitational potential on these scales goes back to the GR form, one can still rely on the IA models developed and tested in the GR case. This is the case for the models of interest here, and for the specific choices of their parameters. This is also qualitatively confirmed  by the results in \cite{Reischke2022} who considered IA for Horndeski models, although only in the linear regime. They indeed found that the GI and II power spectra have the same profile as in the GR case, the only difference being a possible mismatch in the overall amplitude if ellipticity responds to tidal perturbations in a different way. This unknown parameter can be absorbed in our $\mathcal{A}_{\rm IA}$ one and set to 1. 

Once the predictions for all observables are available, we compute the Fisher matrix as the product of their derivatives with respect to the cosmological parameters times the inverse Gaussian covariance of the observables. We refer the interested reader to \citetalias{Euclid:2019clj} for the detailed expressions. We note that following~\citetalias{Euclid:2019clj}, we do not account for non-Gaussian terms like the super sample covariance or the connected non-Gaussian contribution. Moreover, the fraction of sky observed is implemented through a rescaling of the full sky covariance. More realistic analyses accounting for the mask of the survey and non-Gaussian contributions to the covariance are left for future work.

For all the probes considered, we present an optimistic and a pessimistic scenario following~\citetalias{Euclid:2019clj} for comparison purposes. In the optimistic case, all multipoles between 10 and 5000 are considered for weak lensing, while only multipoles between 10 and 3000 are used for photometric galaxy clustering and galaxy-galaxy lensing to account for the biased tracers. For spectroscopic galaxy clustering, we consider all scales from $k=0.001\,h\,$Mpc$^{-1}$ to $k=0.30\,h\,$Mpc$^{-1}$. In the pessimistic scenario, we restrict the maximum multipole to 1500 for weak lensing and 750 for photometric galaxy clustering and galaxy-galaxy lensing. We also limit the spectroscopic analysis to scales of $k=0.25\,h\,$Mpc$^{-1}$. On top of these cases, we consider a quasi-linear scenario for spectroscopic galaxy clustering, where we limit the maximum scale to $k=0.15\,h\,$Mpc$^{-1}$. These scenarios correspond to those chosen in \cite{Euclid:2023rjj} to explore the impact of nonlinear modelling. Results are shown for the spectroscopic probe alone, the combination of all photometric probes, and the combination of spectroscopic and photometric data. We follow \citetalias{Euclid:2019clj} in neglecting the correlation between the spectroscopic and photometric data sets; however, a redshift cut ($z<0.9$) is applied in the pessimistic scenario to photometric galaxy clustering and galaxy-galaxy lensing to further remove possible overlaps between these and the spectroscopic probe. We note that the redshift cut is only applied when all probes are combined.

The analysis method presented in this section neglects the magnification bias on the photometric and spectroscopic galaxy clustering. In \cref{Sec:RelEff}, we will describe how the recipe is modified to include this effect and report  its impact on the \Euclid forecast.

\section{Dark energy and growth} \label{Sec:DEGrowth}

In this section, we review the theory and forecasts that the Euclid Collaboration obtained for a time-varying dark energy component \citepalias{Euclid:2019clj}, its joint analysis with massive neutrinos ($\nu$), the effective number ($N_{\rm eff}$) of neutrinos \citepalias{EUCLID:KP_nu}, the coupling to the electromagnetic sector through the fine-structure constant \citep[$\alpha$; see][]{Euclid:2021cfn}, and the presence of a growth rate different from that of \lcdm, both with \Euclid alone \citepalias{Euclid:2019clj} and in combination with external data such as CMB surveys \citep{Euclid:2022abc}, thus exploring possible deviations from \lcdm in both the background and perturbation sectors.  For these simple extensions of the $\Lambda$CDM model, the approach of \citetalias{Euclid:2019clj} in modelling nonlinear scales is used, thus following what is discussed in \cref{Sec:Analysis} for spectroscopic clustering, while relying on halo model fitting functions for the photometric analysis \citep{Mead:2015yca,Mead:2016zqy}. In these analyses, linear galaxy bias is assumed.

\subsection{Theory}
\subsubsection{\texorpdfstring{$w_0$}{w0}, \texorpdfstring{$w_a$}{wa}}\label{Sec:Theoryw0wa}

A simple generalization of the cosmological constant consists of assuming that dark energy is a fluid with equation of state $w_{\de} \equiv p_{\de} / (\rho_{\de} c^2)$, where $p_{\de}$ and $\rho_{\de}$ are, respectively, the pressure and density of the fluid, and $c$ is the speed of light.

In general, $w_{\de}(z)$ will be a function of redshift. We will start here from the following largely used parameterisation \citep[CPL,][]{Chevallier:2000qy,Linder:2002et}
\begin{equation}\label{eq:weos}
w_{\de}(z) = w_0 + w_a \frac{z}{1+z} \,,
\end{equation}
characterized by two constant free parameters $w_0$ and $w_a$. The first represents the value of the equation of state at present; the second gives an indication of how fast it changes with redshift. The \lcdm model corresponds to $w_0 = -1$ and $w_a = 0$.
For the corresponding background evolution of $\rho_{\de}$ and of the Hubble function, we refer to the notation used in \citetalias{Euclid:2019clj}, see their Sect. 2.1.

\subsubsection{\texorpdfstring{$w_0$, $w_a$ $+\,\gamma$}{w0wagamma}}\label{Sec:Theorygamma}

If gravity is modified with respect to its description within GR, we expect the growth of perturbations to change as well, with respect to a standard scenario. 
One way to parameterize modifications in the growth rate $f(z)$ consists of assuming
\begin{equation}\label{eq:ggrowth}
f(z) \equiv - \frac{{\rm d}\ln D(z)}{{\rm d} \ln(1+z)} = \left[\Omega_{\rm m}(z)\right]^\gamma \,,
\end{equation}
where $\gamma$ is a constant and its value is well approximated by $0.55$ in \lcdm \citep{Lahav:1991wc, Linder:2005in}. Here, $D(z)$ is the growth factor. As discussed more extensively in \citetalias{Euclid:2019clj}, see their Sect. 2.3, this description typically assumes that there are no modifications to the lensing potential, effectively reducing the degrees of freedom to a single one. 

A more general description of modified gravity at the perturbation level, which requires two functions of time and space, will be considered in \cref{sec:MGscaleind}. We still review results for $\gamma$-parameterisation, as they were obtained in \citetalias{Euclid:2019clj}, to allow us to perform a more direct comparison with \citet{Redbook}.

\subsubsection{\texorpdfstring{$w_0$, $w_a$ $+\sum m_\nu$}{w0wanu}}\label{Sec:Theorynu}

Although the previous modifications considered introduce changes that are only redshift dependent, this is not the case for many of the possible extensions of \lcdm. 
Massive neutrinos represent one of the most noticeable cases of scale-dependent growth \citep{ParticleDataGroup:2022pth}.
They decouple from the photon-baryon plasma at a temperature of $T\sim 1$ MeV, while still in the relativistic regime.
After decoupling, their high thermal velocity $v_\mathrm{th}$ defines the free-streaming wavenumber:
\begin{equation}
    k_\mathrm{fs}(a) = \left(\frac{4\pi\, G_{\rm N}\, \bar{\rho}\, a^2}{v_\mathrm{th}}\right)^{1/2}\,,
    \label{eq:free_streaming}
\end{equation}
where $a$ is the scale factor [related to the redshift $z$ as $a=(1+z)^{-1}$], and $\bar\rho$ is the total background density.
At a given redshift, due to their high thermal velocities, neutrinos cannot cluster in regions of size smaller than $\lambda_\mathrm{fs}=2\pi a/k_\mathrm{fs}$.
As temperature drops, neutrinos eventually transition to the non-relativistic regime at some $z_\mathrm{nr}\sim 2\times 10^3 \ (m_\nu/\mathrm{eV})$, where $m_\nu$ is the neutrino mass of a single species, and because thermal velocities scale as $v_\mathrm{th}=\langle p\rangle m_\nu^{-1}\propto a^{-1}$ the free-streaming wavenumber goes through a minimum of $k_\mathrm{nr}=k_\mathrm{fs}(z=z_\mathrm{nr})\approx 0.018 \ \Omega_\mathrm{m,0}^{1/2} \ (m_\nu/\mathrm{eV})^{1/2} \ h \ \mathrm{Mpc}^{-1}$.
All in all, wavenumbers $k<k_\mathrm{nr}$ are never affected by neutrino free-streaming: in this range of scales neutrinos behave like CDM and their perturbations grow accordingly.
However, all scales for which $k>k_\mathrm{nr}$ undergo free-streaming at some point in cosmic history: in this regime not only neutrino density fluctuations are exponentially damped, but neutrinos themselves slow down the CDM perturbations as $\delta_\mathrm{c}\propto a^{1-(3/5)f_\nu}$ \citep{Lesgourgues:2006nd} during matter domination.
Here we have introduced the neutrino fraction
\begin{equation}
    f_\nu \equiv \frac{\Omega_{\nu,0}}{\Omega_\mathrm{m,0}} = \frac{\sum m_\nu}{93.14\ \mathrm{eV} \ h^2 \ \Omega_\mathrm{m,0}}\,,
\end{equation}
with $\Omega_{\nu,0}\equiv 8\pi\bar{\rho}_{\nu,0} G_{\rm N}/3H_0^2$ being the density parameter of neutrinos, and the sum runs over all massive neutrino species.
This double regime in which neutrinos and CDM perturbations grow of course creates a scale dependence in the growth factor $D(k,z)$ and in turn on the growth rate $f(k,z)$.

Cosmology is mainly sensitive to $f_\nu$ and therefore to $\sum m_\nu$.
At the level of the power spectrum, increasing $\sum m_\nu$ while keeping $\Omega_\mathrm{m,0}$ fixed (thus reducing the CDM contribution, $\Omega_\mathrm{cdm,0}$) results in a more pronounced suppression on scales $k>k_\mathrm{nr}$.
The suppression of the power spectrum with respect to a vanilla-$\Lambda$CDM cosmology can be quantified, at the linear level, as
\begin{eqnarray}
    \frac{\Delta P_\mathrm{mm}}{P_\mathrm{mm}}(k\gg k_\mathrm{nr}) \simeq -8 f_\nu\,, \\
    \frac{\Delta P_\mathrm{cc}}{P_\mathrm{cc}}(k\gg k_\mathrm{nr}) \simeq -6 f_\nu\,,
    \label{eq:Pk_suppression_neutrinos}
\end{eqnarray}
where we have introduced the subscripts `mm' and `cc' to indicate the total matter and `cold' matter (i.e. CDM+baryons) power spectra, respectively.
It is of great importance to distinguish the two cases.
It has been shown \citep[e.g.,][]{cosmonuIchiki,cosmonu1,cosmonu2} that while cosmic shear is sensitive to total matter perturbations via the Poisson equation, for galaxy clustering the `fundamental' density field upon which the observables must be built is cold matter.
In fact, using total matter in the presence of neutrinos would result in scale-dependent linear halo bias and non-universal mass functions.

A more detailed discussion of massive neutrino cosmology, nonlinear methods, theoretical predictions, and their expected impact in the \Euclid survey can be found in \citetalias{EUCLID:KP_nu}.

\subsubsection{\texorpdfstring{$w_0$, $w_a$  $+\sum m_\nu+N_{\rm eff}$}{w0wanu}}\label{Sec:TheoryNeff}

Ultra-relativistic species, including neutrinos before the non-relativistic transition, contribute to the background radiation density $\bar\rho_\mathrm{r}$.
This contribution can be parameterized through an effective number of neutrinos $N_\mathrm{eff}$:
\begin{equation}
    \bar\rho_\mathrm{r} = \bar\rho_\gamma \ \left[1+\frac{7}{8} \ \Gamma_\mathrm\nu^4 \ N_\mathrm{eff}\right]\,,
\end{equation}
where $\bar\rho_\gamma$ is the photon background density and $\Gamma_\nu=(4/11)^{1/3}$ is the neutrino-to-photon temperature ratio.
In principle, one may expect $N_\mathrm{eff}=3$ for three Standard Model neutrinos.
However, decoupling is not instantaneous and up-to-date determinations, including high-precision calculation of QED effects during electron-positron annihilation, set the theoretical prediction to $N_\mathrm{eff}=3.043$ \citep{Cielo:2023bqp}.

The number of effective neutrinos $N_\mathrm{eff}$ fully captures the effect of increasing the number of ultra-relativistic species at the background level, but in the most general case it may not be sufficient if, for instance, the additional relics carry small masses or (self-)interaction.
If we assume the additional particles to be fully thermalized, then for fixed $\Omega_\mathrm{m,0}$ the net effect of increasing $N_\mathrm{eff}$ is mainly to delay radiation-matter equality and subsequently induce a suppression in the linear matter power spectrum on scales $k>k_\mathrm{eq}$.
In contrast, if one fixes the time of radiation-matter equality and the angular scale of the sound horizon \cite[well constrained by CMB experiments, see e.g.,][]{2013PhRvD..87h3008H}, to preserve the ratio $\Omega_\mathrm{m,0}/\Omega_\mathrm{r,0}\equiv 1+z_\mathrm{eq}$, with $\Omega_\mathrm{r,0}$ being the radiation density parameter, one must increase the CDM density. 
This, in turn, enhances the power at wavenumbers above $k_\mathrm{eq}$, while shifting the phase and mildly damping BAOs by reason of a smaller baryon fraction.

In analogy with the case of varying neutrino mass, the growth of perturbations becomes scale dependent.
The main difference is that when varying $m_\nu$, scale independence breaks down at $k_\mathrm{nr}$, while when varying $N_\mathrm{eff}$ this occurs at $k_\mathrm{eq}$.

\subsubsection{\texorpdfstring{$w_0$, $w_a + \alpha$}{w0waalpha}}\label{Sec:Theoryalha}

In the description above, the dynamical dark energy model, which is implicitly assumed to stem from a scalar field, is also assumed to be minimally coupled. However, in physically realistic scalar field-based dynamical dark energy models (including, e.g., quintessence) one naturally expects that the field couples to the rest of the model's degrees of freedom, unless unknown symmetries suppress such a coupling. Notably, from the observational point of view is a coupling to the electromagnetic sector, which leads to a time (redshift) dependence of the fine-structure constant ($\alpha$) and a violation of the weak equivalence principle \citep{Carroll,Dvali,Chiba}. 

Denoting a canonical scalar field $\phi$, which takes the value $\phi_0$ at present time, the coupling stems from a gauge kinetic function $B_F(\phi)$\footnote{We denote the magnetic permeability in vacuum with $\tilde{\mu}_0$ rather than with $\mu_0$ to avoid confusion with later notation for $\mu,\Sigma$ parameterization.}:
\begin{equation}
{\cal L}_{\phi F} = - \frac{1}{4\tilde{\mu}_0} B_F(\phi) F_{\mu\nu}F^{\mu\nu}\,,
\end{equation}
which, given the tight local constraints, can be safely linearized,
\begin{equation}\label{eq:linearexp}
B_F(\phi) = 1 - \zeta \kappa (\phi-\phi_0)\,,
\end{equation}
where we have defined $\kappa^2=8\pi\,G_{\rm N}/c$, making $\zeta$ dimensionless. With these assumptions, one can explicitly relate the evolution of $\alpha$ to that of dark energy, as in \citet{Calabrese:2013lga}. The evolution of $\alpha$ can be written as
\begin{equation}\label{eq:zetadef}
\frac{\Delta \alpha}{\alpha} \equiv \frac{\alpha-\alpha_0}{\alpha_0} =B_F^{-1}(\phi)-1=
\zeta \kappa (\phi-\phi_0) \,.
\end{equation}
Defining the fraction of the dark energy density as
\begin{equation}
\Omega_\phi (z) \equiv \frac{\bar{\rho}_\phi(z)}{\bar{\rho}(z)} \simeq \frac{\bar{\rho}_\phi(z)}{\bar{\rho}_\phi(z)+\bar{\rho}_{\rm m}(z)} \,,
\end{equation}
where in the last step we have neglected the contribution from radiation (since we will be interested in low redshifts, $z<5$, where it is indeed negligible), we find 
\begin{equation} \label{eq:dalfa}
\frac{\Delta\alpha}{\alpha}(z) =\zeta \int_0^{z}\sqrt{3\Omega_\phi(z')\left[1+w_\phi(z')\right]}\frac{{\rm d}z'}{1+z'}\,,
\end{equation}
where $\bar{\rho}_\phi$ and $w_\phi$ are, respectively, the energy density and equation of state of the DE fluid associated with the scalar field $\phi$.

This approach assumes a canonical scalar field, but the argument can be repeated for phantom fields, as discussed in \citet{Phantom}, leading to 
\begin{equation} \label{eq:dalfa2}
\frac{\Delta\alpha}{\alpha}(z) =-\zeta \int_0^{z}\sqrt{3\Omega_\phi(z')\left|1+w_\phi(z')\right|}\frac{{\rm d}z'}{1+z'}\,.
\end{equation}
Physically, the change of sign stems from the fact that one expects phantom fields to roll up the potential rather than down. Naturally, the two definitions match across the phantom divide ($w=-1$).

A varying $\alpha$ violates the Einstein equivalence principle, as first discussed by \citet{Dicke} -- we refer the reader to \citet{Damour} for a recent thorough discussion. The key point for our purposes is that the dimensionless E\"{o}tv\"{o}s parameter $\tilde{\eta}$, which describes the level of violation of the weak equivalence principle (WEP), can be written \citep{Dvali,Chiba,Damour}
\begin{equation} \label{eq:eotvos}
\tilde{\eta} \approx 10^{-3}\zeta^2\,,
\end{equation}
therefore, local experimental constraints on the former can be used to constrain the latter. Finally, the current drift rate of the value of $\alpha$, which can be easily found to be
\begin{equation}
\label{eq: drift}
D\equiv\left(\frac{\dot\alpha}{\alpha}\right)_0 = \mp\zeta H_0\sqrt{3 \Omega_{\phi 0} |1+w_0|}\,,
\end{equation}
with the minus and plus signs corresponding, respectively, to the canonical and phantom cases, can be constrained using laboratory experiments that compare atomic clocks based on transitions with different sensitivities to $\alpha$.

 Notice that our modelling assumes that the variation of alpha is entirely connected with the DE field. This means that the integrals of Eqs.\ (\ref{eq:dalfa},\ref{eq:dalfa2}) will not have any contributions from redshifts where the DE energy density is negligible. For such a reason, the amplitude of the variation will stay within the bounds of local measurement. These are well within the bounds achievable from CMB \citep{Planck:2014ylh} and, therefore, this model will not have any observable implication at higher redshifts.

 On its own, \Euclid, at least with its primary probes, is not able to constrain variations of the fine structure constant, since its observables are only weakly sensitive to it, while it provides precise constraints on the DE parameters. Under the assumption that the $\alpha$ variation is produced by the same scalar field responsible for DE, these parameters enter the definition of $w_\phi(z)$. 

 On the other hand, astrophysical and local $\alpha$ measurements cannot provide a constrain for the coupling $\zeta$, as this is degenerate with the scalar field equation of state. Therefore, combining the two sets of observations can allow to break degeneracies and provide a constrain on the coupling $\zeta$. A recent review of further synergies between these astrophysical and local measurements with cosmological observations is given in \citet{ROPP}.

\subsection{\Euclid forecast}\label{Sec:DEGrowthforecasts}

\subsubsection{\texorpdfstring{$w_0$, $w_a$}{w0wa}}\label{Sec:wowaforecasts}
Spearheading the forecasting efforts of the Euclid Consortium, \citetalias{Euclid:2019clj} presented validated Fisher forecasts for constraints on cosmological parameters from the main \Euclid probes. It describes in detail the associated methodology and recipes, as well as providing the community with reliable numerical codes for \Euclid cosmological forecasts. Two specific settings were considered, namely a pessimistic and an optimistic one, characterized by more or less conservative choices for the maximum redshift probed by \Euclid, the number of tomographic redshift bins, and the maximum multipoles considered for the weak lensing and galaxy clustering power spectra (see end of \cref{Sec:Analysis} for more details).

Assuming a flat geometry for our Universe, the combination of the two main probes of \Euclid, weak lensing and galaxy clustering (the latter being measured from both the spectroscopic and photometric samples), yields $1\,\sigma$ constraints on the DE equation of state that range from $(\Delta w_0, \Delta w_a)=(0.086, 0.27)$ in the pessimistic setting to $(\Delta w_0, \Delta w_a)=(0.047, 0.14)$ in the optimistic one. The associated values of the often quoted `Figure of Merit' (FoM) \citep[FoM,][]{Wang:2008zh} of dark energy (defined as $\sqrt{\mathrm{det}\left(\mathrm{\tilde{F}}_{w_0w_a}\right)}$, where $\mathrm{\tilde{F}}_{w_0w_a}$ is the inverse 
of the $w_0$-$w_a$ submatrix of the inverse Fisher matrix) are 123 and 398, respectively. Adding the information contained in the cross-correlation signal between photometric galaxy clustering and weak lensing significantly improves those constraints: $(\Delta w_0, \Delta w_a, {\rm FoM})=(0.040, 0.17, 377)$ in the pessimistic case and $(\Delta w_0, \Delta w_a, {\rm FoM})=(0.025, 0.092, 1257)$ in the optimistic case. Adding curvature as a free parameter predictably degrades these constraints, with the full combination of \Euclid probes now yielding $(\Delta w_0, \Delta w_a, {\rm FoM})=(0.044, 0.30, 128)$ and $(\Delta w_0, \Delta w_a, {\rm FoM})=(0.026, 0.14, 500)$ in the pessimistic and optimistic cases, respectively.

Further illustrating the benefits of multi-probe approaches in constraining DE properties, \citet{Euclid:2022abc} explored forecasts for the joint analysis of \Euclid and CMB experiments. While the CMB by itself provides limited information on the nature of DE, it acts as a strong lever arm in time by tightly constraining the early Universe and some of the global cosmological properties of the Universe, allowing us to break degeneracies between parameters and thus improving constraints on the DE equation of state. Notably, the sensitivity of the CMB to curvature allows us to essentially cancel the degradation of DE constraints mentioned earlier for \Euclid-only results. In practice, the addition of CMB lensing data from a Simons Observatory-like \citep[SO,][]{Ade_2019_SimonsObs} experiment to (pessimistic) \Euclid probes already increases the FoM of DE by 20\%, while a full \Euclid{}$\times$CMB joint analysis (i.e. adding temperature and polarisation data as well) further doubles the FoM. Overall, depending on the CMB and \Euclid specifications considered, the reduction of the $1\,\sigma$ error bars on DE parameters ranges from 23\% to 58\%.

\subsubsection{\texorpdfstring{$w_0$, $w_a+\gamma$}{w0wagamma}}\label{Sec:gammaforecasts}

Exploring additional extensions to the \lcdm paradigm, both \citetalias{Euclid:2019clj} and \citet{Euclid:2022abc} also considered forecasts on the growth index $\gamma$ (cf. \cref{Sec:Theorygamma}). Unlike the introduction of curvature as an additional degree of freedom, a free growth index does not noticeably worsen the constraints on the DE equation of state. \cref{tab:gamma} summarizes the constraints on DE and $\gamma$ in this scenario, obtained from \Euclid only and its combination with CMB probes. Overall, $\gamma$ constraints from all \Euclid probes are expected to lie close to the percent level, while the addition of CMB data could potentially shrink them by up to a factor of 2. 

\begin{table}
\renewcommand{\arraystretch}{1.1}
\centering
\caption{Extracted from \citet{Euclid:2022abc}: predicted constraints on DE parameters and the growth index $\gamma$ (when the latter is left free) from \Euclid only (pessimistic/optimistic) and its joint analysis with an SO-like CMB experiment. For $w_0$ and $\gamma$, the constraints are expressed as the ratio of marginalized $1\,\sigma$ uncertainties over the fiducial values of the parameters considered, namely $w_0=-1$ and $\gamma=0.55$. For $w_a$, the constraints quoted are directly the marginalized $1\,\sigma$ uncertainties.
}
\begin{tabular}{cccc} 
    \hline
      &  \Euclid & \Euclid $+$ CMB & \Euclid $+$ full \\ 
      &  only & lensing & CMB \\ 
    \hline
     & \multicolumn{3}{c}{Flat $w_0$, $w_a$} \\
    $w_0$    & 0.040/0.025  & 0.036/0.021  &  0.027/0.011 \\
    $w_a$    & 0.17/0.092   & 0.13/0.071   &  0.099/0.048 \\
    \hline
     & \multicolumn{3}{c}{Non-flat $w_0$, $w_a$} \\
    $w_0$    & 0.044/0.026  & 0.037/0.021  &  0.026/0.013 \\
    $w_a$    & 0.30/0.14    & 0.13/0.074   &  0.096/0.049 \\
    \hline
     & \multicolumn{3}{c}{Flat $w_0$, $w_a$ $+\gamma$} \\
    $w_0$    & 0.039/0.021  & 0.038/0.021  &  0.028/0.013 \\
    $w_a$    & 0.14/0.073   & 0.14/0.071   &  0.10/0.054 \\
    $\gamma$ & 0.015/0.0077 & 0.013/0.0070  &  0.0088/0.0058 \\
    \hline
     & \multicolumn{3}{c}{Non-flat $w_0$, $w_a$ $+\gamma$} \\
    $w_0$    & 0.039/0.021  & 0.038/0.021  &  0.027/0.015 \\
    $w_a$    & 0.23/0.092   & 0.14/0.074   &  0.10/0.055 \\
    $\gamma$ & 0.016/0.0086 & 0.013/0.0069  &  0.0088/0.0056 \\
    \hline
\end{tabular}
\label{tab:gamma}
\end{table}

\subsubsection{\texorpdfstring{$w_0$, $w_a$ $+\sum m_\nu$}{w0wanu}}\label{Sec:nuforecasts}

In \citetalias{EUCLID:KP_nu}, the authors present sensitivity forecasts of the \Euclid probes to cosmological parameters in a model with CPL dark energy and massive neutrinos with total mass $\sum m_\nu$, assuming a fiducial value $\sum m_\nu = 60\,$meV. Here we call this model $w_0$, $w_a$ $+\sum m_\nu$. These forecasts can be compared with those for free $w_0,w_a$ and a fixed neutrino mass $\sum m_\nu = 60\,$meV, performed in \citetalias{Euclid:2019clj} or in \cite{EUCLID:2023uep}, and summarized in \cref{Sec:wowaforecasts}.

The baseline forecasts of \citetalias{EUCLID:KP_nu} account for the \Euclid spectroscopic and photometric probes, implemented essentially like in \cref{Sec:Analysis}. Some small differences arise from aspects related to the modelling of massive neutrino effects (see \cref{Sec:Theorynu}), like the implementation of a scale-dependent growth factor or the fact that galaxies are assumed to track the power spectrum of the cold dark matter plus baryonic component instead of the total matter. The main results presented in Ref.~\citetalias{EUCLID:KP_nu} are conservative, since they opt for `pessimistic' settings, with a cut-off multipole at $\ell_\mathrm{max}=1500$ for weak lensing and $\ell_\mathrm{max}=750$ for galaxy clustering in the photometric survey (a range in which marginalisation over baryonic feedback parameters can be omitted), and a cut-off wavenumber at $k_\mathrm{max}=0.25\,h\,$Mpc$^{-1}$ for galaxy clustering in the spectroscopic survey. 
The sensitivities obtained in \citetalias{EUCLID:KP_nu} rely on Monte Carlo Markov chain (MCMC) forecasts, which are more suitable than Fisher matrix forecasts given that varying the neutrino mass introduces non-Gaussianity in posterior distributions.

For the combination of the (pessimistic) photometric probes (including their cross-correlations) and spectroscopic probes, the sensitivity of \Euclid to the CPL dark energy parameters was found to be $(\Delta w_0, \Delta w_a)=(0.04, 0.13)$ at the 68\% confidence level (CL) in the $w_0$, $w_a$ $+\sum m_\nu$ model (see \cref{tab:neutrinos}).
 Roughly the same errors are obtained in the $w_0$, $w_a$ model under the same assumptions.
Similarly, there is no significant degradation in the sensitivity to the summed neutrino mass when varying the dynamical dark energy parameters: the constraints only change from a sensitivity $\sigma(\sum m_\nu)=56\,$meV with $(w_0, w_a)=(-1,0)$ to an upper bound at 95\% CL $\sum m_\nu<220\,$meV when marginalizing over $(w_0, w_a)$.
Thus, the primary \Euclid probes can independently resolve  the physical effects associated with dark energy parameters and the summed neutrino mass.

\subsubsection{\texorpdfstring{$w_0$, $w_a$  $+\sum m_\nu+N_{\rm eff}$}{w0wanu}}\label{Sec:Neffforecasts}

\citetalias{EUCLID:KP_nu} further presents forecasts that include the effective neutrino number (described in \cref{Sec:TheoryNeff}) as a free parameter on top of the eight parameters of the $w_0$, $w_a$ $+\sum m_\nu$ model, with a theoretical prior $N_\mathrm{eff}>3.044$. Even in this case  (see Table~\ref{tab:neutrinos}), the predicted sensitivity of the combined (pessimistic) photometric and spectroscopic probes to dark energy parameters remains stable, with $(\Delta w_0, \Delta w_a)=(0.04, 0.14)$ at the 68\% CL even in the $w_0$, $w_a$ $+\sum m_\nu+N_\mathrm{eff}$ case. 
The sensitivity to $(\sum m_\nu, N_\mathrm{eff})$ is also quite stable when switching from a pure cosmological constant to CPL dark energy, with the upper bound changing from $(N_\mathrm{eff}-3.044)<0.746$ to $(N_\mathrm{eff}-3.044)<0.935$ (95\% CL) assuming a value $N_\mathrm{eff}=3.044$ in the fiducial model.

The marginalized two-dimensional probability contours displayed in Fig.~14 of \citetalias{EUCLID:KP_nu} show explicitly that \Euclid data are able to resolve any degeneracy between the CPL dark energy parameters and the neutrino mass and density parameters. 

\begin{table}
\centering
\caption{Marginalized $1\,\sigma$ errors on a selection of cosmological parameters in $w_0$, $w_a+\sum m_\nu$ and in $w_0$, $w_a+\sum m_\nu+N_{\rm eff}$ from \Euclid probes.}
\resizebox{\columnwidth}{!}{
\begin{tabular}{ccc}
\hline
&$w_0$, $w_a+\sum m_\nu$& $w_0$, $w_a+\sum m_\nu+N_{\rm eff}$\\
\hline
$w_0$&0.04& 0.04\\
$w_a$&0.13&0.14\\
$\sum m_\nu \, [{\rm meV}]$& $<220$ & $<220$\\
$\Delta N_{\rm eff}$ &-& 0.935\\
\hline
\end{tabular} 
}
\label{tab:neutrinos}
\end{table}

\subsubsection{\texorpdfstring{$w_0$, $w_a + \alpha$}{w0waalpha}}\label{Sec:alphaforecasts}

The analysis presented in \citet{Euclid:2021cfn} combines simulated \Euclid data products with astrophysical measurements of the fine-structure constant, $\alpha$, and local experimental constraints. Both currently available data and a simulated data set representative of Extremely Large Telescope (ELT) measurements (expected to be available in the 2030s) are considered, with the former providing a benchmark, allowing us to illustrate the gains achieved by next-generation data.
 
For current data, 319 measurements of $\alpha$ are used, of which 293 come from the analysis of archival data by \citet{Webb} and the remaining 26 are more recent dedicated measurements \citep{ROPP,Cooksey,Welsh,Milakovic}, reaching $z\sim4.18$.
In addition to this, the analysis includes available measurements of the current drift rate of $\alpha$, constrained by local comparison experiments between atomic clocks \citep{Lange} and the recent MICROSCOPE bound on the E\"otv\"os parameter of \citet{Touboul}.

As for future data external to \Euclid, a data set is simulated following the specifications of the high-resolution ultra-stable spectrograph currently known as ANDES \citep{SPIE}, which will operate at the ELT. For simplicity a set of 50 measurements for $\alpha$ is assumed, uniformly spaced in the redshift range $0.7\leq z\leq3.2$, each with an uncertainty of 0.03 parts per million, which is commensurate with the assumptions in \citet{Leite} and the top-level requirements for the instrument \citep{HIRES}. 
More details on the assumptions made for current and future data for variations of $\alpha$ can be found in \citet{Euclid:2021cfn}.

In order to combine these astrophysical measurements of $\alpha$ with \Euclid, the results obtained in \citetalias{Euclid:2019clj} for the combination of \Euclid's primary probes (weak lensing alongside photometric and spectroscopic galaxy clustering) in the CPL model are considered for the latter. These results are obtained through the Fisher matrix approach, which cannot be used with the $\alpha$ data, due to the strong non-Gaussian nature of the posterior distribution \citep[see][]{Calabrese:2013lga}. For this reason, the analysis of \citet{Euclid:2021cfn} relies on an MCMC approach, using the publicly available sampler \texttt{Cobaya} \citep{Torrado:2020dgo}, which exploits a Metropolis--Hastings (MH) algorithm \citep{Lewis:2002ah,Lewis:2013hha}. With this method, the matter density parameter $\Omega_{\rm m,0}$, the Hubble constant $H_0$, the two parameters of the CPL parameterisation $w_0$ and $w_a$, and the coupling parameter $\zeta$ that connects the dynamical dark energy scalar field to the electromagnetic sector are sampled (see \cref{Sec:Theoryalha}), from the Gaussian posterior distribution obtained by \citetalias{Euclid:2019clj} when \Euclid is considered \citep[for more details on this approach and on how \Euclid data are combined with $\alpha$ measurements, we refer the reader to][]{Euclid:2021cfn}.

We show in \cref{tab:alphares} the results obtained combining current and forecast $\alpha$ data with \Euclid results, using both the pessimistic and optimistic settings used in \citetalias{Euclid:2019clj}. The first result one can notice is that when \Euclid information is added to the $\alpha$ data (both in the current and forecast case) the constraint on the coupling $\zeta$ relaxes with respect to the case where \Euclid constraints are not included. As pointed out in \citet{Euclid:2021cfn}, this is due to the fact that \Euclid tightly constrains $w_0$ and $w_a$ around their \lcdm values, where the effect of the coupling on observables disappears.

However, it is important to highlight that the combined \Euclid{}$+\alpha$ data improves constraints on $w_0$ and $w_a$, with the FoM increasing between 3\% and 18\% with respect to the results of \citetalias{Euclid:2019clj}. Such an effect is due to the extra information coming from $\alpha$ data, which now contribute to the constraining power on dark energy parameters due to the coupling with the electromagnetic sector that is assumed in this model.

\begin{table}
\centering
\caption{Mean values and $68\%$ CL bounds obtained using current (top rows) and forecast (bottom rows) $\alpha$ measurements and their combination with \Euclid forecasts. The Hubble constant $H_0$ is measured in units of \protect\kmsMpc.}
\resizebox{\columnwidth}{!}{
\begin{tabular}{cccc} 
\hline \rule{0pt}{2ex}
   & current $\alpha$    & +\Euclid pess      &  +\Euclid opt \\
 \hline \rule{0pt}{2ex}
$\Omega_{\rm m,0}$  & ---                 & $0.3200\pm 0.0035$ & $0.3199\pm 0.0018$\\

$w_0$               & $-0.99\pm 0.48$     & $-1.000\pm 0.036$  & $-1.001\pm 0.023$\\

$w_a$               & ---                 & $0.00\pm 0.15$     & $0.002\pm 0.087$\\

$H_0$              & $< 75.2$            & $67.00\pm 0.36$    & $67.00\pm 0.10$\\

$\zeta\times 10^8$       & $0.1^{+3.5}_{-3.9}$ & $-0.1\pm 8.2$      & $0.4\pm 8.9$\\
\hline \rule{0pt}{2ex}
                    & forecast $\alpha$         & +\Euclid pess      &  +\Euclid opt \\
 \hline \rule{0pt}{2ex}
$\Omega_{\rm m,0}$  & ---                      & $0.3201\pm 0.0037$ & $0.3200\pm 0.0018$\\

$w_0$               & $-1.07\pm 0.48$          & $-0.9998\pm 0.038$ & $-1.000\pm 0.023$\\

$w_a$               & ---                      & $0.00\pm 0.15$     & $0.000\pm 0.085$\\

$H_0$               & ---                      & $67.00\pm 0.33$    & $67.000\pm 0.093$\\

$\zeta\times 10^8$  & $0.1\pm 1.9$             & $-0.1\pm 4.8$      & $-0.1\pm 5.9$\\
\hline
\end{tabular} 
}
\label{tab:alphares}
\end{table}

\section{Scale-independent modifications of gravity}\label{sec:MGscaleind}

In this section, we summarize the results obtained on scale-independent modifications of the gravitational interaction, which include: the gravitational couplings ($\mu, \Sigma$) and the effective field theory of DE and MG approaches discussed in \citetalias{Euclid:2025tpw} and specific models such as JBD; the normal branch of DGP (nDGP) gravity and $K$-mouflage (KM) gravity presented in \cite{Euclid:2023rjj}; and the dark scattering model in \cite{kpth1p6:2023}. We first review the theoretical frameworks in \cref{Sec:MGtheory} and then the forecast results in \cref{sec:MGscaleindforecasts}. Here we present only two models among those analyzed in \citetalias{Euclid:2025tpw}, we refer the interested reader to the main paper.

\subsection{Theory}\label{Sec:MGtheory}

\subsubsection{\texorpdfstring{$\mu$, $\Sigma$}{muSigma}}\label{Sec:TheorymuSigma}

A generic prediction in extensions of GR is the non-equivalence of the gravitational metric potential and the lensing potential. Combinations of weak lensing and galaxy clustering data sets allow \Euclid to probe the lensing and gravitational potentials and therefore constrain alternative models of gravity.
A model independent parameterisation based on gravitational potentials can be adopted, as inspired from scalar-tensor type theories.  

Let us define a spatially flat perturbed Friedmann--Lemaître--Robertson--Walker (FLRW) line element in the Newtonian gauge as
\begin{equation}
\label{eq:perturbed_metric}
 \de s^2 = -(1+2\Psi)\,c^2\,\de t^2 + a^2(t)\,(1-2\Phi)\,\delta_{ij} \, \de x^i\,\de x^j\,,
\end{equation}
where $t$ is the cosmic time, $x^i$ are the spatial coordinates, $a(t)$ is the scale factor, and $\Psi$ and $\Phi$ are the two scalar gravitational potentials.

The behaviour of $\Psi$ and $\Phi$ determines how matter perturbations evolve. In MG models the relations between the two scalar gravitational potentials and matter perturbations are modified. In Fourier space we have the following equations \citep{Silvestri:2013ne},
\begin{align}
-k^2\Psi&=\frac{4\pi\,G_{\rm N}}{c^2} \,a^2\mu(a,k)\left[\bar\rho\Delta+3\left(\bar\rho+\frac{\bar{p}}{c^2}\right)\sigma\right]\,, \label{eq:mu}\\ 
-k^2\left(\Phi+\Psi\right) & = \frac{8\pi\,G_{\rm N}}{c^2}\,a^2\left\{\Sigma(a,k)\left[\bar\rho\Delta+3\left(\bar\rho+\frac{\bar{p}}{c^2}\right)\sigma\right]\right. \nonumber \\
& \quad \left.-\frac{3}{2}\mu(a,k)\left(\bar\rho+\frac{\bar{p}}{c^2}\right)\sigma\right\}\,, \label{eq:sigma}
\end{align}
where $\bar\rho\Delta\equiv\bar\rho\delta+3[aH/(kc^2)](\bar{\rho}+\bar{p}/c^2)v$, with $v$ the velocity potential defined through $v_i=-{\rm i} k_i v/k$, $\delta=\rho/\bar{\rho}-1$ is the energy density contrast (with $\rho$ being the matter density and its background value is denoted with an overbar), $p=\bar{p}+\delta p$ is the pressure, $v_i$ is the peculiar velocity and $\sigma$ is the anisotropic stress. We can further define $\eta (a,k)=\Phi/\Psi$ as the ratio of the two potentials. Then $\mu$, $\eta$ and $\Sigma$ are three phenomenological functions which, in the absence of matter anisotropic stress, are related as follows
\begin{equation}
 \Sigma = \frac{1}{2}\mu(1+\eta)\,.
\end{equation}
In GR, all are identically equal to $1$. 

In \citetalias{Euclid:2025tpw} the Euclid Collaboration will provide forecasts (see \cref{Sec:ForecastsmuSigma}) for the following form of the $\mu$ and $\Sigma$ functions that are scale-independent \citep{DES:2018ufa},
\begin{eqnarray}
\mu(z) = 1 + \mu_0\, \frac{\Omega_{\rm DE}(z)}{\Omega_{\rm DE}(z=0)}\, , \label{eq:plkmu}\\
\Sigma(z) = 1 + \Sigma_0\, \frac{\Omega_{\rm DE}(z)}{\Omega_{\rm DE}(z=0)}\, ,\label{eq:plksigma}
\end{eqnarray}
with $\Omega_{\rm DE}(z)$ being the DE density parameter and $\mu_0$ and $\Sigma_0$ are constants defining the deviation from \lcdm at the present time. 
For the evolution of the background DE density, $\Omega_{\rm DE}(z)$, it is assumed that this follows the \lcdm model.

 The constraints on these parameters are  $\mu_0=-0.07^{+0.19}_{-0.32}$ and $\Sigma_0=0.018^{+0.059}_{-0.048}$ using  \emph{Planck} data \citep{Planck:2018vyg}. \citet{DES:2022ccp} reported instead a bound of $\mu_0=-0.4\pm0.4$ and $\Sigma_0=-0.06^{+0.09}_{-0.10}$, inferred from DES Y3, RSDs, BAOs and  SNIa. 
Forecasts from the SKA Observatory (SKAO), show that $\mu_0$ and $\Sigma_0$ can be constrained, respectively, at the $2.7\%$ and $1.8\%$ levels \citep{Casas:2022vik}.  In \cite{Kirk:2013gfc} constraints on MG parameters with same-sky photometry and spectroscopy have been obtained. 

For the linear theoretical prediction of the matter power spectrum we have used \texttt{MGCAMB} \citep{Wang_2023, Zucca:2019xhg, Hojjati:2011ix, Zhao:2008bn} and \texttt{MGCLASS II} \citep{Sakr:2021ylx}. For the nonlinear matter power spectrum, we made use of the halo model reaction approach for which we used the associated public code \texttt{ReACT} \citep[][]{Cataneo:2018cic, Bose:2020wch, Bose:2022vwi}, which has been validated by comparing with $N$-body simulations for specific models \citep{Cataneo:2018cic,Bose:2021mkz,Carrilho:2021rqo,Bose:2022vwi,Parimbelli:2022pmr,Bose:2024qbw,Atayde:2024tnr} and for phenomenological models \citep{Srinivasan_2021, Srinivasan:2022jcap}. In \citetalias{Euclid:2025tpw}, the nonlinear power spectrum has been computed for two cases in order to take into account the effect of screening mechanisms in specific environments that typically force the nonlinear modifications to GR to become small \citep[for a review, see, e.g.,][]{Koyama:2015vza}. The first case is the un-screened (US) case, where any screening mechanism has been neglected in the nonlinear matter power spectrum. This represents the case where MG effects play an important role on all the scales relevant for this analysis. The second case is the super-screened (SS) case, where any possible modification to GR is screened at all nonlinear scales and thus the matter power spectrum returns to the GR prediction as quickly as possible in the transition from linear to nonlinear scales.

\subsubsection{EFT of DE and MG}\label{Sec:TheoryEFT}

In contrast to the more phenomenological $\mu/\Sigma$ parameterisation discussed above, effective field theory-inspired approaches theoretically construct the underlying DE/MG physics using a small set of fundamental assumptions (on the field content, the symmetries at play, and the order in derivatives one is working to). Different formulations of such EFT descriptions exist \citep[see, e.g.][for alternative parameterisations as well as mappings between them]{Gubitosi:2012hu,Bloomfield:2012ff,Battye:2012eu,Gleyzes:2013ooa,Gleyzes:2014rba,Bellini:2014fua,Lagos:2016wyv,Frusciante:2016xoj,Lagos:2017hdr,Frusciante:2019xia}, most commonly focused on providing a common basis encapsulating very wide classes of scalar-tensor theories and parametrising the evolution of their background and (linear) perturbations. Here we will focus on the EFT description in the so called $\alpha$-basis developed by~\cite{Bellini:2014fua}.

When considering scalar-tensor theories and working up to second order in derivatives in the corresponding equations of motion, the freedom in the cosmological background evolution for such models is encoded in a single free function, the Hubble parameter $H(t)$. At the level of linear perturbations, four functions of time suffice to fully specify the dynamics. They are the effective Planck mass $M$ and its running $\aM$, the kineticity $\aK$ that contributes to the kinetic energy of scalar perturbations, the braiding $\aB$ that quantifies the strength of kinetic mixing between scalar and tensor perturbations, and finally the tensor speed excess $\aT$, which is related to the speed of gravitational waves via $c_{\rm GW}^2/c^2 =1+\aT$. Within the Euclid Collaboration we have focused on a sub-set of these functions \citepalias{Euclid:2025tpw}: $\aK$ and $\aB$. We set $\aT=0$ motivated by tests of $c_{\rm GW}$ from GW170817 \citep{LIGOScientific:2017zic} and $\aM=0$ for simplicity. 

Focusing on Horndeski theories with luminal gravitational wave propagation and no coupling with the Ricci scalar, we have the following Lagrangian describing a scalar $\phi$
\begin{align}
\label{Horndeski_simple}
{\cal L}_{\rm H}=G_2(\phi,X) + G_3(\phi,X)\Box \phi + \frac{1}{2}\MPl^2  R\,. %G_4(\phi)
\end{align}
Here $X = -\tfrac{1}{2}\nabla_\mu \phi \nabla^\mu \phi$ is the scalar kinetic term, the $G_i$ are unknown functions of $\phi$ and $X$, and $G_{i,\phi}$ and $G_{i,X}$ denote the partial derivatives of $G_i$ with respect to these arguments. 

For \cref{Horndeski_simple} we have 
\begin{align}
H^2\MPl^2{\alpha}_{\rm K} &=  2X \left[G_{2X}+2XG_{2XX} -2 G_{3\phi} -2X G_{3\phi X}\right. \nonumber \\ 
& \quad \left.+6\dot{\phi}H\left(G_{3X}+XG_{3XX}\right)\right]\,, \nonumber \\
H^2\MPl^2{\alpha}_{\rm B} &= 2X\dot{\phi}HG_{3X}\,,
\label{alphadef}
\end{align}
where $M_{\rm Pl}^2$ is the Planck mass square. 
The kineticity $\aK$ does not affect constraints on other parameters at leading order at the level of linear perturbations \citep{Bellini:2015xja,Alonso:2016suf,Frusciante:2018jzw}, but it is considered to be fixed for stability requirements. 
In \citetalias{Euclid:2025tpw} the parameterisation considered is 
\begin{equation}
    \alpha_{\rm B}=\alpha_{\rm B,0}\,a\,,
\end{equation}
where $\alpha_{\rm B,0}$ is a constant.

In \citetalias{Euclid:2025tpw} using CMB data from Planck 2018 \citep{Planck:2018vyg}, BAO, RSD and the matter power spectrum from SDSS DR4 luminous red galaxies (LRG) we found that the constraints on $\alpha_{\rm B,0}$ are $\alpha_{\rm B,0} = 0.09^{+0.04}_{-0.05}$.

For the linear predictions, we have used \texttt{EFTCAMB} \citep{Hu:2013twa,Raveri:2014cka} and for the nonlinear matter power spectrum \texttt{ReACT} \citep{Cataneo:2018cic,Bose:2021mkz} with two different treatments for the screening mechanisms, as defined in \cref{Sec:TheorymuSigma}.  In this case, {\tt ReACT} has not been validated against simulations, but we do not expect its accuracy to degrade as this parametrisation still results in linear enhancements of structure growth, similar to other specific validated cases and the $\mu$-parametrisation.

\subsubsection{JBD}\label{Sec:TheoryJBD}

The JBD theory of gravity \citep{Brans:1961sx}, first proposed in the early 1960s as an operational implementation of Mach's principle, is the simplest non-minimally coupled scalar-tensor theory. It is of particular interest on cosmological scales because it can, in principle, capture the large-scale behaviour of more general scalar-tensor theories. The action for the metric, $g_{\mu\nu}$, and scalar field (with dimensions of the inverse of Newton's constant), $\phi$, is given by 
\begin{equation}
 S_{\rm BD} = \int \de^4 x\sqrt{-g}\left[\frac{c^4}{16\pi}\left({\phi}R
-\frac{\omega_{\rm BD}}{\phi}g^{\mu\nu}\partial_\mu\phi\partial_\nu\phi - 2\Lambda\right)+\mathcal{L}_{\rm m}\right]\,, 
\label{JBD1}
\end{equation}
where $R$ is the Ricci scalar, $\omega_{\rm BD}$ is the sole, dimensionless, parameter, and the matter Lagrangian, $\mathcal{L}_{\rm m}$, is minimally coupled to the metric. We have also included a cosmological constant $\Lambda$, mimicking a constant potential for the scalar field.

On a homogeneous and isotropic FLRW background, that is \cref{eq:perturbed_metric} with $\Psi=\Phi=0$, we then have that the background equations are given by
\begin{align}
3 H^{2} &=\, {\frac{8\pi}{c^4}} \frac{\bar{\rho}}{\phi} - 3H \frac{\dot{\phi}}{\phi} + \frac{\omega_{\rm{BD}}}{2}\frac{\dot{\phi}^{2}}{\phi^2} +\frac{\Lambda}{\phi}\,, \label{FRWJBD}\\
2\dot{H} + 3H^{2} &= \, -{\frac{8\pi}{c^4}}\frac{\bar{p}}{\phi} - \frac{\omega_{\rm{BD}}}{2}\frac{\dot{\phi}^{2}}{\phi^2} - 2H \frac{\dot{\phi}}{\phi} - \frac{\ddot{\phi}}{\phi}+\frac{\Lambda}{\phi}\,,
\end{align}
where the dot represents a derivative with respect to coordinate time $t$.

At large scales the impact on lensing (via $\Sigma$) and the relativistic Newton--Poisson equation (via $\mu$) is given by
\begin{eqnarray}
\Sigma = \, \frac{1}{G_{\rm N}\phi}, \quad  
 \mu = \, \frac{4 + 2\omega_{\rm BD}}{3 + 2 \omega_{\rm BD} }\Sigma\,.
\end{eqnarray}

 Current constraints on JBD gravity use combinations of different cosmological data sets and  different parameterisations of $\omega_{\rm BD}$.
In \citet{Avilez:2013dxa} it is found  $\omega_{\rm BD} > 1900$ at 95\% CL with {\em Planck} 2013 data. In \citet{Ballardini:2020iws}, using  a combination of {\em Planck} 2018 and Baryon Oscillation Spectroscopic Survey (BOSS) DR12 data, they obtained $\log_{10}\left(1 + 1/\omega_{\rm BD}\right) < 0.0022$ at 95\% CL. In \citet{Joudaki:2020shz}, with a combination of the {\em Planck} 2018 CMB data, the 3\texttimes2\,pt. statistics combination of the Kilo-Degree Survey (KiDS) and the 2 degree Field gravitational Lens survey (2dFLens) data, the Pantheon supernovae data  and BOSS measurements of the BAO, they showed the following constraints at 95\% CL on $\omega_{\rm BD} > \{1540,\, 160,\, 160,\, 350\}$ which are obtained by considering  different choices of parametrization (or priors): $\left\{\log_{10}\left(1/\omega_{\rm BD}\right),\, \log_{10}\left(1+1/\omega_{\rm BD}\right),\, 1/\omega_{\rm BD},\, 1/\log_{10}\omega_{\rm BD}\right\}$. These are more than order of magnitude weaker \citep{SolaPeracaula:2019zsl,SolaPeracaula:2020vpg,Joudaki:2020shz,Ballardini:2021evv} than those obtained from millisecond pulsar observations \citep{Voisin:2020lqi}.

In \cite{Euclid:2023rjj} we have provided forecasts for this model, which are summarized in \cref{Sec:ForecastsJBD}. For the linear regime of the JBD model we used the Einstein--Boltzmann solver \texttt{hi\_class} \citep{Zumalacarregui:2016pph,Bellini:2019syt}. 
For the nonlinear regime, we used a modified version of \texttt{HMCODE} \citep{Mead:2015yca,Mead:2016zqy,Joudaki:2020shz}, an augmented halo model that can be used to make accurate predictions of the nonlinear matter power spectrum over a wide range of cosmologies.

\subsubsection{\texorpdfstring{$K$}{k}-mouflage}\label{Sec:TheoryKM}

KM models are an extension of the $k$-essence models when the coupling of the scalar field to matter is non-trivial \citep{Babichev_2009,Brax:2014yla,Brax_2016}. 
For KM the screening mechanism happens in regions of space where the gravitational acceleration is large enough, typically larger than an acceleration of the order  of $H_0 c$ \citep{Brax:2014yla}. As a result, the Solar System and the Galactic environment are screened whilst galaxy clusters are subject to the extra force induced by the scalar field \citep{Barreira:2015aea,Brax:2015lra}. This could potentially lead to specific signatures in such models. 

The KM models are characterized by two functions $K(X)$ and $A(\varphi)$, where $K$ is a nonlinear function of the rescaled kinetic term $X=-\frac{1}{2\tilde{M}^4}(\tilde\partial \varphi)^2$ with $\tilde{M}^4$ an energy scale of the order of the dark energy now. The coupling function $A(\varphi)$ differs from unity and measures the coupling between the scalar field and matter. Typically, the action is given by
\begin{equation}
S= \int \de^4 x \sqrt{-\tilde g} \left[ \frac{c^4 \tilde R}{16\pi \tilde{G}_{\rm N}}+ c^2 \tilde{M}^4 K(X) \right] + S_{\rm m}(\psi, g_{\mu\nu})\,,
\end{equation}
where the matter action, $S_{\rm m}$, depends on the matter field $\psi$ and on the Jordan frame metric $g_{\mu\nu}= A^2(\varphi) \tilde g_{\mu\nu}$, with $\tilde g_{\mu\nu}$ the Einstein frame metric. Screening occurs when $K'\gg 1$, where $K' = \de K/\de X$.
This makes the inertia of the scalar field $\varphi$ very large, damping its fluctuations, and   GR is recovered in the limit where $\varphi$ is constant.
Locally, the modification of gravity induces a change of Newton's constant
\begin{equation}
   G_{\rm eff}= \tilde{G}_{\rm N} A^2\left(1+ \frac{2\beta^2}{K'}\right)\,,
\end{equation}
where $\beta=\tilde{M}_{\rm Pl}\,\de\ln A/\de\varphi$. This amplifies gravity in low-density regions, where $K' \simeq 1$, as compared with high-density screened regions such as the Solar System where $K' \gg 1$. 
The cosmological dynamics in KM models can be inferred from the time evolution at the background level of the coupling function $\bar A(a)$ and of the kinetic function $\bar K(a)$, seen as functions of the scale factor. 
The modification of gravity is then governed by the functions
\begin{equation}
\epsilon_1(a)= \frac{2\bar\beta^2}{\bar K'}\,, \quad \epsilon_2(a)= \frac{\de\ln \bar A}{\de\ln a}\,,
\end{equation}
which characterize the deviation of Newton's constant from $\tilde{G}_{\rm N} \bar{A}^2$ and its time drift. In particular, the growth of matter perturbations can be inferred from the behaviour of these functions, since we have \citep{Benevento_2019}
\begin{equation} 
\mu(a)=\bar{A}^2(a) \left[1+ \epsilon_1 (a)\right]\,, \quad \Sigma (a)= \bar A^2(a)\,.
\end{equation}

 Current data show the bonds at 95\% CL on the KM parameter $\epsilon_{2,0}$ to be $-0.04 \leq \epsilon_{2,0} \leq 0$ using CMB temperature, polarisation and lensing, SN Ia, BAO and local measurements of $H_0$ \citep{Benevento_2019}. Instead, the other parameters are unconstrained.

\cite{Euclid:2023rjj} provide forecasts for the KM model, which we summarize in \cref{Sec:ForecastsKM}. For the computation of the predictions in the linear regime,  we used the Einstein--Boltzmann solver \texttt{EFTCAMB} \citep{Benevento_2019} and for the nonlinear matter power spectrum we used an analytical approach that combines 1-loop perturbation theory with a halo model \citep{Valageas:2013gba}.  A full comparison with N-body simulations needs to be performed.

\subsubsection{nDGP}\label{Sec:TheorynDGP}

The DGP model assumes that the Universe is described by a five-dimensional brane and that the matter component is, instead, confined to the four-dimensional surface of the brane, which is described by the Minkowski metric \citep{Dvali:2000hr}. 

This model is described by the action
\begin{equation}
S = \frac{c^4}{16\pi G_5}\int_{\cal M} \de^5 x \sqrt{-\gamma} R_5
+ \int_{\partial {\cal M}} \de^4 x \sqrt{-g} 
\left[\frac{c^4}{16\pi G_{\rm N}} R + {\cal L}_{\rm m} \right]\,,
\end{equation}
where $G_5$ and $G_{\rm N}$ are the five- and four-dimensional Newton's constants, respectively, $R_5$ and $R$ the Ricci scalar in five and four dimensions, respectively, $\mathcal{M}$ and $\partial \mathcal{M}$ are the 5D bulk and 4D brane and the matter Lagrangian is denoted by ${\cal L}_{\rm m}$. Gravity on the surface of the brane is represented by the standard four-dimensional Einstein-Hilbert action.
The transition between the five-dimensional modifications and the four-dimensional gravity is given by the cross-over scale 
$r_{\rm c} = G_5/(2 G_{\rm N})$, from which we construct the dimensionless parameter 
$\Omega_{\rm rc} \equiv c^2/(4r_{\rm c}^2H_0^2)$. 
The model we investigated in \cite{Euclid:2023rjj} is the normal branch 
\citep[nDGP,][]{Schmidt:2009sv,Sahni2003} characterized by a \lcdm background.

Rather than solving the perturbations on the brane to investigate the evolution of perturbations, it is helpful to use an effective $3+1$ description based on any two of the functions $\mu$, $\eta$, $\Sigma$. Using, for simplicity, the quasi-static approximation (QSA),  the lensing potential is not modified ($\Sigma = 1$), while for the evolution of density perturbations we find \citep{Lombriser:2013aj}
\begin{equation}\label{eqn:mu}
 \mu(a) = 1 + \frac{1}{3\beta}  \,, \quad 
 \beta(a) \equiv 1+\frac{H}{H_0} \frac{1}{\sqrt{\Omega_{\rm rc}}}   \left(1+\frac{\dot{H}}{3H^2}\right) \,.
\end{equation}

 \cite{2013MNRAS.436...89R}, using measurements of the zeroth- and second-order moments of the correlation function from SDSS DR7 data, found $\Omega_{\rm rc}< 40$ at 95\%. Considering the clustering wedges statistic of the galaxy correlation function and the growth rate values estimated from BOSS DR12 data found $\Omega_{\rm rc}< 0.27$ \citep{Barreira:2016ovx}.

Forecasts with galaxy cluster abundance from the Simons Observatory and galaxy correlation functions from a Dark Energy Spectroscopic Instrument (DESI)-like experiment found $\delta(\Omega_{\rm rc}) \sim$ 0.038 around a fiducial value of 0.25 \citep{Liu:2021weo}.  Forecasts for \Euclid-like constraints on a fiducial $\Omega_{\rm rc} \sim 0.0625$ found a $1\,\sigma$ accuracy of 0.0125 from combining the 3D matter power spectrum and the probability distribution function of the smoothed three-dimensional matter density field probes \citep{Cataneo:2021xlx}.  \cite{Bose:2020wch} also forecast for a Large Synoptic Survey Telescope (LSST)-like survey a $1\,\sigma$ accuracy of 0.08 using cosmic shear alone on a fiducial $\Omega_{\rm rc} \sim 0$.

\subsection{\Euclid forecast}\label{sec:MGscaleindforecasts}
\begin{figure*}[ht!]
 \centering
 \includegraphics[scale=0.6]{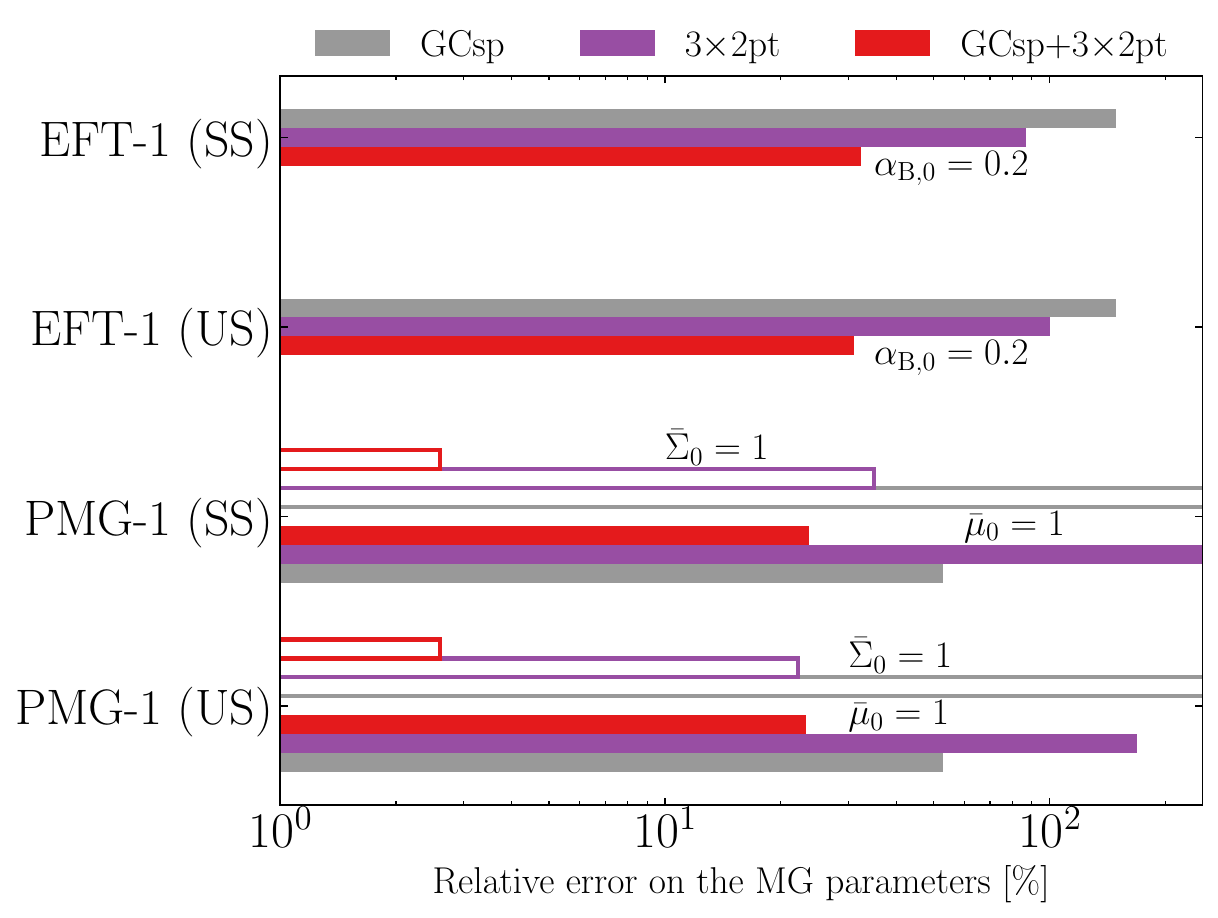}\\
 \caption{ Marginalized $1\,\sigma$ errors on the model parameters relative to their corresponding fiducial values in the optimistic setting, for PMG-1,  and EFT-1 in the US and SS settings. We refer the reader to \cref{Sec:ForecastsmuSigma,Sec:ForecastsEFT} for the values of the free parameters for each case.}
 \label{fig:PMGEFTforecast}
\end{figure*}

\cite{Euclid:2023rjj} forecast the nDGP model and the results are summarized in \cref{Sec:ForecastsDGP}. In this work we have used for the linear predictions a modified version of the Einstein--Boltzmann code \texttt{MGCAMB} and for the nonlinear matter power spectrum we used \texttt{ReACT}.

\subsubsection{\texorpdfstring{$\mu$, $\Sigma$}{muSigma}}\label{Sec:ForecastsmuSigma}

The expected constraints for the $\mu/\Sigma$ parameterisation were obtained in \citetalias{Euclid:2025tpw}, and we show them here in terms of the present-time values of Eqs.~\eqref{eq:plkmu} and \eqref{eq:plksigma}:
\begin{equation}
\bar{\mu}_0\equiv\mu(z=0) = 1+\mu_0\,; \quad 
\bar{\Sigma}_0\equiv\Sigma(z=0) = 1+\Sigma_0\,.
\end{equation}

When considering a fiducial cosmology coinciding with the \lcdm\ limit ($\mu_0=\Sigma_0=0$), it is found that the constraints significantly depend on the degree of confidence one has on the accuracy of the nonlinear prescription used. 
 In \cref{fig:PMGEFTforecast}, we present the relative error on the $\bar{\mu}_0$ and $\bar{\Sigma}_0$ parameters for the different combinations of \Euclid primary probes.
In the conservative case where the scale cuts [implemented following \citetalias{Euclid:2025tpw}] for the photometric and spectroscopic probes, are, respectively, $k_{\rm max}^{\rm 3\times2pt}=0.25\,{\rm Mpc}^{-1}$ and $k_{\rm max}^{\rm GCsp}=0.1\,{\rm Mpc}^{-1}$, the 3\texttimes2\,pt. statistics  does not yield tight constraints, with the relative uncertainty on $\bar{\mu}_0$ and $\bar{\Sigma}_0$ being 278\% (169\%) and 35\% (22\%), when the SS (US) nonlinear prescription is assumed (see \cref{Sec:TheorymuSigma} for the definition of US and SS). While not sensitive to $\Sigma(z)$, \GCsp\ is able to constrain $\bar{\mu}_0$ with a bound of 53\%, and this allows us to break the degeneracy between the MG parameters. Thus, the full combination 3\texttimes2pt+\GCsp\ yields constraints on the present value of $\bar{\mu}_0$ and $\bar{\Sigma}_0$ of 24\% (23\%) and 3\% (3\%), in the SS (US) case, highlighting how crucial is the synergy between \Euclid\ primary probes when investigating this kind of model.

In \citetalias{Euclid:2025tpw}, it is shown how the impact of 3\texttimes2\,pt. statistics on the constraints increases when more optimistic values of $k_{\rm max}^{\rm 3\times2pt}$ are chosen, bringing constraints of the order of 1\% for the two functions in the full 3\texttimes2pt+\GCsp\ combination. However, such an improvement requires ensuring that the modelling of nonlinear scales can reach a high level of accuracy in this regime.

Qualitatively, similar results are found when a fiducial cosmology deviating from \lcdm\ is chosen ($\mu_0=-0.5$ and $\Sigma_0=0.5$). However, here the change in the fiducial values of the MG parameters leads to improved constraints in the 3\texttimes2\,pt. statistics, with constraints on $\bar{\mu}_0$ ($\bar{\Sigma}_0$) being $55\%$ ($47\%$) and $5\%$ ($5\%$) in the SS (US) case, while \GCsp\ yields a bound on $\bar{\mu}_0$ of 103\%. This change in the constraining power expected from \Euclid\ probes can be explained by the changes in the degeneracies between the cosmological and MG parameters, found with this new fiducial cosmology. Such a change also reduces the constraining power of the full combination 3\texttimes2pt+\GCsp, as less complementarity between photometric and spectroscopic probes is found in this case, thus reducing the efficiency of the full combination in breaking degeneracies. However, we still find that 3\texttimes2pt+\GCsp\ improves the bounds on the MG parameters, with constraints on $\bar{\mu}_0$ ($\bar{\Sigma}_0$) being $34\%$ ($36\%$) and $3\%$ ($3\%$) in the SS (US) case. 

Despite this change in the strength of the bounds, the trend of the constraining power with $k_{\rm max}^{\rm 3\times2pt}$ is similar to what is found in a \lcdm\ fiducial cosmology, again showing the need to explore nonlinear scales to significantly improve the results that one can obtain from \Euclid.

 Our results will improve the constraints on these parameters by one order of magnitude with respect to the currently available constraints, e.g. \cite{Planck:2018vyg} and \cite{DES:2022ccp}, while they are comparable to the forecasts from \cite{Casas:2022vik} using a different combination of Stage IV experiments.

\begin{figure*}[ht!]
 \centering
 \includegraphics[scale=0.6]{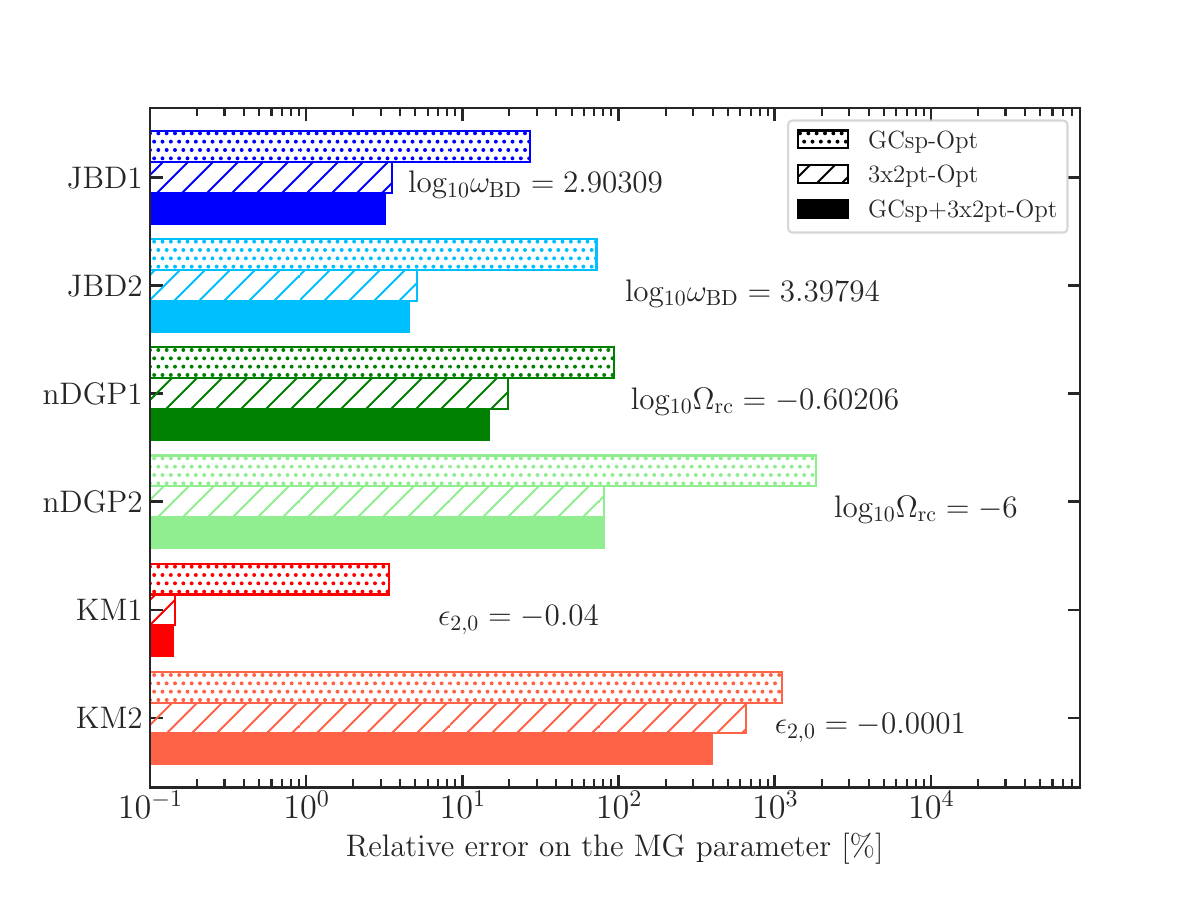}\\
 \caption{Marginalized $1\,\sigma$ errors on the model parameters relative to their corresponding fiducial values in the optimistic setting, for JBD (case 1 in blue and case 2 in cyan), nDGP (case 1 in green and case 2 in light green), and KM (case 1 in red and case 2 in light red). We refer the reader to \cref{Sec:ForecastsJBD,Sec:ForecastsKM,Sec:ForecastsDGP} for the values of the free parameters for each case.}
 \label{fig:JBDKMDGPforecast}
\end{figure*}

\subsubsection{EFT of DE and MG}\label{Sec:ForecastsEFT}

In \citetalias{Euclid:2025tpw} we present the forecast constraints for the EFT model with $\alpha_{\rm B}=\alpha_{\rm B,0}\,a$ on a \lcdm\ background. The fiducial value we considered was $\alpha_{\rm B,0}=0.2$.  In \cref{fig:PMGEFTforecast} we present the relative error on the $\alpha_{\rm B,0}$ parameter for the different combinations of \Euclid primary probes for US and SS.

When examining the standard cosmological parameters, namely $\{\Omega_{\rm m,0}, \Omega_{\rm b,0}, h, n_{\rm s}, \sigma_8\}$, the constraints remain consistent using \GCsp\ alone and comparable to what was found for the $\mu/\Sigma$ parameterisation. The parameter $\sigma_8$ is particularly well-constrained, up to an error of $2.1\%$, effectively aiding in breaking degeneracies between the EFT parameter. For the 3\texttimes2\,pt. statistics case, (except for $\Omega_{\rm m,0}$ for which the error is reduced to half), the constraints are considerably weaker across all cosmological parameters compared to \GCsp. Moreover, the errors remain unchanged between the SS and US cases, both independently and when combining \GCsp\ with the 3\texttimes2\,pt. statistics data.

The specific EFT parameter $\alpha_{\rm B,0}$ is constrained at $148.5\%$ using \GCsp\ alone. When using 3\texttimes2\,pt. statistics data, the constraints are of the same order, namely $86.9\%$ (SS) and $100.6\%$ (US). Combining \GCsp\ and 3\texttimes2\,pt. statistics data significantly reduces the error to $32.3\%$ (SS) and $31.1\%$ (US). In particular, the correlations between $\alpha_{\rm B,0}$ and the cosmological parameters do not change between the SS and US scenarios.
In summary,  the forecast on the EFT model shows a competitive constraining power on both standard cosmological parameters and the EFT parameter $\alpha_{\rm B,0}$, especially when combining data from the photometric and spectroscopic surveys, which effectively reduces errors and breaks parameter degeneracies. 

In \citetalias{Euclid:2025tpw} we also assess how the constraint on $\alpha_{\rm B,0}$ depends on the nonlinear modelling and on the choice of cut-off scales, represented by $k^{\rm 3\times2pt}_{\rm max}$. We found that different cut-offs have a varying impact on the $\alpha_{\rm B,0}$ error between SS and US. At large $k^{\rm 3\times2pt}_{\rm max}$, the photometric probe dominates the constraint, while for smaller $k^{\rm 3\times2pt}_{\rm max}$, the difference between US and SS becomes small, and it effectively disappears at $k_{\rm max} \simeq 0.3\, h\,{\rm Mpc}^{-1}$. Regarding the error on $\sigma_8$, for the US prescription we still found that the constraint is dominated by the photometric probe at large $k^{\rm 3\times2pt}_{\rm max}$. On the other hand, for the SS case, both 3\texttimes2\,pt. statistics alone and the \GCsp+3\texttimes2\,pt. statistics combination provide similar constraints for $k_{\rm max} > 1\, h\,{\rm Mpc}^{-1}$.
 
Furthermore, we have investigated another EFT model whose form is inspired by shift-symmetric Galileon models, even though here we do not discuss the results, but refer the reader to \citetalias{Euclid:2025tpw}.

\subsubsection{JBD}\label{Sec:ForecastsJBD}

%[Pedro, Emilio, Nicola, Daniela, Mario ]

Current constraints from cosmology at 68.3\% CL are of order $\omega_{\rm BD}>10^3$. 
In \cite{Euclid:2023rjj} we focused on fiducial values $\omega_{\rm BD}=800$ (JBD1) and $\omega_{\rm BD}=2500$ (JBD2), which are either similar to current constraints or within current bounds \citep{Ballardini:2020iws,Joudaki:2020shz,Ballardini:2021evv}. In \cref{fig:JBDKMDGPforecast} we present the relative error on the $\omega_{\rm BD}$ parameter for the different combinations of \Euclid primary probes we used in the optimistic setting. We refer the reader to \cite{Euclid:2023rjj} for a full discussion in the pessimistic and quasi-linear settings.

We found that $\omega_{\rm BD}$ in the JBD1 model can be measured with high statistical significance even in the pessimistic case, while this is not true for JBD2. More specifically, \Euclid primary probes will constrain the $\omega_{\rm BD}$ parameter in the JBD1 case to the following precision:
$\omega_{\rm BD}= 800^{+4100}_{-670}$ for \GCsp, $\omega_{\rm BD}=800^{+210}_{-170}$ for 3\texttimes2pt, and $\omega_{\rm BD}=800^{+200}_{-160}$ for 3\texttimes2pt+\GCsp, corresponding to relative errors of $513\%$, $27\%$ and $24\%$, respectively.
In the case JBD2 and for optimistic settings the precision will be: 
$\omega_{\rm BD}=2500^{+1200}_{-820}$ for 3\texttimes2pt and 
$\omega_{\rm BD}=2500^{+1070}_{-750}$ for 3\texttimes2pt+\GCsp\ corresponding to relative errors of $49\%$ and $43\%$, respectively. We note that JBD2 is unconstrained by \GCsp\ alone.  Thus, given current constraints on $\omega_{\rm BD}$, \Euclid could detect the presence of the BD field at a statistically significant level. We note that in our analysis we found an upper $1\,\sigma$ bound, while current constraints can set only lower bonds. The origin of it can be traced back to the Fisher analysis method we employ for which we use  fiducial cosmologies which are not $\Lambda$CDM.

\subsubsection{\texorpdfstring{$K$}{k}-mouflage}\label{Sec:ForecastsKM}

The deviations from \lcdm in the KM models are mainly governed by the parameter $\epsilon_{2,0}$, which is the value of $\epsilon_2$ today. For smooth models, this sets the magnitude of $\bar A$ today, as well as the current value of $\epsilon_1$ (because in the linear regime of scalar field perturbations we have $\epsilon_1=-\epsilon_2$). This determines the deviations from \lcdm both at the background and linear perturbation levels. The current data, including CMB temperature and polarization anisotropies, CMB lensing, SNIa, and BAO from galaxy surveys, provide the lower bound \citep{Benevento_2019} $-0.04 < \epsilon_{2,0}<0$ ($\epsilon_{2,0}$ must be negative for theoretical consistency).

\cite{Euclid:2023rjj} show the forecasts for the fiducial values $\epsilon_{2,0}=-0.04$ (KM1) and $\epsilon_{2,0}=-0.0001$ (KM2).
In case the largest value $\epsilon_{2,0}=-0.04$ is realized in our Universe, it would be detected and measured by \Euclid with a very high accuracy, of $0.14\%$ in the optimistic case for the 3\texttimes2pt+\GCsp\ probe combination, and of $0.15\%$ with 3\texttimes2pt and of $3.4\%$ for \GCsp\ alone. 
In the pessimistic case, these results would only degrade to $0.22\%$, $0.22\%$ and $3.75\%$, respectively. On the other hand, if $\epsilon_{2,0}$ is much smaller (KM2), making the model indistinguishable from \lcdm, \Euclid would be able to provide the upper bound $|\epsilon_{2,0}| < 4 \times 10^{-4}$, for the 3\texttimes2pt+\GCsp{} combination in the optimistic case.  Therefore, \Euclid will improve the constraint on $\epsilon_{2,0}$ by almost two orders of magnitude. This takes advantage of  measurements on small scales and at redshifts $z\gtrsim 1$, where the background shows small deviations from \lcdm. In \cref{fig:JBDKMDGPforecast} we present the relative error on the $\epsilon_{2,0}$ parameter for the different combinations of \Euclid primary probes we used in the optimistic setting. We refer the reader to \cite{Euclid:2023rjj} for a complete discussion in the pessimistic and quasi-linear settings.

\subsubsection{nDGP}\label{Sec:ForecastsDGP}

\cite{Euclid:2023rjj} show the forecasts for the fiducial values $\log_{10}{\Omega_{\rm rc}}=-0.60206$ (nDGP1) and $\log_{10}{\Omega_{\rm rc}=-6}$ (nDGP2). In the optimistic setting, we found that the relative errors for \GCsp, 3\texttimes2pt, and 3\texttimes2pt+\GCsp{} on the parameter $\Omega_{\rm rc}$ in the optimistic scenario for nDGP1 can reach $264\%$, $32\%$ and $24\%$ respectively. For nDGP2 we obtain an upper bound that is $\Omega_{\rm rc} < 0.07$
 which
significantly improves the current constraints. For the pessimistic case $\Omega_{\rm rc}$ is unconstrained, regardless of the probe we consider because the fiducial parameter is very close to \lcdm and we do not have enough nonlinear information, indicating that most of the power of \Euclid in constraining this model would come from its ability to precisely probe these scales. This is further seen when we perform a cut in the maximum multipole and scale at lower values, as it is for the pessimistic (or quasi-linear) scenario, and found larger uncertainties on the model parameter ($\log_{10}\Omega_{\rm rc}$) compared to the optimistic setting while the relative errors on the cosmological parameters only slightly deteriorate. 
We summarize in \cref{fig:JBDKMDGPforecast} the relative error on the $\log_{10}\Omega_{\rm rc}$ parameter for the different combinations of \Euclid primary probes we used in the optimistic setting. We refer the reader to \cite{Euclid:2023rjj} for a complete discussion in the pessimistic and quasi-linear settings.

\section{Scale-dependent modifications of gravity:  \texorpdfstring{$f(R)$}{f(R)} gravity}\label{sec:MGscaledep}

In this section we consider the modification to the gravitational interaction that includes a scale dependence in the growth of structures. We consider the Hu \& Sawicki $f(R)$ model \citep{Hu:2007nk} for which we review the theory in \cref{Sec:TheoryMGscaledep} and the forecasts in \cref{Sec:ForecastMGscaledep}. The analysis considers first the case in which the gravitational interaction is altered by a Lagrangian considering an extra $f(R)$ term beyond the standard simple $R$ form \citep{Euclid:2023tqw}.

\subsection{Theory}\label{Sec:TheoryMGscaledep}

The $f(R)$ theory of gravity \citep{1970MNRAS.150....1B} is characterized by the following action:
\begin{align}
 S = \frac{c^4}{16\pi G_{\rm N}} \int{\de^4 x \sqrt{-g} \left[R+f(R)\right]} \,, \label{eq:EHaction}
\end{align}
where $f(R)$ is a general function of the Ricci scalar, $R$. 
This theory introduces an additional `fifth force’, which is screened at the Solar System scale by the so-called `chameleon mechanism' \citep{Khoury:2003aq}. 

We consider the Hu--Sawicki model \citep{Hu:2007nk} with $n=1$ and in the limit $|f_R|\ll1$ (with $f_R=\de f/\de R$), we have
\begin{equation}
f(R) = - 6 \OmegaDE \frac{H_0^2}{c^2} + |f_{R0}| \frac{\bar{R}_0^2}{R}\,,\label{eq:fR}
\end{equation}
where $f_{R0} < 0$ is the free parameter of the model, $\bar R_0$ is the Ricci scalar evaluated at background at present time, and $\OmegaDE$ is the dark energy density parameter today. \lcdm is recovered in the limit of $f_{R0}\rightarrow0$.

At the perturbation level, considering the phenomenological functions defined in \cref{Sec:TheorymuSigma} and for negligible matter anisotropic stress, one finds \citep{Pogosian:2007sw}
\begin{align}
 \mu(a,k) & = \frac{1}{1+f_R(a)}\frac {1+4k^2a^{-2}m_{f_R}^{-2}(a) }{1+3k^2a^{-2}m_{f_R}^{-2}(a)}\,,\label{eq:mu_fR}
\end{align}
where the characteristic scale of the fifth force is 
\begin{equation}
 m_{f_R}(a) = \frac{H_0}{c\sqrt{2|f_{R0}|}}\frac{\left(4\OmegaDE+\Omegam a^{-3}\right)^{3/2}}{4\OmegaDE+\Omegam}\,,
\end{equation}
and 
\begin{equation}\label{eq:SigmaHS}
 \Sigma(a)=\frac{1}{1+f_{R}(a)}\,.
\end{equation}

 Current constraints use CMB data from \emph{Planck} 2015, BAO, SN Ia from JLA WiggleZ, and CFHTLenS data sets and obtained an upper bound of $|f_{R0}| < 6.3 \times 10^{-4}$ at the 95\% CL. A similar bound is obtained in \citet{2017JCAP...01..005N}, \citet{2013PhRvD..87j3002O}, and \citet{2018PhRvD..97b3525P}. Astrophysical tests are capable of constraining $|f_{R0}|$ at the level of $10^{-7}$ \citep{2019arXiv190803430B}. Furthermore \citet{Desmond2020} constrained $|f_{R0}|<1.4\times 10^{-8}$ using galaxy morphology.

In the work developed by the Euclid Collaboration \citep{Euclid:2023tqw}, the factor $1/(1+f_R)$ is assumed to be unity \citep{Hojjati:2015ojt}, which is a valid approximation for viable values of $f_{R0}$ \citep{Hu:2016zrh,2017JCAP...01..005N,2013PhRvD..87j3002O,2018PhRvD..97b3525P,Lombriser:2014dua,2019arXiv190803430B,Desmond2020}.
Given the scale dependence in the $\mu$ function, this theory is characterized by a scale-dependent growth of structure \citep{Zhang:2005vt,Song:2006ej}. 
In \cite{Euclid:2023tqw} \texttt{MGCAMB} was used for the linear regime while the fitting formula of \citet{Winther:2019mus} was used for the nonlinear power spectrum.

\subsection{\Euclid forecast}
\label{Sec:ForecastMGscaledep}

\citet{Euclid:2023tqw} presented the forecasts for the fiducial values $|f_{R0}| =5\times 10^{-5}, 5\times 10^{-6},  5\times 10^{-7}$.  Since the $f(R)$ model introduces a scale dependence in the growth of perturbations, they used a phenomenological approach in the terms related to baryon acoustic oscillations and redshift-space distortions for the spectroscopic probe, while for weak lensing and the combination of photometric probes, they used a fitting formula that captures the modifications in the nonlinear matter power spectrum \citep[see][for details]{Euclid:2023tqw}.

\begin{figure}[t!]
 \centering
 \includegraphics[width=\linewidth]{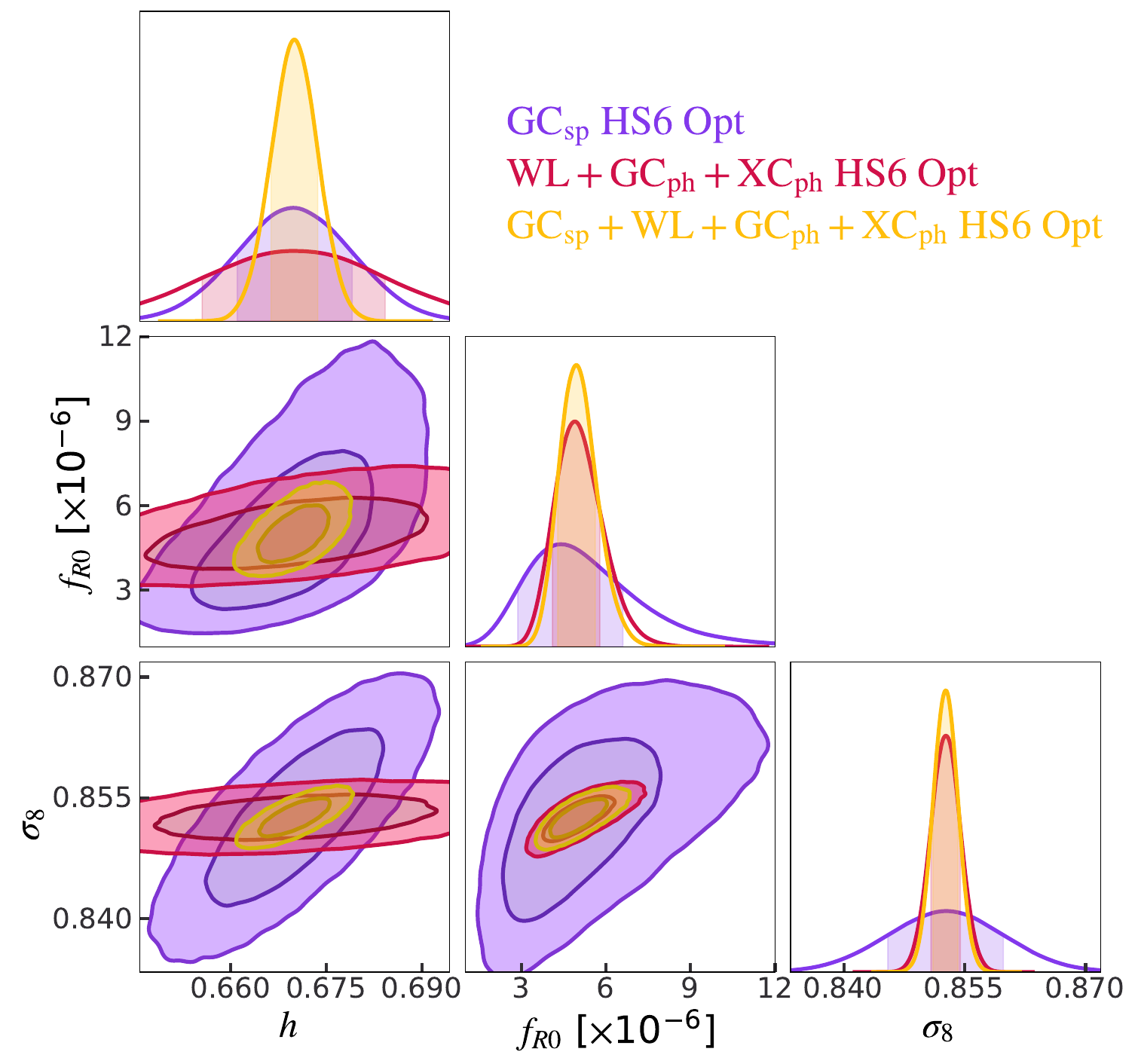}\\
 \caption{Marginalized  $68\%$ and $95\%$ uncertainties on the cosmological parameters $\fr$, $h$, and $\sigma_8$ for a flat $f(R)$ model with $\fr =5\times 10^{-6}$ in the optimistic scenario. Purple is for \GCsp, pink for all the photometric probes including their cross-correlation \XCph, and yellow when this is combined with \GCsp, namely \GCsp+WL+\GCph+\XCph.}
 \label{fig:fofRforecast}
\end{figure}

In \cref{fig:fofRforecast} we show the marginalized posterior contours for the cosmological parameters $\fr$, $h$, and $\sigma_8$ for a flat $f(R)$ model with $\fr =5\times 10^{-6}$ (labelled HS6) in the optimistic scenario. The non-Gaussianity of the forecasted posteriors has been achieved by re-sampling the Fisher matrices, originally computed for the logarithmic parameter $\logfr$, back into the fundamental model parameter $\fr$.
Here we summarize the results only for $ \fr =5\times 10^{-6}$, for which in the optimistic (pessimistic) setting the additional parameter will be constrained: at  $1\,\sigma$ at the 3\% (7\%) level using \GCsp\ alone; at the 1.4\% (3\%) using 3\texttimes2pt; and at the 1\% (1.8\%) level using 3\texttimes2pt+\GCsp.  We found that in the optimistic scenario, \Euclid will be able to distinguish the $f(R)$ model from \lcdm at more than $3\,\sigma$.

\section{Impact of nonlinear methods in clustering and photometric probes } \label{Sec:NLmethods}

In this section we examine the effects of various choices for modelling nonlinear physics in beyond-\lcdm scenarios. The results shown here are drawn from two \Euclid papers, \cite{kpth1p6:2023} and \cite{Euclid:2024xfd} which focus on the spectroscopic and photometric probes, respectively. 
These papers have considered models that offer an improvement over those used in previous forecasts [see for example \cref{eq:GC:pk-ext}] in order to safely probe smaller scales without incurring biases in the inferred cosmology.

\subsection{Spectroscopic galaxy clustering}

Two prominent nonlinear clustering models for spectroscopic galaxy clustering, based on 1-loop perturbation theory, are the effective field theory of large-scale structure \citep[EFTofLSS; see][]{Baumann:2010tm,Carrasco:2012cv} and the Taruya--Nishimichi--Saito (TNS) model \citep{Taruya:2010mx}. Both of these models rely on solving for the perturbative density and velocity kernels up to third order. For concordance cosmology, these kernels can be approximated extremely well by their analytic Einstein--de Sitter (EdS) forms. However, the validity of this approximation is less clear for non-standard models \citep{Bose:2018orj,Aviles:2020wme,Fasiello:2022lff}. 

The EdS approximation is extremely useful because it allows us to make use of invaluable fast Fourier transforms, which significantly speed up the computation of the predictions \citep{McEwen:2016fjn,DAmico:2020kxu}. In the absence of such analytic forms, one must solve for the kernels using other, more computationally expensive and numerically difficult methods \citep{Taruya:2016jdt,Bose:2016qun}, which are largely unfeasible for use in parameter inference analyses on data. 

Beyond the difficulty in calculating the perturbation theory kernels, one must also consider the applicability of other standard modelling approaches for clustering to non-standard cosmologies and gravitational theories. These include the modelling of tracer bias, the effects of massive neutrinos, and the effects of resumming infrared modes. In summary, we would like to consider the following effects on our final predictions for the spectroscopic probe in beyond-$\Lambda$CDM scenarios: 
\begin{enumerate}
    \item 
    impact of perturbation theory kernel approximations;
    \item 
    impact of bias model;
    \item 
    impact of massive neutrino modelling;
    \item 
    impact of infrared-resummation method. 
\end{enumerate}

In \citet{kpth1p6:2023} the authors investigate these by computing the minimum $\chi^2_{\rm red}$ statistic as a function of the maximum scale, $k_{\rm max}$, included in the fit. The $\chi^2_{\rm red}$ is computed by fitting the nuisance parameters of the various modelling approaches to $N$-body measurements of the first two redshift-space clustering multipoles for halos, $P_0$ and $P_2$, at $z=1$. The nuisance parameters include redshift-space distortion and bias degrees of freedom. All cosmological and gravitational parameters are fixed to their fiducial values when performing the fit. The authors further employ a Gaussian covariance that assumes the corresponding simulation volume, which is typically lower than the volume \Euclid will probe. Further tests on mock data assuming a volume closer to that of \Euclid are also performed, and outlined later. 

The $\chi^2_{\rm red}$ statistic for all modelling is then compared to a benchmark case to identify any degradation in accuracy. The benchmark is typically the modelling that uses the exact computation of the perturbative kernels, i.e., including screening effects and not assuming the EdS solutions. Each beyond-$\Lambda$CDM case is also compared to the $\Lambda$CDM case.  

The paper also uses three independent codes to produce clustering predictions: {\tt PBJ}  \citep{Moretti:2023drg}, {\tt PyBird} \citep{DAmico:2020kxu} and {\tt MG-Copter} (\citealp{Bose:2016qun}, see Table~1 of \citealp{kpth1p6:2023}). The former two employ fast Fourier transforms and can compute the EFTofLSS predictions, while the latter employs a numerical algorithm to calculate the perturbative kernels exactly. This makes {\tt MG-Copter} the most versatile but the slowest and most subject to numerical issues. The paper provides a validation of these codes amongst themselves. 

We split the summary of the paper's results into those concerning models that induce a scale-independent modification to the linear growth of structure, and those that induce a scale-dependent modification, as well as general results for both classes of model.

\subsubsection{Scale-independent growth modifications}

The models investigated are $\{w_0,w_a\}$ and nDGP (see \cref{Sec:DEGrowth,sec:MGscaleind}). Additionally, the dark scattering model \citep{Simpson:2010vh, Pourtsidou:2013nha, Pourtsidou:2016ico, Baldi:2014ica, Baldi:2016zom, Bose:2017jjx,Carrilho:2021rqo, kpth1p6:2023} is considered to explore this kind of modification.

For these models, negligible degradation of the EdS approximation is observed for modifications within current data constraints, with the $\chi^2_{\rm red}$ of all approximations being highly comparable, at least below scales where $\chi^2_{\rm red}$ is significantly greater than 1. Further, the standard Eulerian bias expansion \citep{McDonald:2009dh} and the wiggle-no-wiggle resummation method \citep{Baldauf:2015xfa,Vlah:2015sea,delaBella:2017qjy} perform just as well as for $\Lambda$CDM. In the nDGP case, the effects of screening on the perturbative kernels can also be neglected while remaining within the percent-level accuracy of the full solutions. 

It is also found that both EdS and the omission of screening provide $\chi^2_{\rm red} \approx 1$ for $k_{\rm max}\leq 0.3~h \, {\rm Mpc}^{-1}$ when comparing to a mock data vector produced without these approximations and assuming a \Euclid volume in the Gaussian covariance.  Similar results were found in \cite{Barreira:2016ovx}. The same is found when excluding the effects of massive neutrinos in the modeling but fitting to a data vector that assumes a sum of neutrino masses $\sum m_\nu = 0.16~{\rm eV}$. This last statement was shown to be true both in $\Lambda$CDM and in $\{w_0,w_a\}$ models. 
These results suggest that the spectroscopic probe in a single redshift bin, and using only $P_0$ and $P_2$, may offer limited information on Vainshtein-type \citep{Vainshtein:1972sx} screening, as well as massive neutrinos effects. 

 The extraction of cosmological parameters for beyond $\Lambda$CDM models with scale-independent growth modifications using simulated dark matter halo mocks will be the focus in an upcoming Euclid paper, Euclid collaboration: D'Amico et al. in preparation.
 %\textcolor{red}{{D'Amico et al. (2025)}.

\subsubsection{Scale-dependent growth modification: \texorpdfstring{$f(R)$}{fR}}

Hu--Sawicki $f(R)$ gravity (see \cref{sec:MGscaledep}) induces a scale dependency of the linear growth function, as well as higher-order scale dependencies from screening effects. This was shown to lead to a significant break down of both the EdS and omission of screening approximations at the level of the dark matter redshift-space multipoles. The failure of these approximations for $f(R)$ would make it a problematic model to implement in the fast Fourier transform based codes, \texttt{PBJ} and \texttt{PyBird}. 

To solve this, an approximate treatment was developed in \cite{kpth1p6:2023}, which matches the large- and small-scale behaviors of the full kernel solutions, while remaining implementable in a fast Fourier transform formalism. This approximation was called the `$\Lambda$CDM-screened approximation'. It was shown to produce predictions for all dark matter clustering multipoles, including $P_4$, that were within $1\%$ of the exact kernel-based predictions down to $k=0.25~h \, {\rm Mpc}^{-1}$ at $z=1$. The $\Lambda$CDM-screened approximation also performed equally well to the exact solution in terms of the $\chi^2_{\rm red}$ fit to the $N$-body simulations, for $k_{\rm max}\leq 0.3\,h \, {\rm Mpc}^{-1}$. 

This being said, at the level of the halo multipoles, all approximations (EdS, omission of screening, and $\Lambda$CDM-screened) produce comparable $\chi^2_{\rm red}$ to the exact solution for the kernels over all scales considered ($k_{\rm max}\leq 0.3\,h\,{\rm Mpc}^{-1}$). This is observed both when comparing to the $N$-body measurements and mock data produced using the exact solution. This result also suggests that screening information for scale-dependent theories will be hard to extract from a single redshift bin, or using the clustering probe alone. 

It was also found that the phenomenological bias prescription presented in \cite{Song:2015oza} performed equally well to the Eulerian bias expansion that introduces standard scale-independent coefficients for $f(R)$, indicating that the use of a standard bias prescription is sufficient for this theory and will not bias cosmological constraints. 

\subsubsection{General results}

For both scale-independent and scale-dependent models the following results were found.

\begin{itemize}
    \item 
    The standard Eulerian bias expansion \citep[see][for example]{McDonald:2009dh} works well, with all models producing comparable $\chi^2_{\rm red}$ to the $\Lambda$CDM case.
    \item 
    Similarly, the local Lagrangian relation \citep[see][for example]{Sheth:2012fc,Baldauf:2012hs,Saito:2014qha} does not degrade the $\chi^2_{\rm red}$ relative to the $\Lambda$CDM case.
    \item 
    TNS and EFTofLSS both give similar goodness-of-fits to the simulations, with TNS giving a slightly better fit to the simulations over all beyond-$\Lambda$CDM scenarios. The rough range of scales where $\chi^2_{\rm red} \approx 1$ is found to be $0.19 \, h \, {\rm Mpc}^{-1} \leq k_{\rm max} \leq 0.21\, h \, {\rm Mpc}^{-1}$ for both TNS and EFTofLSS, as well as in both $\Lambda$CDM and beyond-$\Lambda$CDM cases.
\end{itemize}

\subsection{Photometric probes}
Unlike spectroscopic galaxy clustering, to predict weak lensing, galaxy clustering and their correlation for photometric probes, we need to model the nonlinear matter power spectrum at higher wavenumbers and perturbative approaches are not applicable. For photometric probes, a key quantity that needs to be modelled is the ratio of the power spectrum in beyond-$\Lambda$CDM models to that in a corresponding $\Lambda$CDM model,
\begin{equation}
\Xi(k, z) = \frac{P(k, z)}{P_{\Lambda {\rm CDM}}(k, z)}\,.
\end{equation}
This quantity is particularly useful if the model has a $\Lambda$CDM background, such as $f(R)$ and nDGP models, since this ratio captures the information of modifications to $\Lambda$CDM in structure formation. The advantage of using this ratio is that we can exploit emulators for $P_{\Lambda {\rm CDM}}$ that are calibrated by high resolution $N$-body simulations. $N$-body simulations in modified gravity models are much more expensive to run and it is difficult to emulate the power spectrum itself. Another advantage of using the ratio is that cosmic variance and numerical inaccuracies (such as low mass and force resolution) are largely cancelled out and we can accurately estimate the ratio with less expensive simulations \citep{Brando:2022gvg}. 

\subsubsection{Scale-dependent growth modification: \texorpdfstring{$f(R)$}{fR}}
 This case was investigated in~\cite{Euclid:2024xfd}. Several semi-analytic predictions and emulators for $\Xi(k, z)$ have been developed in Hu--Sawiki $f(R)$ models.

\paragraph{Fitting formula:} a fitting formula is obtained for $\Xi(k, z)$ in \citet{Winther:2019mus}. This is based mainly on the measurements from ELEPHANT simulations \citep{Cautun:2017tkc} performed by the \texttt{ECOSMOG} \citep{Li:2011vk} code in WMAP9 cosmology. The fitting formula was tested and validated using the DUSTGRAIN simulations suite \citep{Giocoli:2018gqh} using \Planck 2015 cosmology performed by the \texttt{MG-Gadget} code \citep{Puchwein:2013lza}. It was shown that the cosmological dependence of $\Xi(k, z)$ was weak and the formula does not include any dependence on cosmological parameters. \citet{Winther:2019mus} additionally provided the fitting formula for the corrections coming from different $\sigma_8$ and $\Omega_{\rm m,0}$, which are the two main cosmological parameters that affect $\Xi(k, z)$. This nonlinear model for $\Xi(k, z)$ was used in our Fisher forecast.
\paragraph{\texttt{FORGE}:} this emulator is based on simulations performed by the \texttt{MG-Arepo} code \citep{Arnold:2019vpg}. Such simulations explore the cosmological parameter space around the \Planck 2018 cosmology with a Latin hypercube for 50 combinations of $f_{R0}$, $\Omega_{\rm m,0}$, $\sigma_8$, and $h$ with all other parameters fixed. Emulation was performed via Gaussian processes. Note that \texttt{FORGE} emulates the ratio to the analytic \texttt{halofit} power spectrum in a corresponding \lcdm model. In order to obtain $\Xi(k, z)$, it is required to apply a correction calibrated in a \lcdm simulation $P_{\Lambda {\rm CDM}}/P_{\rm halofit}$. This is provided only in the reference cosmology. As such the cosmology dependence in this correction is not taken into account when we construct $\Xi(k, z)$ from the \texttt{FORGE} emulator.
\paragraph{\texttt{mgemu}:} this emulator is based on \texttt{COLA} simulations \citep{Valogiannis:2016ane}, which use an approximate method for screening to increase speed \citep{Ramachandra:2020lue}. These simulations cover $\Omega_{\rm m,0}\, h^2$, $\sigma_8$, $n_{\rm s}$, and $\logten f_{R0}$. Emulations are based on Gaussian processes. Since it is based on approximate simulations, the recommended range of $k$ is limited to $k < 1\, h\,{\rm Mpc}^{-1}$. 
\paragraph{\texttt{e-MANTIS}:} this emulator \citep{Saez-Casares:2023olw} is based on simulations performed by the \texttt{ECOSMOG} code. These simulations sample 3D space spanned by $f_{R0}$, $\Omega_{\rm m,0}$, and $\sigma_8$. The emulation was done using Gaussian processes. In \citet{Saez-Casares:2023olw}, comparisons of $\Xi(k, z)$ with the fitting formula, \texttt{FORGE} and \texttt{mgemu} were shown. For $|f_{R0}| = 10^{-6}$ and $10^{-5}$ at $z=0$, the agreement with the fitting formula is better than 3$\%$ at $k< 10\, h\,{\rm Mpc}^{-1}$, while the agreement with \texttt{mgemu} is at the 1$\%$ level at $k< 1\, h\,{\rm Mpc}^{-1}$. On the other hand, the agreement with \texttt{FORGE} is at the 1$\%$ (3$\%$) level for $|f_{R0}| =10^{-6}\, (10^{-5})$  at $k< 10~h\,{\rm Mpc}^{-1}$.
\paragraph{\texttt{ReACT}:} a halo-model based approach to model the reaction defined as ${\cal R} = P(k)/P^{\rm pseudo}(k)$ where the pseudo-power spectrum is obtained by adjusting the initial conditions in $\Lambda$CDM so that the linear clustering matches that of the target beyond-$\Lambda$CDM cosmology \citep{Cataneo:2018cic}. $P^{\rm pseudo}(k)$ and $P_{\Lambda {\rm CDM}}(k)$ can be predicted using semi-analytic formulae such as \texttt{halofit} or \texttt{HMcode} \citep{Takahashi:2012em,Mead:2020vgs}, and the reaction can be used to predict $\Xi(k, z)$. In $f(R)$ models, a comparison with $N$-body simulations showed that the formalism is 4$\%$ accurate out to at least $k=3\,h\,{\rm Mpc}^{-1}$ for $|f_{R0}|\leq10^{-5}$ \citep{Bose:2021mkz}  in the presence of massive neutrinos. In the absence of massive neutrinos, or a very low neutrino mass, this improves to $2\%$~\citep{Cataneo:2018cic}.

A systematic comparison between these different nonlinear models in terms of \Euclid observables and a study of the effect on parameter constraints due to the use of different nonlinear models are crucial to assess the validity of using these nonlinear models for actual data. It is also important to develop a way to include theoretical errors in the nonlinear prediction.  These issues have been investigated in \cite{Euclid:2024xfd}. The main results of this work are:

\begin{itemize}
    \item 
    A few percent difference in the nonlinear matter power spectrum prescription can bias the inference of the model parameter, $f_{R0}$, at the $2\,\sigma$-level. In the case of the $N$-body-based approaches, differences can arise from both $N$-body accuracy settings and emulation errors. 
    \item 
    Due to its limited accuracy, it was found that the semi-analytic {\tt ReACT} approach can not be applied to the optimistic choice of scale cuts, $\ell_{\rm max}=5000$ for weak lensing alone, and can only hope to be applied in the pessimistic weak lensing only setting of $\ell_{\rm max}=1500$.
    \item 
    Baryonic effects worsen the 1D constraints on $f_{R0}$ due to degeneracies between baryons and modified gravity. 
    \item 
    An implementation of theoretical errors into the covariance can mitigate biases, but come at the cost of significantly worse constraints. Improving the accuracy of the emulators can reduce this cost.  
\end{itemize}

%--------------
\section{Relativistic effects in models beyond \texorpdfstring{$\Lambda$CDM}{LCDM}} \label{Sec:RelEff}

The observable in galaxy clustering, the galaxy number counts, is affected by relativistic effects due to the geodesic motion of photons in a clumpy Universe, which affects the observed redshift, and the observed solid angle within which we count sources.
In linear theory and for any metric theory of gravity, these relativistic effects have been computed for the first time in \citet{yoo_general_2010}, \citet{Challinor:2011bk}, \citet{Bonvin:2011bg}, and \citet{Jeong:2011as}.
In addition to the Kaiser redshift-space distortions (\citealp{Kaiser:1987qv}, see also \citealp{Hamilton_review}), the observed galaxy counts include extra terms such as lensing magnification, Doppler effects, gravitational redshift, Shapiro time-delay and integrated Sachs--Wolfe effects. 

A key objective is to evaluate how relativistic effects generating additional contributions to the observable will influence the analysis of \Euclid data. More specifically, how much they could influence or be degenerate to the parameters of the various DE or MG models that we have considered so far.
In particular, we want to understand: a) whether they can provide additional information and constraining power on a specific theory of gravity or model under consideration; and b) if neglecting them in the analysis will lead to systemic biases in the estimation of cosmological parameters. 
The rest of this section is a summary of the studies carried out within the Euclid Collaboration to address these points. These works focus on the effect of lensing magnification. In \cref{sec:magn_photometric,sec:magn_spectroscopic} we discuss the impact of magnification on the analysis of the photometric and spectroscopic surveys, respectively.
The comprehensive analysis is detailed in \cite{Euclid:2021rez} for the photometric survey and \cite{Euclid:2023qyw} for the spectroscopic survey. A study of relativistic effects in \lcdm for the Euclid spectroscopic will be found in \cite{Euclid:2024ufa}.

\subsection{Impact of lensing magnification in \texorpdfstring{$w_0,w_a$}{w0wa} for the photometric analysis of \Euclid}\label{sec:magn_photometric}

Lensing magnification denotes the contribution to the galaxy counts due to gravitational lensing. Lensing stretches the solid angle of observation, so that the number density of galaxies behind an overdensity appears diluted. Moreover, as a consequence of the conservation of surface brightness, the observed fluxes appear magnified. Since a galaxy survey can only find objects with fluxes above a threshold, the magnified sources can be detected, even if the intrinsic luminosity is below the limit of the survey.
While the amplitude of the first effect is fully determined by the geometry of our Universe, the second effect depends on the properties of the detected objects, and the flux cut adopted for selecting the galaxy catalogue. 
The amplitude of this second effect is regulated by a redshift-dependent function $s(z, F_{\rm lim})$, called the `local counts slope' or `magnification bias', defined as ~\citep{Challinor:2011bk}
\be
\frac{5}{2}s(z, F_{\rm lim}) \equiv - 
\frac{\partial\logten N(z,F> F_{\rm lim})}{\partial\logten F_{\rm lim}}\,,
\label{e:sbias}
\ee
where $N(z,F> F_{\rm lim})$ is the cumulative number of objects in a redshift bin $z$ with fluxes above the threshold $F_{\rm lim}$. 

 Lensing magnification has been extensively studied in the literature~\citep{Menard:2002vz,  Menard:2002da, Menard:2002ia, Matsubara:2004fr, Hui:2007tm}, and it has been detected by cross-correlating background discrete tracers at high redshift, such as quasars or Lyman-break galaxies, with foreground lens galaxies~\citep{Scranton:2005ci, Hildebrandt:2009}. More recently, this effect has also been observed in the cross-correlation of background galaxy shapes with foreground galaxy positions, known as galaxy-galaxy lensing~\citep{Liu:2021gbm}. Magnification has been studied both as an independent probe~\citep{VanWaerbeke:2010yp, Heavens:2011ei} and as a systematic effect that must be accounted for in the analysis of photometric galaxy surveys~\citep{Duncan:2013haa, Cardona:2016qxn, Unruh:2019pyk, LSSTDarkEnergyScience:2021bah}.  
In the context of modified gravity models, its impact on the cosmological analysis of upcoming surveys has been investigated, for example, in \citet{Lorenz:2017iez} and \citet{Villa_2018}. While these works strongly suggest that neglecting magnification would significantly bias the cosmological analysis, the robustness of this statement for specific surveys has to be tested case by case, given the strong dependence of the results on the assumptions adopted for the local count slope $s(z)$. 

An adapted Fisher forecast analysis for the photometric catalogue of the \Euclid survey was carried out in~\citet{Euclid:2021rez}, where for the first time the local count slope has been measured from the \Euclid Flagship simulation \citep{EuclidSkyFlagship}. Here we will report the main findings, focusing on the \Euclid baseline cosmological model: the dynamical dark energy parameterisation $w_0,w_a$.

The recipe used in the forecast follows closely the method adopted in \citetalias{Euclid:2019clj} (pessimistic setting), with the difference that the number of tomographic bins has been chosen to optimize the constraining power of galaxy clustering and galaxy-galaxy lensing \citep[see][]{Pocino2021}. 
Two types of analyses are considered: an analysis containing only galaxy clustering (\GCph); and the probe combination of galaxy clustering, weak lensing, and galaxy-galaxy lensing (\GCph{} + WL + GGL). 
Magnification is included in the recipe at linear order in the lensing expansion, thus it affects only \GCph{} (in both analyses) and $\text{GGL}$ (in the probe combination).
The spectra for the photometric galaxy clustering observable $\Delta$ contain extra contributions, to be added to the recipe outlined in \cref{Sec:Analysis}:
\begin{align}
&C^{\Delta\Delta}_{ij}(\ell) = C^{\rm GG}_{ij}(\ell)+\left(5s_j-2\right)\,C^{\rm G \kappa}_{ij}(\ell)\label{eq:ClDelta}\\
&+\left(5s_i-2\right)\,C^{\kappa \rm G}_{ij}(\ell)+\left(5s_i-2\right)\left(5s_j-2\right)\,C^{\kappa\kappa}_{ij}(\ell)\nonumber\,, 
\end{align}
where $C^{\rm GG}_{ij}(\ell)$ are the spectra used in the standard recipe that neglects magnification, the terms proportional to $C^{\text{G}\kappa}_{ij}$ and $C^{\kappa \rm G}_{ij}$ are the two cross-correlations of the galaxy density and magnification contributions for the two redshift bins $i$ and $j$, $C^{\kappa\kappa}_{ij}$ represents the cross-correlation of the magnification correction for the two photometric bins, and $s_{i/j}$ are the local count slope parameters in the redshift bins $i$ and $j$.
Similarly, the spectra for galaxy-galaxy lensing is corrected to include magnification as follows:
\be
C^{\Delta L}_{ij}(\ell) = C^{\text{G} L}_{ij}(\ell)
+[5s_i-2]\,C^{\kappa L}_{ij}(\ell) \, , \label{eq:crossCl}
\ee
where $C^{\text{G} L}_{ij}(\ell)$ are the cross-spectra of galaxy-galaxy lensing without magnification, and $C^{\kappa L}_{ij}$ is the cross-correlation of the magnification contributions to the galaxy counts and lensing shear. 
The full expression for the extra contributions can be found in \cite{Euclid:2021rez}. 
The observable angular power spectra have been computed using the \texttt{class} code \texttt{v2.9.4} \citep{class2, CLASSgal}.

\subsubsection{Cosmological constraints}

The impact of magnification on cosmological constraints strongly depends on our knowledge of the local count slope. 
Three scenarios are considered: an optimistic scenario, where the local count slope is exactly known; a pessimistic scenario, where we assume no prior knowledge of $s(z)$ and marginalize over its value in each redshift bin; and a realistic scenario, where the local count slope is measured independently of the cosmological analysis with $10\%$ precision, that is we marginalize over the local count slopes in each bin and we add a $10\%$ prior on each parameter. The latter case is considered as `realistic', since $s(z)$ can in principle be measured directly from the luminosity function of the sample. 
For the analysis of \GCph{} alone, we find that including magnification in the analysis leads to a significant information gain for the $w_0,w_a$ parameters; in the optimistic treatment of $s(z)$, the $1\,\sigma$ errors on $w_0,w_a$ are reduced by about $30\%$; in the pessimistic case, this improvement in constraining power is halved to about $15\%$; and in the realistic scenario, the information gain is only partially retrieved and we find that magnification reduces the $1\,\sigma$ error bars by about $20\%$. 
For the joint analysis \GCph{} + WL + GGL, including magnification in the analysis does not affect the cosmological constraints, both in the optimistic and realistic treatment of the local count slope. 
This is due to the fact that magnification and cosmic shear, being both lensing effects, carry similar cosmological information. On the other hand, assuming no prior knowledge on $s(z)$, neglecting magnification will provide constraints slightly too optimistic: including local count slope as free parameter in the analysis increases the $1\,\sigma$ constraints by approximately $15\%$ and $10\%$ on $w_0,w_a$, compared to an analysis without magnification. This is due to the fact that $s(z)$ is degenerate with cosmological parameters and can therefore degrade the constraints if we do not have any a priori knowledge about it.

\subsubsection{Shift in best-fit estimation}
\label{subsec:shift}
The shift in the best-fit estimation in a cosmological analysis that neglects magnification has been studied using the model comparison method described in \cite{Taylor:2006aw}, and widely adopted in the literature. 
The amplitude of magnification can be modulated by a fixed parameter, which is identically zero in the `wrong' model, where magnification is neglected, and which is instead set to $1$ in the `correct' model, where magnification is taken into account. Therefore, `wrong' and `correct' models have a common set of parameters to be measured, and a fixed parameter that differs in the two cases. Since the amplitude of magnification is shifted in the `wrong' model, the maximum of the likelihood will also be shifted, leading to biases in the estimation of the common set of parameters. 

We find that neglecting magnification leads to large biases in the estimation of cosmological parameters, both in the \GCph{} alone analysis and the probe combination \GCph{} + WL + GGL. 
The estimation of $w_0,w_a$, for example, would be highly biased at the level of approximately $1\,\sigma$ (\GCph) and 2--5$\,\sigma$ (\GCph{} + WL + GGL). Since the shifts are estimated using a first-order Taylor expansion of the likelihood around the wrong model, large values of the shifts cannot be quantitatively trusted. Nevertheless, they robustly indicate that it is imperative to model magnification in the analysis of the photometric sample of \Euclid.

\subsection{Impact of lensing magnification beyond \texorpdfstring{\lcdm}{LCDM} in the spectroscopic analysis of \Euclid}\label{sec:magn_spectroscopic}
\newcommand{\ftildebtilde}{\ensuremath{\{\tilde{b}, \tilde{f}}\}}

\cite{Euclid:2023qyw} perform a similar analysis as in \cref{sec:magn_photometric}, but this time applied to \Euclid's spectroscopic galaxy sample. See also \cite{Euclid:2024ufa} where they performed a similar analysis using two-point clustering statistics.

We use the multipoles of the 2-point correlation function (2PCF) as summary statistics. The correlation function in configuration space is more suitable than the Fourier space power spectrum to study non-local effects such as lensing magnification. In fact, lensing effects are only well defined on our past light-cone, while the Fourier transform of a field requires knowing its value on the whole set of equal-time hypersurfaces between the observed galaxies and our telescope \citep{Tansella_2018}. The magnification contribution to the observed Fourier power spectrum can be computed for specific estimators \citep{Castorina:2021xzs}. Nevertheless, it is straightforward to work in configuration space directly. For this reason, our recipe to include magnification in the analysis cannot be directly related to the standard recipe in \cref{Sec:Analysis}. However, the magnification contributions to the observable, similar to the photometric clustering case, can be expressed as the sum of the cross-correlation between galaxy density and magnification, along with a magnification-magnification term. The expressions can be found in \citet{Euclid:2023qyw}.
We employ two approaches in our forecast: the Fisher matrix formalism and the Markov chain Monte Carlo (MCMC) method.
We perform the analysis for two cases: a model-dependent analysis where we consider two specific models (the concordance \lcdm and a dynamical dark energy model); and a model-independent analysis, where we leave as free parameters $b(z)\sigma_8(z)$ and $f(z)\sigma_8(z)$ in each of the redshift bins. Here $b$ denotes the bias and $f$ the growth rate. We call this second analysis the $\ftildebtilde$ parameterisation.

\subsubsection{Cosmological constraints}

We find that magnification has little impact on the cosmological constraints, and the gain or loss of information depends on our prior knowledge of the local count bias parameters. For a $w_0 w_a$CDM model the maximum improvement reaches 11\% for the dynamical dark energy model parameters, assuming perfect knowledge of the local count slope.

\subsubsection{Shift in best-fit estimation}

We estimated the shift in the best-fit parameters following the same procedure described in \cref{subsec:shift}, applied to the multipoles of the 2PCF.
For a $w_0 w_a$CDM dynamical dark energy model, we find shifts below 0.5 standard deviations  for all model parameters, when lensing magnification is not included in the model. In particular, $w_0$ and $w_a$  are only mildly affected, with shifts at the level of $0.1\,\sigma$.
 For the $\ftildebtilde$ parameterisation, the impact is redshift dependent. Due to the combination of lensing magnification being an integrated effect, and the values of the local count slope, we find that the estimation of parameters when neglecting lensing magnification can lead to biases up to $1\,\sigma$ in the estimation of the growth factor in the highest redshift bin.
Since measuring the growth rate is one of the main targets for \Euclid as a test of our theory of gravity, shifts above $1\,\sigma$ cannot be accepted since they would lead to wrong a detection of deviations from general relativity.

\subsubsection{Modelling lensing magnification as a systematic effect}

We also employ a method for modelling lensing magnification as a systematic effect to remove any shifts in best-fit parameter estimates \citep{Martinelli:2021ahc}.
We pre-compute the lensing magnification with fixed cosmological parameters, e.g., determined via CMB experiments and only vary the contributions from density and redshift-space distortions (RSDs) in our analysis.
We find that this results in the shift being reduced to $0.13\,\sigma$ for $w_0w_a$CDM, and $0.1\,\sigma$ for the $\ftildebtilde$ parameterisation.

\section{Non-standard simulations} \label{Sec:NSsim}

This section is devoted to describing how predictions for observable quantities in beyond-$\Lambda $CDM cosmologies can be extended to fully nonlinear scales by means of $N$-body simulations suitably modified to include the extra physics characterizing each specific class of extended models. The numerical methods reviewed here are presented in more detail  in \cite{Euclid:2024jej} and a suite of available simulations developed to test future \Euclid data analysis is described  in \cite{Euclid:2024guw}.  Although the discussion presented here is a general overview on numerical methods, the main classification of models, their corresponding assumptions, and the resulting equations that represent the key modifications of standard $N$-body codes match the suite of available numerical codes and simulations that have been validated and post-processed within the {\em Euclid} collaboration \citep[see again][]{Euclid:2024jej,Euclid:2024guw} and that provide simulated mock observables to test the {\em Euclid} analysis pipeline (see e.g. \cite{Euclid:2024xfd}.

\subsection{Non-standard background evolution}

As discussed above, the simplest extension to the standard \lcdm cosmological scenario for an alternative description of the DE component consists of $w$CDM models, where the DE equation of state takes values different from $w=-1$ and possibly evolving in time as $w(a)$. This class of models includes both parameterized shapes of the equation of state -- such as in the widely-used $w_{0}-w_{a}$ CPL parameterisation -- and more physically motivated models where $w$ evolves according to the underlying dynamics of a new degree of freedom (typically a classical scalar field) playing the role of DE. In all these cases, the DE component is still assumed to be smooth at sub-horizon scales (i.e., with negligible density fluctuations, $\delta_{\rm DE} = 0$), so that its only effect on the cosmological evolution stems from the different background expansion history associated with the specific $w$ shape. Nonetheless, a modified expansion history indirectly impacts on the evolution of density fluctuations and ultimately on the formation of nonlinear structures by modifying the Hubble friction term in the linear perturbation evolution equation, which in Fourier space and in Newtonian gauge reads:
\begin{equation}
\ddot{\delta}_{\rm m} + 2H\dot{\delta}_{\rm m} = 4\pi G_{\rm N} \left(\rho_{\rm m}\,\delta_{\rm m} + \rho_{\rm DE}\,\delta_{\rm DE}\right)\,.
\label{eq:linear_growth}
\end{equation}
Consequently, the linear growth rate and the formation history of collapsed objects is modified with respect to the standard cosmological model.

These effects have to be taken into account consistently in $N$-body simulations by replacing the standard (analytical) calculation of the Hubble function $H(z)$ with the specific expansion history characterising each particular $w$CDM model. While the latter can be computed analytically for simple parameterized shapes of $w(a)$, a full solution of the background dynamics needs to be obtained in the case of scalar field models. This is typically done by numerically integrating the background equations (e.g., by means of suitable modified versions of linear Boltzmann solvers such as \texttt{CAMB} or \texttt{class}) beforehand and then providing $N$-body simulation codes with a tabulated solution of the Hubble function. Once the Hubble function at every timestep is interpolated from the provided pre-computed table, the $N$-body algorithm can be executed following the standard implementation because the DE effects on the gravitational dynamics are already consistently taken into account.

\subsection{Linearized DE perturbations}
While the homogeneity of DE at sub-horizon scales is a common feature of a wide range of models, such an assumption no longer holds for DE models characterized by a low sound speed of the DE fluid ($c_{\rm s}^{2}\ll c^2$) or by non-minimal couplings of the DE field. In the former case, the contribution of the DE density fluctuations as an additional source term of the gravitational potential in \cref{eq:linear_growth} can no longer be neglected and DE perturbations should be modelled in the $N$-body implementation. Nonetheless, in these models the DE contribution to the gravitational potential becomes relevant only at late cosmic epochs. In the latter case, the presence of a direct coupling between the DE degree of freedom and massive particles results in a more complex modification of the dynamics governing the evolution of structures \citep[see][]{Amendola:2003wa} that requires substantial modifications to standard $N$-body algorithms. More specifically, these models generically predict the existence of a fifth-force acting on the coupled matter species, which can be in general expressed as an additional term on the acceleration equation of massive particles besides the standard Newtonian force:
\begin{equation}
    \dot{\vec{v}} = -c^2\vec{\nabla }\Phi -c^{2}\beta(\phi )\vec{\nabla}\delta \phi\,,
\end{equation}
where $\vec{v}$ is the spatial velocity of a mass element, $\Phi$ is the standard Newtonian potential, and $\beta(\phi)$ is the coupling function between the DE scalar degree of freedom $\phi $ and the massive particles. If the interaction involves only dark matter particles while leaving the baryonic component of the Universe uncoupled -- which implies a non-universal coupling of the DE scalar field -- it is generally possible to avoid non-linearities in the scalar field equation of motion and obtain linearized solutions for the perturbations of the scalar field $\delta \phi$ that can be expressed as a function of the Newtonian potential 
\begin{equation}
\delta \phi \approx 2\beta _{i}(\phi)\Phi_{i}\,,    
\end{equation}
where $i$ indicates the matter species involved by the interaction, and thus $\Phi _{i}$ is the Newtonian potential generated by the $i$-th  species. Therefore, the fifth-force can be implemented in $N$-body simulations by appropriately rescaling the standard gravitational accelerations only for the particle types involved in the interaction, while other uncoupled matter species follow the standard Newtonian dynamics. This requires us to implement in the codes a different dynamical evolution for different particle types, such as dark matter and baryons \citep[see e.g.,][]{Baldi:2008ay}.

Besides the fifth-force, the interaction also induces other effects that directly impact on the evolution of perturbations. First of all, the mass of coupled particles will depend on time as a consequence of the evolution of the scalar field. Such an evolution can be computed beforehand by solving the background evolution equations for the scalar field and providing the $N$-body codes with a tabulated evolution of the associated mass variation of particles. In turn, the mass variation of particles induces an additional velocity-dependent acceleration that arises from momentum conservation in the coordinate frame of the other (i.e., minimally coupled) matter species. Such an acceleration term typically takes the form (in comoving coordinates)
\begin{equation}
\vec{a}_{\rm friction} = \beta(\phi) \dot{\phi} \vec{v}\,,
\label{eq:friction}
\end{equation}
and has therefore been named the `friction' or `drag' term. If the background evolution of the field $\phi$ is known, the coefficient $\beta(\phi) \dot{\phi}$ can also be tabulated and provided to an $N$-body code, so that the term in \cref{eq:friction} can be added to the acceleration of every particle \citep[see][]{Baldi:2008ay,Baldi:2012ky}. Note that in the previous equations the scalar field $\phi$ is dimensionless.

A further class of interacting DE models is given by the so-called `dark scattering' scenario, where the DE scalar field and matter interact through the exchange of momentum rather than through the exchange of energy \citep[see][]{Simpson:2010vh}. For these models, the acceleration equation is modified by the presence of a velocity-dependent term similar to the one described in \cref{eq:friction}, but taking the form (still in comoving coordinates)
\begin{equation}
\vec{a}_{\rm scattering} = - (1+w) \frac{3H^{2}\Omega_{\rm DE}}{8\pi G_{\rm N}}\frac{c\sigma}{m_{\rm CDM}}\vec{v}\,,
\label{eq:dark_scattering}
\end{equation}
where $w$ is the DE equation of state parameter, $\sigma$ is the scattering cross section between the DE field and dark matter particles, and $m_{\rm CDM}$ is the dark matter particle mass. Also in this case, as for the friction term arising in coupled DE models, the pre-factor in \cref{eq:dark_scattering} can be pre-computed and tabulated for $N$-body codes to interpolate its value at every time step and compute the additional acceleration \citep[see e.g.,][]{Baldi:2014ica,Baldi:2016zom}.
 
\subsection{Nonlinear scalar field perturbations}

In models where a scalar field interacts universally with matter, screening mechanisms are employed to mitigate this interaction and conform to rigorous local tests of gravity. These screening mechanisms introduce non-linearity into the scalar field equation, which often takes the form of a wave equation:
\begin{equation}
\square \phi = {S}(\phi, \nabla_{\mu} \phi, \nabla_{\mu} \nabla_{\nu} \phi, \rho_{\rm m})\,.
\end{equation}
In this equation $\square \equiv \nabla_{\mu} \nabla^{\mu}$ represents the D'Alembert operator and $S$ is a source function that depends on the matter density, the scalar field, and its derivatives. To address the non-linearity in scalar field equations encountered in cosmological simulations, various methods have been developed  \citep[see e.g.][]{Khoury_2009,Schmidt:2009sg,Li:2011vk, Puchwein:2013lza, Llinares:2013jza,Barreira:2015xvp,Arnold:2019vpg}. For a detailed description of methods and a comparison of various codes we refer to \cite{Winther:2015wla}. These approaches are crucial for ensuring both numerical accuracy and computational efficiency in different scenarios.

In models with non-linearity arising from potential or coupling functions, such as the $f(R)$ gravity model \citep{Oyaizu_2008}, the scalar field equation is adapted to prevent unphysical behaviour. Techniques like redefining the scalar field help maintain numerical stability while avoiding blow-ups in high density regions.
For models characterized by non-linearity in the kinetic terms, for example, in models with the Vainshtein screening mechanism, the equation exhibits non-linearity in the second derivatives of the scalar field \citep[see again][]{Schmidt:2009sg}. To mitigate numerical errors and prevent the emergence of imaginary solutions, a common strategy involves the use of an operator-splitting trick \citep{Chan_2009}. This approach simplifies equations, enhancing code performance, especially in models like the DGP braneworld, cubic, and quartic Galileons \citep{Winther_2015a, Li:2013tda}.

Approximate methods provide alternative solutions for dealing with non-linearity. These include introducing screening factors or linearizing equations. These methods are used in scenarios involving effective mass densities or modified Poisson equations in Fourier space \citep{Khoury_2009,Winther_2015a, Winther:2017jof,Brando:2023fzu}.

\subsection{Quasi-static approximations}

The quasi-static approximation, widely employed in modified-gravity simulations, entails disregarding the time-evolution of scalar field perturbations. This simplification reduces the scalar field equation to a purely spatial partial differential equation, implying that the scalar field configuration at any given time is solely determined by the prevailing matter distribution. This approximation holds true within the sound horizon of the scalar field perturbations \citep{Sawicki:2015zya}. 
While most modified gravity simulations adopt the quasi-static approximation, non-static cosmological simulations have been conducted for various modified gravity models employing alternative approaches, such as the explicit leap-frog method and the implicit Newton--Gauss--Seidel method \citep{Llinares_2013,Bose_2015,Winther_2015b,Christiansen:2023tfy,Christiansen:2024vqv}.

\section{Outlook} \label{Sec:Out}

In this review, we have presented the efforts of the Euclid Collaboration in forecasting beyond-\lcdm models, specifically dark energy and modified gravity models, as a result of the activity carried out by different science working groups. The reviewed analyses considers all, or only part of, the \Euclid primary probes, that is the spectroscopic and photometric galaxy clustering and weak lensing cosmic shear, as well as the cross-correlation between the two photometric probes, which produce the so-called 3\texttimes2\,pt. statistics correlation. 
We have summarized and discussed four aspects that are relevant for preparation to real data.

\paragraph{i) Forecast constraints on cosmological parameters for different cosmological models:} a representative sample of the most popular DE and MG models was investigated, exploring both specific models and parameterized deviations from \lcdm. Depending on the details of the model under investigation, the results achievable with \Euclid will have varying degrees of constraining power, and they will strongly depend on the prescription used to model nonlinear scales, as well as on the possibility to remove systematic effects. Nevertheless, all results obtained highlight the significant improvement that \Euclid will achieve with respect to currently available constraints on these models.  While \Euclid\ has the potential to detect deviations from the standard cosmological models with the approaches we highlighted, we want to stress that such a detection could also be due to the inaccurate modelling of some cosmological or astrophysical effect. For this reason, we dedicate significant efforts to the improvement of the theoretical recipe used for the modelling.
  
\paragraph{ii) Nonlinear clustering  methods:} different theoretical approximations, which are used in \Euclid galaxy clustering analyses and different RSD models, were analysed and compared in four beyond-\lcdm scenarios. The results show that the adopted approximation methods do not require any additional effort in either improving computational efficiency to calculate the exact beyond-\lcdm standard perturbation theory (SPT) kernels or theoretically developing more accurate kernel approximations. 

 \paragraph{iii) Nonlinear lensing methods:} different prescriptions for the nonlinear matter power spectrum in $f(R)$ gravity were analysed in a \Euclid photometric analysis. These included high quality emulators, fitting formulae, and the semi-analytic halo model reaction method. The results show that even in the pessimistic setting, a few percent disagreement between the methods can lead to significantly biased inference of the model parameter. Including baryonic effects can worsen this due to degeneracies. This necessitates an improvement in the accuracy of such methods or the introduction of theoretical errors in the covariance.

\paragraph{iv) Relativistic corrections:} the increase in sensitivity brought by the \Euclid mission and its possibility to access scales that were not probed by previous experiments pose new challenges for the theoretical modelling of the observables. In particular, at the largest scales observable with \Euclid, we expect relativistic corrections to become non-negligible. The Euclid Consortium investigated their impact on the achievable results, focusing in particular on lensing magnification and quantifying its effect on galaxy clustering and its cross-correlation with shear for the photometric sample of \Euclid, as well as on the spectroscopic survey. We found that magnification significantly affects both the best-fit estimation of cosmological parameters and the constraints in the galaxy clustering analysis of the photometric sample. Similarly, it cannot be neglected in the spectroscopic survey since it is crucial to estimate the growth factor. Therefore, for accurate results with upcoming \Euclid data, magnification has to be included in the analysis, or approaches able to mitigate its effect need to be considered, in order to avoid biased results on cosmological parameters.

\paragraph{v) Theoretical predictions at fully nonlinear scales by means of $N$-body simulations:} while SPT methods allow us to obtain theoretical predictions on mildly nonlinear scales, in order to fully exploit the constraining power of \Euclid, photometric surveys require to explore the deeply nonlinear regime. Here, it is crucial to produce accurate $N$-body simulations to understand the behaviour of matter at very small scales in beyond-\lcdm theories. We reviewed the effort of the Euclid Collaboration in producing reliable algorithms to obtain $N$-body simulations that include all modifications brought by non-standard theories, which can be used to extract theoretical predictions at nonlinear scales.

In conclusion, in this review we have summarized available \Euclid forecasts on DE and MG models, highlighting the significant potential of the mission to tightly constrain their parameters, thus possibly ruling out the deviations from \lcdm under investigation, or detecting signatures of a \lcdm failure. In order to achieve such a result a significant effort in the theoretical modelling of the observables is needed, with particular attention paid to the largest and smallest scales that can be accessed by the survey. Here, given the sensitivity of \Euclid, the approaches currently used in the cosmological community might provide inaccurate results and the research lines outlined in this work aim at mitigating such effects.

 While the problem of correctly modelling small scales is already present in $\Lambda$CDM, for extended models this is even more challenging. Given that extensions of standard cosmology affect the growth of perturbations, the parameters of extended models can be degenerate with those controlling nonlinear corrections, such as the nonlinear galaxy bias or the parameters controlling baryonic feedback at even smaller scales. In order to avoid issues with these effects, one can think of exploiting scale cuts to avoid poorly modelled regimes. Even in this case, the scale at which one needs to cut will depend on the investigated model, introducing a further complication in the analysis.

Further efforts on  the topic of this work are still ongoing within the Euclid Collaboration, such as moving from a Fisher matrix approach to more realistic data analysis pipelines, and developing new reliable methods for theoretical predictions at small scales. This effort will be crucial to be ready to fully exploit the first cosmological data release of the \Euclid mission.

\begin{acknowledgements}
\AckECol 
%--
I.S.A. is supported by the FCT PhD fellowship grant with ref. number 2020.07237.BD.
L.A. is supported by FCT  grants UIDB/04434/2020, UIDP/04434/2020 and 2022.11152.BD.
E.B. has received funding from  the Marie Sklodowska-Curie grant agreement No 754496.
G.B. acknowledges support from ASI/INFN grant n. 2021-43-HH.0.
%--
D.B. acknowledges support from the COSMOS network  through the ASI  Grants 2016-24-H.0, 2016-24-H.1-2018 and 2020-9-HH.0.
%--
C.B. acknowledges financial support from the SNSF and ERC grant agreement No.~863929.
%--
B.B. was supported by a UK Research and Innovation Stephen Hawking Fellowship (EP/W005654/2). 
%--
N.F. is supported by MUR through the Rita Levi Montalcini project with ref. PGR19ILFGP. N.F., I.S.A, L.A., F.P., E.M.T.  also acknowledges the FCT project with ref. PTDC/FIS-AST/0054/2021.
%--
F.L. acknowledges financial support from the SNSF. 
%--
L.L. acknowledges support by SNSF Professorship grant (Nos. 170547 \& 202671).
%--
M.M. acknowledges funding by ASI under agreement no. 2018-23-HH.0 and support from INFN/Euclid Sezione di Roma.
%--
C.J.A.P.M. is financed by FCT  project 2022.04048.PTDC and also acknowledges FCT and POCH/FSE (EC) support through Investigador FCT Contract 2021.01214.CEECIND/CP1658/CT0001. 
%--
D.F.M. thanks the Research Council of Norway for their support and the UNINETT Sigma2.
%--
J.N. is supported by an STFC Ernest Rutherford Fellowship (ST/S004572/1). 
%--
F.P. acknowledges  the INFN grant InDark and  MUR, PRIN 2022 `EXSKALIBUR  Grant No.\ 20222BBYB9, CUP C53D2300131 0006, and from the European Union -- Next Generation EU.
%----
Z.S. acknowledges funding from DFG project 456622116 and support from the IRAP and IN2P3 Lyon computing centers.
%----
E.M.T. acknowledges financial support from the ERC grant agreement no. 101076865.
For the purpose of open access, the author(s) has applied a Creative Commons Attribution (CC BY) licence to any Author Accepted Manuscript version arising.

\end{acknowledgements}

\bibliographystyle{aa}
\bibliography{biblio}

\label{LastPage}

\end{document}